\documentclass[aps,prl,twocolumn,amsmath,amssymb,nofootinbib,pdffig]{revtex4-2}

\usepackage{graphicx}
\usepackage{dcolumn}
\usepackage{bm,psfrag}
\usepackage{dsfont}

\usepackage{xcolor} 

\newcommand{\ben}{\begin{equation}}
\newcommand{\een}{\end{equation}}
\newcommand{\be}{\begin{equation}}
\newcommand{\ee}{\end{equation}}
\newcommand{\bea}{\begin{eqnarray}}
\newcommand{\eea}{\end{eqnarray}}
\newcommand{\ba}{\begin{eqnarray}}
\newcommand{\ea}{\end{eqnarray}}

\newcommand{\beq}{\begin{equation}}
\newcommand{\eeq}{\end{equation}}
\newcommand{\beqa}{\begin{eqnarray}}
\newcommand{\eeqa}{\end{eqnarray}}
\newcommand{\beqar}{\begin{eqnarray*}}
\newcommand{\eeqar}{\end{eqnarray*}}

\newcommand{\cO}{{\cal O}}

\definecolor{brown}{RGB}{139,69,19}
\definecolor{darkgreen}{rgb}{0,0.4,0}

\def\t6 {T_\mt{D6}}

\newcommand{\mt}[1]{\textrm{\tiny #1}}

\def\cale         {{\cal E}}

\def\calh         {{\cal H}}

\def\calo         {{\cal O}}

\def\ee           {{\rm e}}

\def\sqr#1#2{{\vcenter{\vbox{\hrule height.#2pt
 \hbox{\vrule width.#2pt height#1pt \kern#1pt
 \vrule width.#2pt}\hrule height.#2pt}}}}

\def\ee{\cale}

\def\aa1{\phi}
\def\cc1{\psi}

\def\vx{\vec{x}}

\def\ben{\begin{equation}}
\def\een{\end{equation}}
\def\bea{\begin{eqnarray}}
\def\eea{\end{eqnarray}}

\usepackage[bookmarks=false]{hyperref} 
\hypersetup{pdfstartview=FitH,pdfhighlight=/O,colorlinks=false}

\begin{document}

\preprint{arXiv:1312.nnnn [hep-th]; UWO-TH-13/xx}

\title{Non-relativistic Floquet Conformal Field Theory}

\author{Diptarka Das$^{1}$, Sumit R. Das,$^{2}$ Arnab Kundu$^{3,4}$ and Krishnendu Sengupta $^5$}
\affiliation{$^1$ Department of Physics, Indian Institute of Technology, Kanpur, UP 208016, INDIA.}
\affiliation{$^2$Department of Physics and Astronomy, University of Kentucky, Lexington, KY 40506, U.S.A.}
\affiliation{$^3$Saha Institute of Nuclear Physics, 1/AF Bidhannagar, Kolkata 700064, INDIA.}
\affiliation{$^4$Homi Bhabha National Institute, Training School Complex, Anushaktinagar, Mumbai 400094, INDIA.}
\affiliation{$^5$School of Physical Sciences, Indian Association for the Cultivation of Science, Jadavpur, Kolkata 700032, INDIA.}


\begin{abstract}
We develop a formalism for studying Floquet dynamics of systems with {\it non-relativistic} conformal invariance in $d$ spatial dimensions. Our analysis indicates the existence of two dynamical phases, hyperbolic and elliptic, separated by a parabolic transition surface. We demonstrate this by studying the fidelity of the driven state and the expectation value of a conformal generator in the many body ground state during the drive. Stroboscopically, they behave exponentially in the hyperbolic phase, show oscillatory behavior in the elliptic phase, and exhibit power-law on the transition surface. Our analysis is completely universal and can be directly applied to several systems including trapped fermions near unitarity and resonant anyons. The former can provide experimental signatures of these dynamical phases. We also comment on a holographic perspective of such driven non-relativistic CFTs and demonstrate that the hyperbolic phase is associated with a timelike stationary-limit surface, such as an ergosphere, in the bulk, while the parabolic phase corresponds to an extremal Killing horizon.
\end{abstract}

\maketitle


{\it Introduction:} Periodically driven strongly correlated quantum systems host a plethora of phenomena that have no equilibrium analogue \cite{rev1,rev2,rev3,rev4,rev5,rev6,rev7,rev8,rev9,rev10,rev11,rev12,rev13}. Generically, a periodic drive results in heating, leading to an infinite temperature steady state \cite{rigol1,rev5,rev11}. However, for a large drive amplitude or frequency, this thermalization is preceded by a long prethermal timescale during which the driven system exhibits approximate emergent symmetries \cite{rev12,rev13} leading to several phenomena that have no equilibrium analogue (see Ref.\ \cite{rev13} and references therein). Some of these phenomena have been observed using ultracold atom systems \cite{rev7,rev9,rev10,bloch1}. 

Usually, it is difficult to treat such driven systems exactly \cite{rev3,rev6,rev8}. A notable exception to this rule, as realized recently, involves systems with relativistic conformal symmetry. These systems are described by  ``Floquet Conformal Field Theories" (FCFTs). Such FCFTs display dynamical phases and corresponding phase transitions that separate them. Initial works on FCFTs were restricted to  $1+1$ dimensional systems \cite{2dpapers1,2dpapers2,2dpapers3,2dpapers4,2dpapers5,2dpapers6,
2dpapers7,2dpapers8,2dpapers9,2dpapers10,2dpapers11,dynph2} which can be viewed as CFT on curved space-time having holographic duals \cite{2dpapers12,2dcurved1,2dcurved2,2dcurved3, holo1,holo2,holo3,holo4,holo5}. In \cite{hd1,hd2} it was shown that FCFTs in two and three spatial dimensions, can be studied using quaternionic representations of the conformal algebra, leading to a richer dynamical phase structure. The $1+1$ dimensional results have received recent confirmation in a digital quantum processor \cite{experiments}. However to the best of our knowledge such relativistic FCFTs have not been realized in laboratory experiments. This motivates us to study Floquet dynamics in non-relativistic systems which are more readily experimentally realizable.

Such non-relativistic CFTs can be applied to a wide variety of systems. These include $n$ interacting fermions at unitarity, where a contact four-Fermi interaction can be tuned to a critical value in $d=2+\epsilon$ \cite{Mehen,werner,son1,son2,son3, Boisvert:2025hex, fermirev1, fermirev2, fermirev3, fermirev4, moroz1, maki2019, maki2020, maki2024, deng1,deng2}; this system can be experimentally realized  using ultracold fermions in a trap and near a Feshbach resonance \cite{fermirev1,fermirev2,fermirev3,fermirev4,feshrev}. Other interesting setups involve breathing modes and dynamics of two-dimensional (2D) Bose gases \cite{pitaevski1,dalibard1}, Tonks-Girardeau gases \cite{tonks1}, Calogero model \cite{calogero1} and anyons with contact interactions which can also be tuned to achieve conformality \cite{manuel,chou, bergman}. Moreover, such CFTs may also describe properties of low-energy neutrons in nuclear matter \cite{nuc1,son3} and have gravity duals with two additional non-compact dimensions \cite{dual1,dual2}. 

In this letter, we initiate a study of Floquet dynamics in systems with {\em non-relativistic} conformal invariance in $d$ space dimensions. Our analysis is based entirely on conformal symmetry and is independent of a particular microscopic realization.  We derive its consequences—including exact many-body fidelity and experimentally accessible phase diagnostics. We also provide a holographic geometric classification which is not transparent in earlier system-specific treatments \cite{tonks1, moroz1,scopa1,deng1,maki2019,deng2,maki2020,maki2024,pra1,pra2}.

As in relativistic CFTs, we will first consider a periodic drive with a square pulse protocol with time period $T=2 \pi/\omega_D$, where $\omega_D$ is the drive frequency. Each period contains steps $T_1$ and $T_2= T-T_1$ with Hamiltonians ${\cal H}_1$ and ${\cal H}_2$ respectively. Each of these Hamiltonians is a linear combination of the generators of a $SO(2,1)$ subalgebra of the full conformal group and chosen to be bounded from below.  When the initial state for such a protocol is a CFT primary, we use the $SO(2,1)$ symmetry to calculate the time dependence of the fidelity  of the driven state and the  one-point functions (Heisenberg picture) exactly. 

The stroboscopic response of all these quantities reveals three distinct dynamical phases for both square pulse and continuous protocols, characterized by the types of conjugacy class of the Floquet Hamiltonian. There is a hyperbolic phase where the fidelity decays exponentially while the one-point functions grow exponentially as a function of the number of time periods. There is an elliptic phase, where they oscillate in the stroboscopic time. Finally, these two phases are separated by a parabolic critical transition surface, where a power-law decay (growth) is observed. The system can be driven between these phases by tuning the parameters in ${\cal H}_1, {\cal H}_2$ and $T_1,T_2$. Numerical calculations with protocols with continuous time dependence lead to the same dynamical phase diagram. 

Floquet dynamics in ultracold trapped fermions at unitarity \cite{moroz1} and 1D hard-core Bose gas \cite{scopa1} have been studied earlier using Ermakov-Pinney equations \cite{Pinney1950, Lewis1968} (see Supplementary Material (SM) \cite{sicit} for a general discussion); similar studies for quench and ramp dynamics have also been carried out \cite{maki2019,maki2020,maki2024,deng1,deng2,dalibard1}. In contrast to these, our work provides a representation-independent framework using conformal symmetry within which the earlier results can be recovered as special cases. Moreover, the results obtained are consistent with existing experiments in unitary fermions and provide new predictions for Floquet dynamics for such systems. They might also motivate experiments in other systems with conformal symmetry.

Our analysis also indicates a novel holographic realization of the dynamical phases, in contrast to relativistic CFTs. For the latter, there exist multiple $SO(2,1)$ subalgebras within the conformal algebra, whereas for the non-relativistic case, there is a unique $SO(2,1)$ subalgebra. For the non-relativistic case, the bulk Killing vector corresponding to the Floquet Hamiltonian is constructed as an appropriate linear combination of the corresponding generators of the unique $SO(2,1)$-algebra. In the hyperbolic class, this Killing vector gives rise to a timelike stationary-limit surface (equivalent to an ergo-surface where the Killing norm vanishes, but the surface remains timelike) in the bulk. The elliptic conjugacy class, on the other hand, corresponds to a globally non-degenerate timelike Killing vector; whereas in the parabolic class, the vanishing Killing norm bulk surface becomes an extremal horizon. This structure is markedly different from what emerges in the relativistic case, based on a particular given $SO(2,1)$ subalgebra, see {\it e.g.}~\cite{hd2}.

{\it Non-relativistic Conformal Symmetry}: In $d$ space and one time dimensions, the non-relativistic conformal algebra consists of the standard Hamiltonian $H$, the special conformal transformation $C$, dilatation $D$, momenta $P_i$ and boosts $K_i$, together with a number operator $N$ which is a central term commuting with all generators \cite{son2, Boisvert:2025hex}. In this paper, we are interested in a $SO(2,1)$ subalgebra obeyed by $H,D,C$,
\ben
[H,C] = -iD \ , ~~~[D,C] = -2iC \ , ~~~~~[D,H]=2iH \ .
\label{2-1}
\een
A primary operator $\cO_{\Delta,n}$ for this conformal algebra commutes with $C, K_i$, has a scaling dimension $\Delta$ given by $[D, \cO_{\Delta,n}] = i \Delta \cO_{\Delta,n}$, and a number $n$ is given by $[N , \cO_{\Delta, n}] = n \cO_{\Delta,n}$ \cite{son2, Boisvert:2025hex}. The details of transformation of these primary operators under a finite transformation of the $SO(2, 1)$ group are discussed in the SM \cite{sicit}.

The invariant vacuum corresponding to this conformal theory $|0 \rangle$ obeys $D | 0 \rangle = C | 0 \rangle = N | 0 \rangle = 0$. A general primary state $|\Delta\rangle_{\omega,n}$, for a real parameter $\omega$, is then constructed as $|\Delta\rangle_{\omega,n} = e^{-H/\omega} \cO^\dagger_{\Delta,n} |0 \rangle$. Since $n$ is a fixed number, we will omit the subscript $n$ in the following. Defining the operators
\ben
L_{\pm,\omega} \equiv \frac{1}{2\omega} (H-\omega^2C \pm i\omega D )~~~L_{0,\omega} \equiv \frac{1}{2\omega} (H + \omega^2 C),
\label{2-11}
\een
for any real $\omega$, we find that a primary state $|\Delta\rangle_\omega$ obey $L_-|\Delta \rangle_\omega = {_\omega}\langle \Delta | L_+ = 0, L_0 |\Delta \rangle_\omega = \frac{\Delta}{2}|\Delta \rangle_\omega $. This is therefore an eigenstate of the Hamiltonian $H_{osc} = H + \omega^2 C$ with eigenvalue $\omega \Delta$. This fact will be crucial in the subsequent study of dynamics. Finally, we will need the expectation values given by $~~_\omega \langle \Delta | D |\Delta \rangle_\omega = 0$ and 
\bea
_\omega \langle \Delta | H |\Delta \rangle_\omega &=& \omega^2\,  _\omega \langle \Delta | C |\Delta \rangle_\omega = \Delta \omega/2,  \label{2-16}
\eea

{\it Periodic drive}: We will begin by studying a periodic square-pulse drive, starting from $t=0$, given by
\bea
{\cal H} &= &{\cal H}_1 = H + \mu_1^2 C \equiv {\cal H}_1~~~~\ell T \leq t \leq \ell T+T_1  \label{2-6} \\
{\cal H} &= & {\cal H}_2 = H + \mu_2^2 C \equiv {\cal H}_2~~~~\ell T+T_1\leq t \leq (\ell+1)T,  \nonumber
\eea
where $\ell$ is zero or a positive integer representing the number of periods. Since the Hamiltonians  ${\cal H}_i $ are operators in the $SO(2,1)$ algebra, the time evolution operator can always be expressed as exponential of a linear combination of $H$, $C$ and $D$; this follows entirely from the conformal algebra. 

The time evolution over a single time period is then governed by the Floquet Hamiltonian ${\cal H}_F$, which is defined as $U_T = e^{-i\calh_FT} \equiv e^{-i{\cal H}_2 T_2}e^{-i{\cal H}_1 T_1}$. $H_F$ depends only on the $SO(2,1)$ algebra and can be therefore calculated by using a Pauli matrix representation of the generators given by  $D=i\sigma_z, H = \frac{i}{2}(\sigma_x + i\sigma_y), C = -\frac{i}{2}(\sigma_x - i \sigma_y)$. In this representation, we find
\ben
\calh_F  =  \alpha_+ H + \gamma D +  \alpha_- C \label{2-7}
\een
Here, the parameters are
\bea
\alpha_+ &= & \frac{\rho}{T \sin \rho} \left[ {\mathcal A}_1/ \mu_2 +  {\mathcal A_2}/\mu_1 \right] , \nonumber \\
\alpha_-  & = &  \frac{\rho}{T \sin \rho} \left[\mu_2 {\mathcal A}_1 +  \mu_1 {\mathcal A}_2 \right] , \nonumber \\
\gamma  & = &  \frac{\rho}{T \sin \rho} \frac{\mu_2^2-\mu_1^2}{2 \mu_1 \mu_2}\sin (\mu_1 T_1)\sin (\mu_2 T_2)  , \nonumber \\
\cos (\rho)  & = &  \cos (\mu_1T_1)\cos (\mu_2 T_2) \nonumber\\
&& - \frac{\sin (\mu_1 T_1) \sin (\mu_2 T_2)(\mu_1^2 + \mu_2^2)}{2 \mu_1\mu_2}, 
\label{2-8}
\eea
where ${\mathcal A}_1= \cos (\mu_1 T_1) \sin (\mu_2 T_2)$ and ${\mathcal A}_2= \cos (\mu_2 T_2) \sin (\mu_1 T_1)$.
From these definitions it follows that
\ben
\rho^2 = T^2 \left( \alpha_+\alpha_- - \gamma^2\right) \equiv T^2 \kappa^2.
\label{2-10}
\een
As we shall see in the following discussion, $\rho$ will play a central role in determining the dynamical phases of the driven system. 

Finally, in the Pauli matrix representation, the time evolution operator for $\ell$ time periods is given by
\bea
U(\ell T) &=& e^{-i\calh_F \ell T} \sim \left( \begin{array}{cc}
a_\ell & b_\ell \\
c_\ell & d_\ell \end{array} \right)
\label{2-20} 
\eea
where $a_\ell ,b_\ell, c_\ell, d_\ell$ are given by
\bea
a_\ell & = & \cos (\rho \ell ) + \frac{T\gamma}{\rho}\sin (\rho\ell )~~~~b_\ell = \frac{T\alpha_+}{\rho}\sin(\rho \ell )  \label{2-5}\\
c_\ell & = & -\frac{T\alpha_-}{\rho}\sin(\rho \ell )~~~~d_\ell =\cos (\rho\ell ) -\frac{T\gamma}{\rho}\sin (\rho\ell )\nonumber
\eea
As we show in the SM \cite{sicit}, the same coefficients appear in the transformation properties of a local primary operator $\calo'_{\Delta,n}(\vx,t)= U^{-1}(\ell T)\calo_{\Delta,n}(t,\vx)U (\ell T)$:  
\bea
\calo'_{\Delta,n}(\vx,t) &=& \frac{e^{-\frac{in c_\ell \vx^2}{2(c_\ell t +d_\ell)}}}
{(c_\ell t +d_\ell)^\Delta}  \calo_{\Delta,n} (\frac{a_\ell t +b_\ell}{c_\ell t +d_\ell}, \frac{\vx}{c_\ell t +d_\ell})
\eea
The dependence on stroboscopic time $\ell$ is completely contained in the coefficients $a_\ell \cdots d_\ell$.

\begin{figure}
\rotatebox{0}{\includegraphics*[width= 0.48 \linewidth]{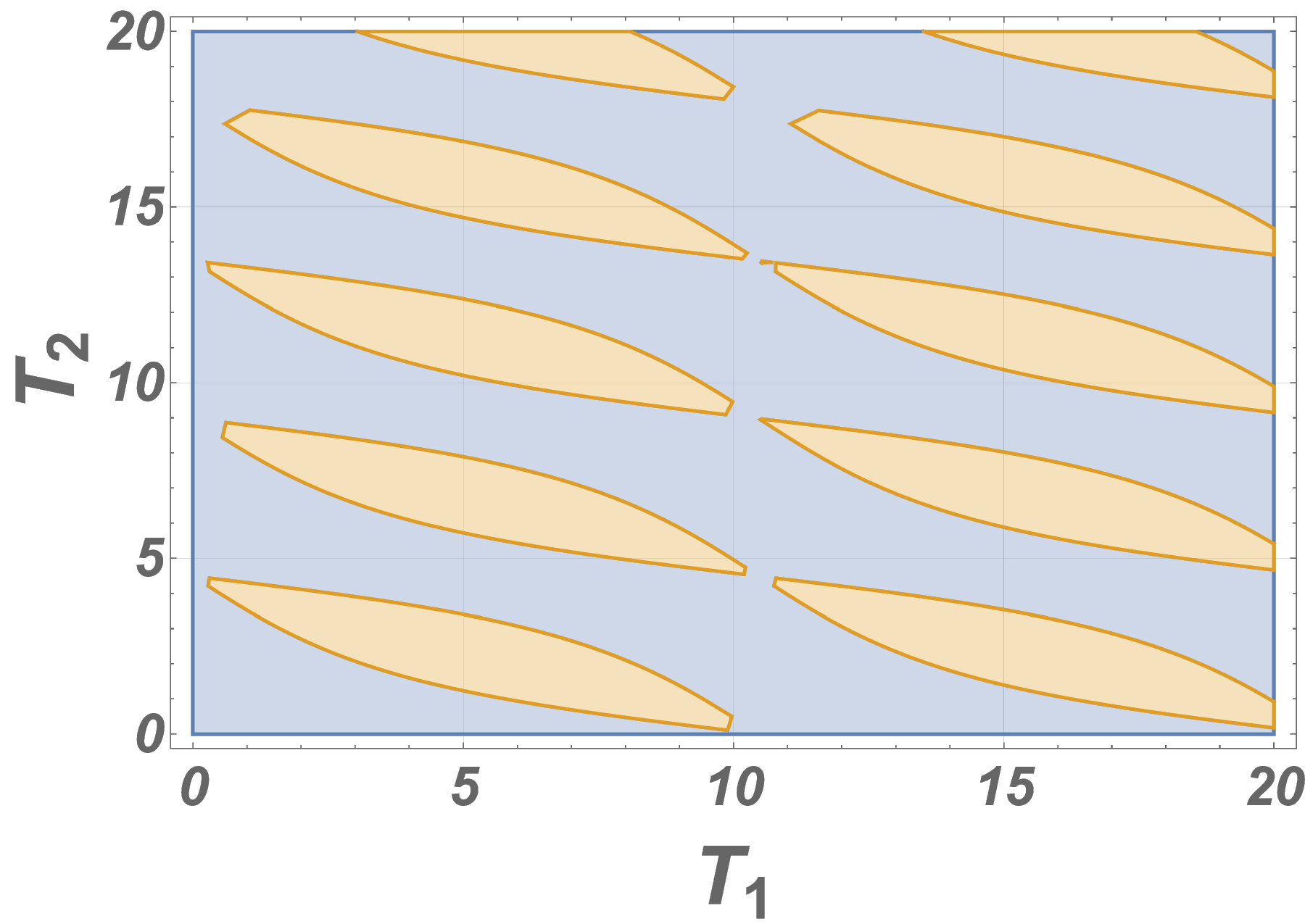}}
\rotatebox{0}{\includegraphics*[width= 0.48 \linewidth]{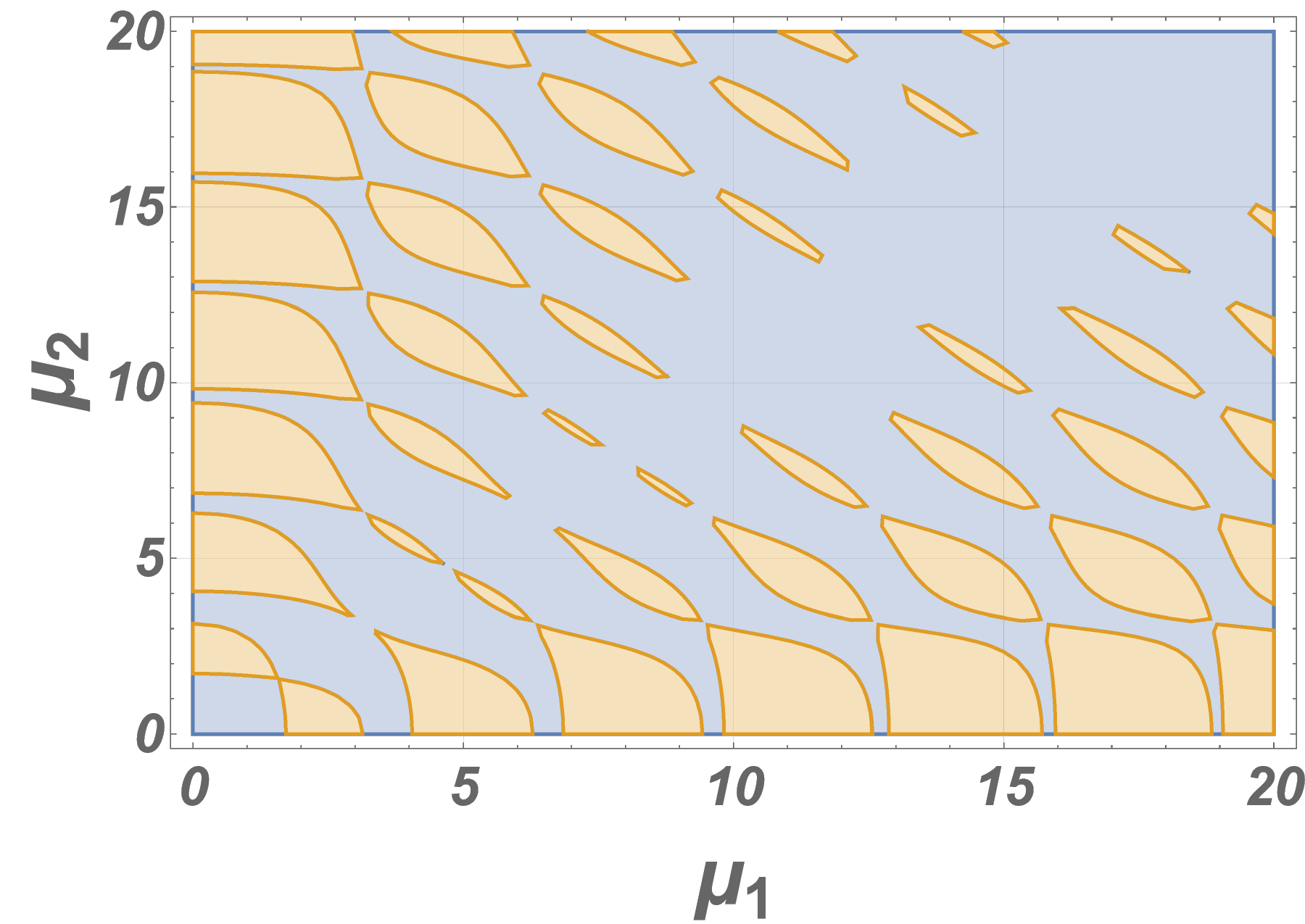}}
\caption{Plot of ${\mathcal T}$ as a function of $T_1$ and $T_2= T-T_1$. The blue(yellow) regions indicate phases of oscillatory (exponential) dependence on $T$ . The left panel corresponds to $\mu_1 = 0.3, \mu_2 = 0.7$ with $T_1, T_2$ varied  while the right panel corresponds to $T_1 = T_2 = 1$, and $\mu_1, \mu_2 $ are varied. }  \label{fig1}
\end{figure}

{\it Fidelity and Dynamical Phases}: The fidelity of the driven state is defined as ${\mathcal F}= |\langle\Delta|U(\ell T,0)|\Delta\rangle|/\langle \Delta|\Delta\rangle$, where $U(\ell T)$ is given by Eq.\ \ref{2-20}. Here the initial state $|\Delta\rangle$ is a primary state of ${\mathcal H}_1$. The state-operator correspondence described above then implies that this is a primary state $|\Delta_0\rangle_{\mu_1}$ where $\Delta_0$ is the smallest scaling dimension of a primary operator with particle number $n$. Using the standard Baker-Campbell-Hausdorff (BCH) relations \cite{bch1}, one can express ${\mathcal F}$ in the form
 \begin{eqnarray} 
{\mathcal F} &=&{_{\mu_1}}\langle \Delta| e^{\Lambda_+ L_+} e^{-2\Lambda_0 L_0} e^{\Lambda_- L_-}|\Delta\rangle_{\mu_1} = e^{-\Lambda_0 \Delta}, \label{fid1} 
\end{eqnarray} 
where we have used Eq.\ \ref{2-11} to express $H$, $C$ and $D$ in terms of $L_0$ and $L_{\pm}$. We note that $\Lambda_{\pm}$ do not enter the expression of ${\mathcal F}$ due to the relation $L_{-}|\Delta \rangle= \langle \Delta| L_+=0$ and a straightforward calculation using Eqs.\ \ref{2-7}, \ref{2-20}, and \ref{2-5} leads to 
\begin{eqnarray} 
\Lambda_0 &=&\frac{1}{2}  \ln \Big|\frac{1}{4} \left( (a_{\ell}+ d_{\ell})^2 + (\mu_1 b_{\ell} - c_{\ell}/\mu_1)^2 \right)  \Big|. \label{fid2} 
\end{eqnarray} 
Note that ${\mathcal F}=1$ when $\rho \ell=m \pi$ for integer $m$.   

\begin{figure}
\rotatebox{0}{\includegraphics*[width= 0.49 \linewidth]{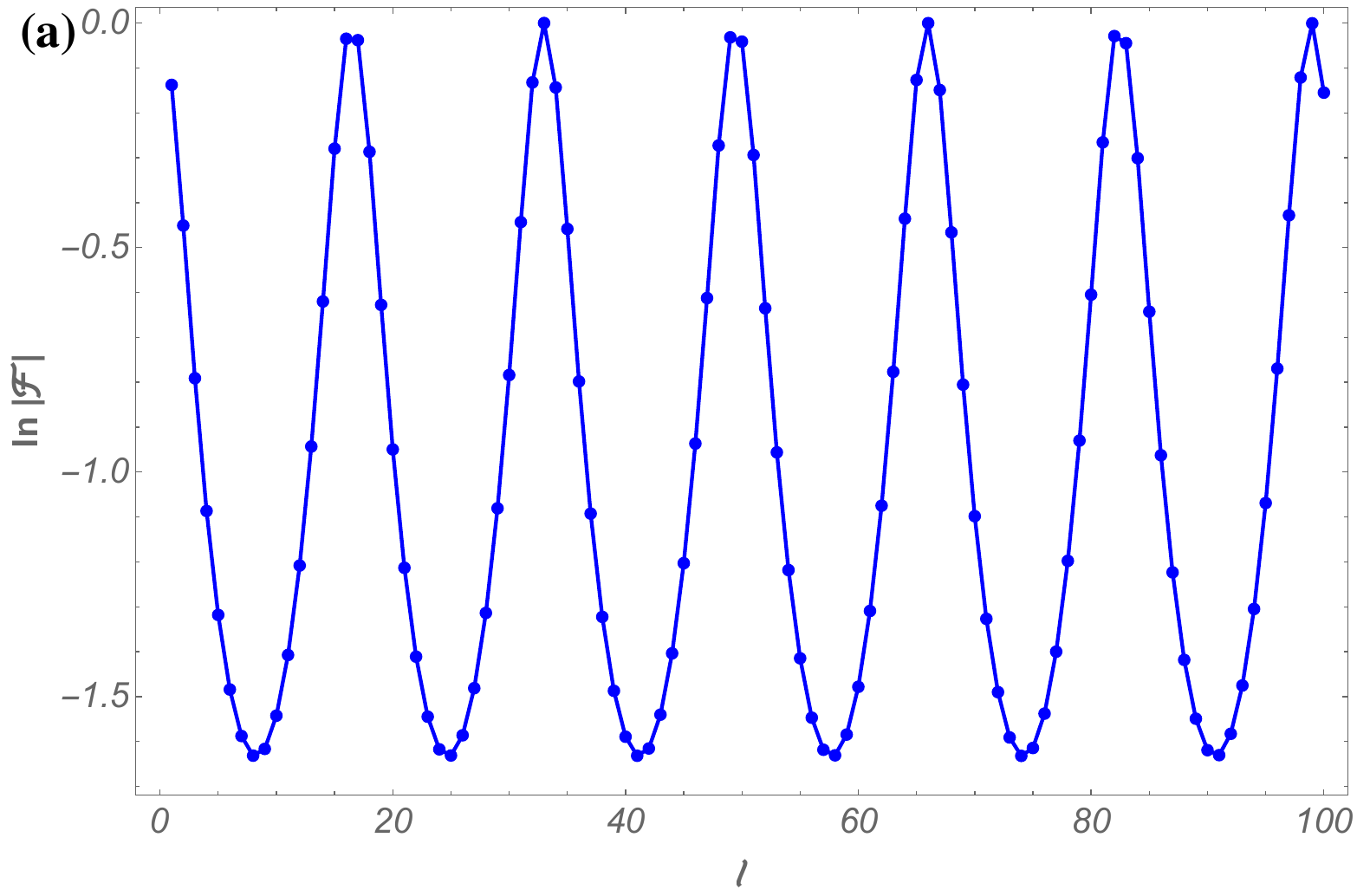}}
\rotatebox{0}{\includegraphics*[width= 0.49 \linewidth]{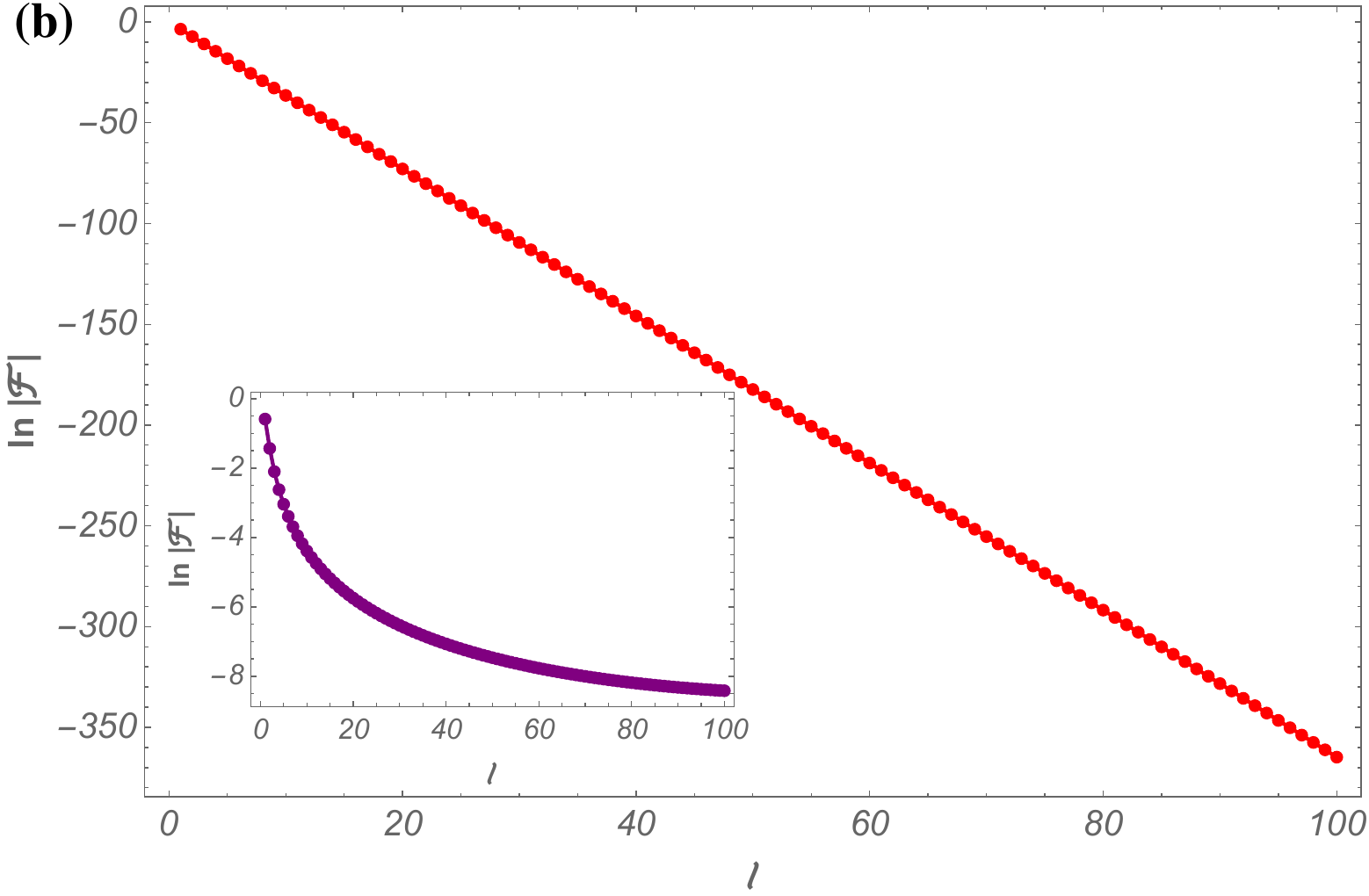}}
\rotatebox{0}{\includegraphics*[width= 0.49 \linewidth]{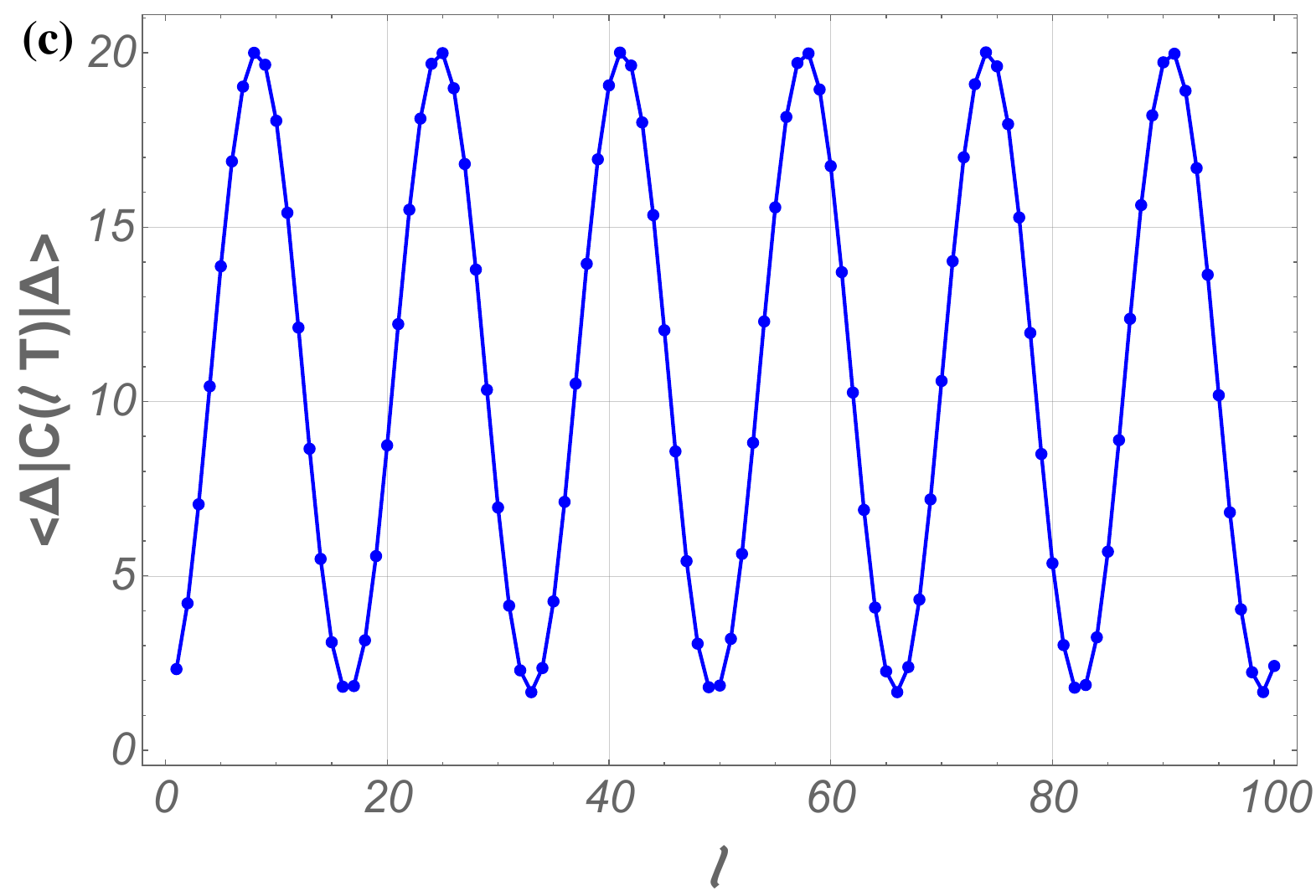}}
\rotatebox{0}{\includegraphics*[width= 0.49 \linewidth]{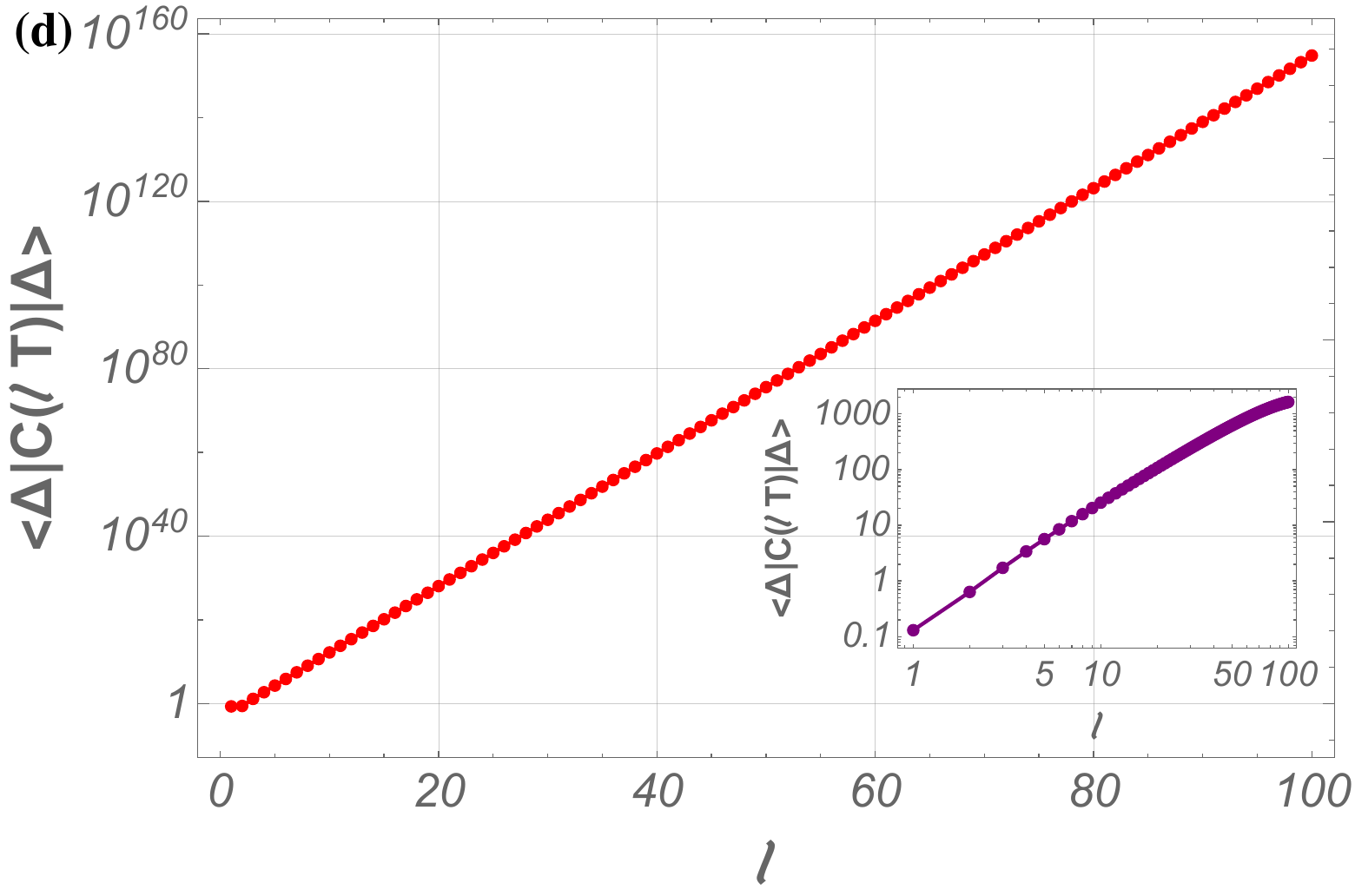}}
\caption{(a) Plot of $\ln |{\mathcal F}|$  as a function of $\ell$ in the elliptic phase.  (b) Same as in (c) but in the hyperbolic phase. The inset of (d) shows a slower, almost logarithmic, decay of $\ln |{\mathcal F}|$ near the transition surface. (c) Plot of $C_{\ell}$ (blue curve) as a function of $\ell$ in the elliptic phase. The parameters for the plot are $T=0.3$, $T_1=0.2$, $\mu_1=0.6$ and $\mu_2=0.7$.  (d) Exponential growth of $C_{\ell}$ as a function of $\ell$ in the hyperbolic phase corresponding to $T=9$, $T_1=5$, $\mu_1=0.1$ and $\mu_2=1.3$. The inset  (log-log plot) shows the near-linear growth of $C_{\ell}$ as a function of $\ell$ near 
the transition surface. The parameters for the inset are $T=2$, $T_1=1$, $\mu_1=4.833$ and $\mu_2=15.011$. (c)  All parameters in (c) and (d) are same as those in (a) and (b) respectively. See text for details.}  \label{fig2}
\end{figure}

Eq.\ \ref{fid2} is one of the central results of this letter. The large stroboscopic time (i.e. large $\ell$) behavior of ${\mathcal F}$ is determined by the nature of the quantity $\rho$ (Eq.\ \ref{2-10}). This leads to three distinct dynamical phases: (1) An elliptic oscillatory phase when $\rho$ is real; (2) A hyperbolic phase with purely imaginary $\rho$ when the expectation value behaves exponentially in $\ell$ for large $\ell$ (3) A parabolic phase which separates these two phases when $\rho=0$ and the expectation value increases quadratically in $\ell$. The nomenclature comes from the fact that these phases correspond to the three types of conjugacy classes of the $SO(2,1)$ matrix (Eq.\ \ref{2-20}) characterized by the single-cycle trace 
\bea
{\mathcal T} &=& \frac{1}{2}{\rm Tr} [U(T)] = (a_1+ d_1)/2 = \cos ( \rho) \label{2-18b}
\eea
whose absolute value is less than, greater than, or equal to unity for the elliptic, hyperbolic, and parabolic class ($\rho$ real, purely imaginary and $\pi$ times an integer), respectively. We indicate in Fig.\ref{fig1} the oscillatory and exponential phases as a function of drive frequencies (left panel) and amplitudes (right panel).

A numerical plot of $\ln |{\mathcal F}|$ as a function of $m$ is shown in the upper panels of Fig.\ \ref{fig2}. Fig.\ \ref{fig2}(a) indicates oscillatory behavior of $\ln |{\mathcal F}|$ in the elliptic phase while Fig.\ \ref{fig2}(b) exhibits its linear decay in the hyperbolic phase. The inset of Fig.\ \ref{fig2}(b) shows a slower, almost logarithmic, decay of $\ln |{\mathcal F}|$ near the transition surface. This indicates an exponentially fast scrambling of the driven system in the hyperbolic phase and periodic revivals in the elliptic phases; at the transition surface, the scrambling turns out to be linear in $\ell$ \cite{Emerson:2002lgj}.

{\it One-point functions}: These dynamical phases are also reflected in the behavior of one-point functions of the conformal generators. For example, consider the operators $C$ after $\ell$ time periods given by  
\ben
C(\ell) = e^{i\calh_F \ell T} C e^{-i\calh_F \ell T}
\label{2-13}
\een
and similarly for $D(\ell),H(\ell)$. Clearly $C(\ell), D(\ell),H(\ell)$ depend {\em only on the algebra} (\ref{2-1}). We can therefore use the Pauli matrix representation (\ref{2-20}) to get
\bea
C(\ell) & = & b_\ell^2 H + a_\ell^2 C + a_\ell  b_\ell D \nonumber \\
H(\ell) & = & d_\ell^2 H + c_\ell^2 C + c_\ell d_\ell D \nonumber \\
D(\ell) & = &  (a_\ell d_\ell+b_\ell c_\ell) D +2(c_\ell a_\ell C + d_\ell b_\ell H)
\label{2-14}
\eea
Using (\ref{2-14}) and (\ref{2-16}) and identifying $\omega \to \mu_1$ such that $|\Delta\rangle_{\mu_1}$ is a primary state 
that satisfies $H_{\rm osc} |\Delta\rangle_{\mu_1} = \Delta \mu_1 |\Delta\rangle_{\mu_1}$, we obtain the following expectation values
\bea
_{\mu_1}\langle \Delta_0| C(\ell) |\Delta_0\rangle_{\mu_1} & = & \frac{\Delta}{2} (\mu_1 b_\ell^2 + a_\ell^2/\mu_1) \nonumber \\
_{\mu_1}\langle \Delta_0| H (\ell) |\Delta_0\rangle_{\mu_1} & = & \frac{\Delta}{2}(\mu_1 d_\ell^2 + c_\ell^2/\mu_1) \nonumber \\
_{\mu_1}\langle \Delta_0| D (\ell) |\Delta_0\rangle_{\mu_1} &=& \Delta (\mu_1 d_\ell b_\ell + c_\ell a_\ell/\mu_1)
\label{2-17}
\eea
For the square pulse protocol, these expressions 
can be obtained analytically and we get (\ref{2-5}), for example, 
\bea
_{\mu_1}\langle \Delta_0| C(\ell) |\Delta_0\rangle_{\mu_1} & = & \frac{\Delta \mu_1}{2} (T\alpha_+)^2 \left(\frac{\sin (\rho \ell)}{\rho} \right)^2 \nonumber \\ & + &
\frac{\Delta}{2\mu_1} \left (\cos (\rho \ell) + \frac{\gamma T}{\rho} \sin (\rho \ell) \right)^2
\label{2-18}
\eea
The same framework extends to continuous protocols of direct experimental relevance, which need to be studied numerically. This is described in the End Matter.

The behavior of $C({\ell})$ as a function of $\ell$, shown in the bottom panels of Fig.\ \ref{fig2}, confirms this picture.  
Fig.\ \ref{fig2}(c) shows the oscillation of $C(\ell)$ as a function of $\ell$ for the elliptic phase ($\cos \rho \simeq 0.982$); this corresponds to the oscillation of the density of fermions in the trap about their mean value. Fig.\ \ref{fig2}(d) displays an exponential growth of $C (\ell) $ ($\cosh \rho \simeq 3.181$). The inset of Fig.\ \ref{fig2}(d) shows an almost linear growth of $C (\ell)$ ($\cos \rho \simeq 0.99992$); this is due to proximity to the parabolic critical line. The behavior of $C (\ell)$ described here has a direct correspondence with 
the dynamics of density profile of trapped ultracold fermions; this is discussed in the End Matter. Analogous methods can be used to compute the auto-correlation functions $\langle \Delta | C(\ell) C |\Delta \rangle$. This is discussed in the SM \cite{sicit}.

{\it Holography}: It is well-known that some Schr\"{o}dinger-invariant systems can have holographic duals. They form a special class of strongly coupled systems with a large number of degrees of freedom that have a dual weakly-coupled (Einstein) gravitational description. The $(d+1)$-dimensional Schr\"{o}dinger symmetry is realized as the isometry of a classical $(d+3)$-dimensional geometry, where the two additional directions are a radial direction $r$ and a null direction $\xi$. Such geometric solutions are obtained by solving {\it e.g.}~an effective Einstein-Proca theory in $(d+3)$-dimensions\cite{dual1, dual2}, which can sometimes be obtained as a consistent truncation of a full ten or eleven dimensional supergravity\cite{Adams_2008, Herzog:2008wg, Maldacena:2008wh}.

In this class of examples, the Schr\"{o}dinger isometry is generated by the corresponding Killing vectors of the geometry. We can focus on the $SO(2,1)$ sub-algebra and construct a Killing vector that is a general linear combination of the $SO(2,1)$-algebra generating Killing vectors. 
This provides a geometric description of the Floquet Hamiltonian that is considered here, {\it e.g.}~in equation (\ref{2-7}). It is interesting to explore where this Killing vector becomes null. The norm is given by (see \cite{sicit} for more details)

\begin{eqnarray}
||\mathcal{K}_X\|^2
\sim
\left[
-
\left(\alpha_{+}+2\gamma t+\alpha_{-}t^2\right)^2
-\kappa^2
r^2\left(x^2+r^2\right)
\right] \ .
\end{eqnarray}
which always has a positive real root for the radial location, $r$, in the hyperbolic class, {\it i.e.}~when $- \kappa^2>0$; and no real solution in the elliptic class. As shown in \cite{sicit}, for the hyperbolic class, ${\cal K}_X$ becomes null on a time-like hypersurface, which is similar to an ergosurface in the geometry.

The parabolic class is more subtle. In this case, the Killing norm is given by
\begin{eqnarray}
|| \mathcal{K}_X\|^2 = - \frac{L^2}{r^4}(\beta^2\alpha_-^2)\left( t- t_0\right)^4 \ , \quad t_0 = - \frac{\gamma}{\alpha_-} \ ,
\end{eqnarray}
which is also a null hypersurface, and therefore a Killing horizon at $t=t_0$. The corresponding surface gravity vanishes (see \cite{sicit} for a detailed calculation) and therefore becomes an extremal horizon.  

Note that the discussion above is {\it kinematic} and provides a geometric picture of the three phases. To obtain the genuine holographic dual we need to calculate the back-reacted metric. (For other holographic Floquet systems, see \cite{Rangamani2015, Hashimoto2016, murata} and for $1+1$ dimensional relativistic  Floquet CFT see \cite{holo4,2dcurved2})
It is tempting to conjecture that these geometric features persist in the full back-reacted geometry in which the hyperbolic class is equipped with dynamical instabilities, amplifications, and non-equilibrium energy-extraction due to the ergo-region in the bulk. Compared to similar studies in relativistic $1+1$ dimensional CFTs the emergence of an ergo-region in the non-relativistic case is a novel feature.

{\it Experiments}: As mentioned earlier, non-relativistic CFTs may be applied to a wide variety of systems. Typical experiments involving measurement of Loschmidt echo for a driven state in ultracold atom platforms \cite{echo1,echo2} may provide information about ${\mathcal F}$ and hence distinguish between the dynamical phases. However, such measurements may be experimentally difficult. As an alternative, we consider trapped fermions near unitarity which provide a concrete example of a system with such conformal symmetry. Our analysis is consistent with experiments studying dynamics of such fermions \cite{pra1,pra2}. A detailed analysis of the application of our formalism to such systems is presented in the End Matter. We consider fermions whose trapping potential is subjected to a periodic drive till $t=t_s$ followed by their Hamiltonian evolution with a fixed trap potential amplitude for $t>t_s$ \cite{pra1}. The dynamics for this latter evolution is oscillatory. Our sharp prediction, detailed in the End Matter, is that the amplitude of these oscillations provides a diagnostic for the dynamical phases; they grow exponentially with $t_s$ in the hyperbolic phase and are bounded in the elliptic phase. 

{\it Conclusion}: In this work, we have developed a general formalism for studying non-relativistic FCFTs. Our analysis is completely universal and predicts distinct dynamical phases for these FCFTs; it also provides a sharp diagnostic for these phases using fidelity and one-point functions of the driven system. This provides a generic framework for understanding Floquet dynamics applicable to a wide range of systems. This feature provides a distinction of our study from Floquet dynamics of relativistic CFTs studied earlier.  
For holographic systems our treatment naturally leads to a classification of the dynamical phases in terms of properties of Killing vectors representing Floquet Hamiltonians in the bulk. We expect our work to lead to several further studies, including other signatures of dynamical phases for trapped fermions, the relevance of similar phases in other systems described by such CFTs, and studies involving FCFTs in the presence of quasiperiodic and aperiodic drives.

\noindent {\bf Acknowledgements:} S.R.D. would like to thank D.T. Son for a conversation, and NCTS, National Taiwan University and Indian Association for the Cultivation of Science, Kolkata for hospitality. D.D. would like to thank SINP, Kolkata for hospitality during May 2026, when a part of this work was completed. The work of SRD is partially supported by the National Science Foundation grants NSF-PHY-2410547. DD is partially supported by an ANRF grant ANRF/ARG/2025/001338/PS. KS thanks DST
for support through the ANRF grant JCB/2021/000030. A.K. acknowledges the support of the Humboldt Research Fellowship for Experienced Researchers by the Alexander von Humboldt Foundation. The numerical coding and draft-preparation workflow made use of AI coding and reasoning assistants, including
OpenAI Codex and ChatGPT, under human supervision. All physics assumptions, numerical choices, and conclusions are the responsibility of the authors.

{\bf Data and code availability.}
The numerical code and generated figure outputs supporting the cosine-drive calculations are publicly available at
\url{https://github.com/diptarkadas/cosine-drive-response}.
The repository contains the Trotterized cosine-drive evolution code, scripts to reproduce the response plots and phase diagram, a README with reproduction instructions, and the generated PDF/PNG figures. No additional datasets are required; all plotted numerical data are generated by the scripts in the repository.

\bibliography{nrfl}

\begin{thebibliography}{83}%
\makeatletter
\providecommand \@ifxundefined [1]{%
 \@ifx{#1\undefined}
}%
\providecommand \@ifnum [1]{%
 \ifnum #1\expandafter \@firstoftwo
 \else \expandafter \@secondoftwo
 \fi
}%
\providecommand \@ifx [1]{%
 \ifx #1\expandafter \@firstoftwo
 \else \expandafter \@secondoftwo
 \fi
}%
\providecommand \natexlab [1]{#1}%
\providecommand \enquote  [1]{``#1''}%
\providecommand \bibnamefont  [1]{#1}%
\providecommand \bibfnamefont [1]{#1}%
\providecommand \citenamefont [1]{#1}%
\providecommand \href@noop [0]{\@secondoftwo}%
\providecommand \href [0]{\begingroup \@sanitize@url \@href}%
\providecommand \@href[1]{\@@startlink{#1}\@@href}%
\providecommand \@@href[1]{\endgroup#1\@@endlink}%
\providecommand \@sanitize@url [0]{\catcode `\\12\catcode `\$12\catcode
  `\&12\catcode `\#12\catcode `\^12\catcode `\_12\catcode `\%12\relax}%
\providecommand \@@startlink[1]{}%
\providecommand \@@endlink[0]{}%
\providecommand \url  [0]{\begingroup\@sanitize@url \@url }%
\providecommand \@url [1]{\endgroup\@href {#1}{\urlprefix }}%
\providecommand \urlprefix  [0]{URL }%
\providecommand \Eprint [0]{\href }%
\providecommand \doibase [0]{https://doi.org/}%
\providecommand \selectlanguage [0]{\@gobble}%
\providecommand \bibinfo  [0]{\@secondoftwo}%
\providecommand \bibfield  [0]{\@secondoftwo}%
\providecommand \translation [1]{[#1]}%
\providecommand \BibitemOpen [0]{}%
\providecommand \bibitemStop [0]{}%
\providecommand \bibitemNoStop [0]{.\EOS\space}%
\providecommand \EOS [0]{\spacefactor3000\relax}%
\providecommand \BibitemShut  [1]{\csname bibitem#1\endcsname}%
\let\auto@bib@innerbib\@empty
\bibitem [{\citenamefont {Dziarmaga}(2010)}]{rev1}%
  \BibitemOpen
  \bibfield  {author} {\bibinfo {author} {\bibfnamefont {J.}~\bibnamefont
  {Dziarmaga}},\ }\bibfield  {title} {\bibinfo {title} {Dynamics of a quantum
  phase transition and relaxation to a steady state},\ }\href
  {https://doi.org/10.1080/00018732.2010.514702} {\bibfield  {journal}
  {\bibinfo  {journal} {Advances in Physics}\ }\textbf {\bibinfo {volume}
  {59}},\ \bibinfo {pages} {1063–1189} (\bibinfo {year} {2010})}\BibitemShut
  {NoStop}%
\bibitem [{\citenamefont {Polkovnikov}\ \emph {et~al.}(2011)\citenamefont
  {Polkovnikov}, \citenamefont {Sengupta}, \citenamefont {Silva},\ and\
  \citenamefont {Vengalattore}}]{rev2}%
  \BibitemOpen
  \bibfield  {author} {\bibinfo {author} {\bibfnamefont {A.}~\bibnamefont
  {Polkovnikov}}, \bibinfo {author} {\bibfnamefont {K.}~\bibnamefont
  {Sengupta}}, \bibinfo {author} {\bibfnamefont {A.}~\bibnamefont {Silva}},\
  and\ \bibinfo {author} {\bibfnamefont {M.}~\bibnamefont {Vengalattore}},\
  }\bibfield  {title} {\bibinfo {title} {Colloquium: Nonequilibrium dynamics of
  closed interacting quantum systems},\ }\href
  {https://doi.org/10.1103/RevModPhys.83.863} {\bibfield  {journal} {\bibinfo
  {journal} {Rev. Mod. Phys.}\ }\textbf {\bibinfo {volume} {83}},\ \bibinfo
  {pages} {863} (\bibinfo {year} {2011})}\BibitemShut {NoStop}%
\bibitem [{\citenamefont {Bukov}\ \emph {et~al.}(2015)\citenamefont {Bukov},
  \citenamefont {D'Alessio},\ and\ \citenamefont {Polkovnikov}}]{rev3}%
  \BibitemOpen
  \bibfield  {author} {\bibinfo {author} {\bibfnamefont {M.}~\bibnamefont
  {Bukov}}, \bibinfo {author} {\bibfnamefont {L.}~\bibnamefont {D'Alessio}},\
  and\ \bibinfo {author} {\bibfnamefont {A.}~\bibnamefont {Polkovnikov}},\
  }\bibfield  {title} {\bibinfo {title} {Universal high-frequency behavior of
  periodically driven systems: from dynamical stabilization to {F}loquet
  engineering},\ }\href {https://doi.org/10.1080/00018732.2015.1055918}
  {\bibfield  {journal} {\bibinfo  {journal} {Advances in Physics}\ }\textbf
  {\bibinfo {volume} {64}},\ \bibinfo {pages} {139} (\bibinfo {year} {2015})},\
  \Eprint {https://arxiv.org/abs/https://doi.org/10.1080/00018732.2015.1055918}
  {https://doi.org/10.1080/00018732.2015.1055918} \BibitemShut {NoStop}%
\bibitem [{\citenamefont {D'Alessio}\ \emph {et~al.}(2016)\citenamefont
  {D'Alessio}, \citenamefont {Kafri}, \citenamefont {Polkovnikov},\ and\
  \citenamefont {Rigol}}]{rev4}%
  \BibitemOpen
  \bibfield  {author} {\bibinfo {author} {\bibfnamefont {L.}~\bibnamefont
  {D'Alessio}}, \bibinfo {author} {\bibfnamefont {Y.}~\bibnamefont {Kafri}},
  \bibinfo {author} {\bibfnamefont {A.}~\bibnamefont {Polkovnikov}},\ and\
  \bibinfo {author} {\bibfnamefont {M.}~\bibnamefont {Rigol}},\ }\bibfield
  {title} {\bibinfo {title} {From quantum chaos and eigenstate thermalization
  to statistical mechanics and thermodynamics},\ }\href
  {https://doi.org/10.1080/00018732.2016.1198134} {\bibfield  {journal}
  {\bibinfo  {journal} {Advances in Physics}\ }\textbf {\bibinfo {volume}
  {65}},\ \bibinfo {pages} {239} (\bibinfo {year} {2016})},\ \Eprint
  {https://arxiv.org/abs/https://doi.org/10.1080/00018732.2016.1198134}
  {https://doi.org/10.1080/00018732.2016.1198134} \BibitemShut {NoStop}%
\bibitem [{\citenamefont {Oka}\ and\ \citenamefont {Kitamura}(2019)}]{rev5}%
  \BibitemOpen
  \bibfield  {author} {\bibinfo {author} {\bibfnamefont {T.}~\bibnamefont
  {Oka}}\ and\ \bibinfo {author} {\bibfnamefont {S.}~\bibnamefont {Kitamura}},\
  }\bibfield  {title} {\bibinfo {title} {Floquet engineering of quantum
  materials},\ }\href
  {https://doi.org/10.1146/annurev-conmatphys-031218-013423} {\bibfield
  {journal} {\bibinfo  {journal} {Annual Review of Condensed Matter Physics}\
  }\textbf {\bibinfo {volume} {10}},\ \bibinfo {pages} {387–408} (\bibinfo
  {year} {2019})}\BibitemShut {NoStop}%
\bibitem [{\citenamefont {Blanes}\ \emph {et~al.}(2009)\citenamefont {Blanes},
  \citenamefont {Casas}, \citenamefont {Oteo},\ and\ \citenamefont
  {Ros}}]{rev6}%
  \BibitemOpen
  \bibfield  {author} {\bibinfo {author} {\bibfnamefont {S.}~\bibnamefont
  {Blanes}}, \bibinfo {author} {\bibfnamefont {F.}~\bibnamefont {Casas}},
  \bibinfo {author} {\bibfnamefont {J.}~\bibnamefont {Oteo}},\ and\ \bibinfo
  {author} {\bibfnamefont {J.}~\bibnamefont {Ros}},\ }\bibfield  {title}
  {\bibinfo {title} {The magnus expansion and some of its applications},\
  }\href {https://doi.org/10.1016/j.physrep.2008.11.001} {\bibfield  {journal}
  {\bibinfo  {journal} {Physics Reports}\ }\textbf {\bibinfo {volume} {470}},\
  \bibinfo {pages} {151–238} (\bibinfo {year} {2009})}\BibitemShut {NoStop}%
\bibitem [{\citenamefont {Eckardt}(2017)}]{rev7}%
  \BibitemOpen
  \bibfield  {author} {\bibinfo {author} {\bibfnamefont {A.}~\bibnamefont
  {Eckardt}},\ }\bibfield  {title} {\bibinfo {title} {Colloquium: Atomic
  quantum gases in periodically driven optical lattices},\ }\href
  {https://doi.org/10.1103/RevModPhys.89.011004} {\bibfield  {journal}
  {\bibinfo  {journal} {Rev. Mod. Phys.}\ }\textbf {\bibinfo {volume} {89}},\
  \bibinfo {pages} {011004} (\bibinfo {year} {2017})}\BibitemShut {NoStop}%
\bibitem [{\citenamefont {Sen}\ \emph {et~al.}(2021)\citenamefont {Sen},
  \citenamefont {Sen},\ and\ \citenamefont {Sengupta}}]{rev8}%
  \BibitemOpen
  \bibfield  {author} {\bibinfo {author} {\bibfnamefont {A.}~\bibnamefont
  {Sen}}, \bibinfo {author} {\bibfnamefont {D.}~\bibnamefont {Sen}},\ and\
  \bibinfo {author} {\bibfnamefont {K.}~\bibnamefont {Sengupta}},\ }\bibfield
  {title} {\bibinfo {title} {Analytic approaches to periodically driven closed
  quantum systems: methods and applications},\ }\href
  {https://doi.org/10.1088/1361-648X/ac1b61} {\bibfield  {journal} {\bibinfo
  {journal} {Journal of Physics: Condensed Matter}\ }\textbf {\bibinfo {volume}
  {33}},\ \bibinfo {pages} {443003} (\bibinfo {year} {2021})}\BibitemShut
  {NoStop}%
\bibitem [{\citenamefont {Bloch}\ \emph {et~al.}(2008)\citenamefont {Bloch},
  \citenamefont {Dalibard},\ and\ \citenamefont {Zwerger}}]{rev9}%
  \BibitemOpen
  \bibfield  {author} {\bibinfo {author} {\bibfnamefont {I.}~\bibnamefont
  {Bloch}}, \bibinfo {author} {\bibfnamefont {J.}~\bibnamefont {Dalibard}},\
  and\ \bibinfo {author} {\bibfnamefont {W.}~\bibnamefont {Zwerger}},\
  }\bibfield  {title} {\bibinfo {title} {Many-body physics with ultracold
  gases},\ }\href {https://doi.org/10.1103/RevModPhys.80.885} {\bibfield
  {journal} {\bibinfo  {journal} {Rev. Mod. Phys.}\ }\textbf {\bibinfo {volume}
  {80}},\ \bibinfo {pages} {885} (\bibinfo {year} {2008})}\BibitemShut
  {NoStop}%
\bibitem [{\citenamefont {Tarruell}\ and\ \citenamefont
  {Sanchez-Palencia}(2018)}]{rev10}%
  \BibitemOpen
  \bibfield  {author} {\bibinfo {author} {\bibfnamefont {L.}~\bibnamefont
  {Tarruell}}\ and\ \bibinfo {author} {\bibfnamefont {L.}~\bibnamefont
  {Sanchez-Palencia}},\ }\bibfield  {title} {\bibinfo {title} {Quantum
  simulation of the hubbard model with ultracold fermions in optical
  lattices},\ }\href {https://doi.org/10.1016/j.crhy.2018.10.013} {\bibfield
  {journal} {\bibinfo  {journal} {Comptes Rendus. Physique}\ }\textbf {\bibinfo
  {volume} {19}},\ \bibinfo {pages} {365–393} (\bibinfo {year}
  {2018})}\BibitemShut {NoStop}%
\bibitem [{\citenamefont {Ho}\ \emph {et~al.}(2023)\citenamefont {Ho},
  \citenamefont {Mori}, \citenamefont {Abanin},\ and\ \citenamefont
  {Dalla~Torre}}]{rev11}%
  \BibitemOpen
  \bibfield  {author} {\bibinfo {author} {\bibfnamefont {W.~W.}\ \bibnamefont
  {Ho}}, \bibinfo {author} {\bibfnamefont {T.}~\bibnamefont {Mori}}, \bibinfo
  {author} {\bibfnamefont {D.~A.}\ \bibnamefont {Abanin}},\ and\ \bibinfo
  {author} {\bibfnamefont {E.~G.}\ \bibnamefont {Dalla~Torre}},\ }\bibfield
  {title} {\bibinfo {title} {Quantum and classical floquet prethermalization},\
  }\href {https://doi.org/10.1016/j.aop.2023.169297} {\bibfield  {journal}
  {\bibinfo  {journal} {Annals of Physics}\ }\textbf {\bibinfo {volume}
  {454}},\ \bibinfo {pages} {169297} (\bibinfo {year} {2023})}\BibitemShut
  {NoStop}%
\bibitem [{\citenamefont {Mori}\ \emph {et~al.}(2018)\citenamefont {Mori},
  \citenamefont {Ikeda}, \citenamefont {Kaminishi},\ and\ \citenamefont
  {Ueda}}]{rev12}%
  \BibitemOpen
  \bibfield  {author} {\bibinfo {author} {\bibfnamefont {T.}~\bibnamefont
  {Mori}}, \bibinfo {author} {\bibfnamefont {T.~N.}\ \bibnamefont {Ikeda}},
  \bibinfo {author} {\bibfnamefont {E.}~\bibnamefont {Kaminishi}},\ and\
  \bibinfo {author} {\bibfnamefont {M.}~\bibnamefont {Ueda}},\ }\bibfield
  {title} {\bibinfo {title} {Thermalization and prethermalization in isolated
  quantum systems: a theoretical overview},\ }\href
  {https://doi.org/10.1088/1361-6455/aabcdf} {\bibfield  {journal} {\bibinfo
  {journal} {Journal of Physics B}\ }\textbf {\bibinfo {volume} {51}},\
  \bibinfo {pages} {112001} (\bibinfo {year} {2018})}\BibitemShut {NoStop}%
\bibitem [{\citenamefont {Banerjee}\ and\ \citenamefont
  {Sengupta}(2025)}]{rev13}%
  \BibitemOpen
  \bibfield  {author} {\bibinfo {author} {\bibfnamefont {T.}~\bibnamefont
  {Banerjee}}\ and\ \bibinfo {author} {\bibfnamefont {K.}~\bibnamefont
  {Sengupta}},\ }\bibfield  {title} {\bibinfo {title} {Emergent symmetries in
  prethermal phases of periodically driven quantum systems},\ }\href
  {https://doi.org/10.1088/1361-648x/ada860} {\bibfield  {journal} {\bibinfo
  {journal} {Journal of Physics: Condensed Matter}\ }\textbf {\bibinfo {volume}
  {37}},\ \bibinfo {pages} {133002} (\bibinfo {year} {2025})}\BibitemShut
  {NoStop}%
\bibitem [{\citenamefont {D'Alessio}\ and\ \citenamefont
  {Rigol}(2014)}]{rigol1}%
  \BibitemOpen
  \bibfield  {author} {\bibinfo {author} {\bibfnamefont {L.}~\bibnamefont
  {D'Alessio}}\ and\ \bibinfo {author} {\bibfnamefont {M.}~\bibnamefont
  {Rigol}},\ }\bibfield  {title} {\bibinfo {title} {Long-time behavior of
  isolated periodically driven interacting lattice systems},\ }\href
  {https://doi.org/10.1103/PhysRevX.4.041048} {\bibfield  {journal} {\bibinfo
  {journal} {Phys. Rev. X}\ }\textbf {\bibinfo {volume} {4}},\ \bibinfo {pages}
  {041048} (\bibinfo {year} {2014})}\BibitemShut {NoStop}%
\bibitem [{\citenamefont {Wintersperger}\ \emph {et~al.}(2020)\citenamefont
  {Wintersperger}, \citenamefont {Braun}, \citenamefont {{\~A}{\oe}nal},
  \citenamefont {Eckardt}, \citenamefont {Liberto}, \citenamefont {Goldman},
  \citenamefont {Bloch},\ and\ \citenamefont {Aidelsburger}}]{bloch1}%
  \BibitemOpen
  \bibfield  {author} {\bibinfo {author} {\bibfnamefont {K.}~\bibnamefont
  {Wintersperger}}, \bibinfo {author} {\bibfnamefont {C.}~\bibnamefont
  {Braun}}, \bibinfo {author} {\bibfnamefont {F.~N.}\ \bibnamefont
  {{\~A}{\oe}nal}}, \bibinfo {author} {\bibfnamefont {A.}~\bibnamefont
  {Eckardt}}, \bibinfo {author} {\bibfnamefont {M.~D.}\ \bibnamefont
  {Liberto}}, \bibinfo {author} {\bibfnamefont {N.}~\bibnamefont {Goldman}},
  \bibinfo {author} {\bibfnamefont {I.}~\bibnamefont {Bloch}},\ and\ \bibinfo
  {author} {\bibfnamefont {M.}~\bibnamefont {Aidelsburger}},\ }\bibfield
  {title} {\bibinfo {title} {Realization of an anomalous floquet topological
  system with ultracold atoms},\ }\href
  {https://doi.org/10.1038/s41567-020-0949-y} {\bibfield  {journal} {\bibinfo
  {journal} {Nature Physics}\ }\textbf {\bibinfo {volume} {16}},\ \bibinfo
  {pages} {1058} (\bibinfo {year} {2020})}\BibitemShut {NoStop}%
\bibitem [{\citenamefont {Wen}\ and\ \citenamefont {Wu}(2018)}]{2dpapers1}%
  \BibitemOpen
  \bibfield  {author} {\bibinfo {author} {\bibfnamefont {X.}~\bibnamefont
  {Wen}}\ and\ \bibinfo {author} {\bibfnamefont {J.-Q.}\ \bibnamefont {Wu}},\
  }\href {https://arxiv.org/abs/1805.00031} {\bibinfo {title} {Floquet
  conformal field theory}} (\bibinfo {year} {2018}),\ \Eprint
  {https://arxiv.org/abs/1805.00031} {arXiv:1805.00031 [cond-mat.str-el]}
  \BibitemShut {NoStop}%
\bibitem [{\citenamefont {Lapierre}\ \emph
  {et~al.}(2020{\natexlab{a}})\citenamefont {Lapierre}, \citenamefont {Choo},
  \citenamefont {Tauber}, \citenamefont {Tiwari}, \citenamefont {Neupert},\
  and\ \citenamefont {Chitra}}]{2dpapers2}%
  \BibitemOpen
  \bibfield  {author} {\bibinfo {author} {\bibfnamefont {B.}~\bibnamefont
  {Lapierre}}, \bibinfo {author} {\bibfnamefont {K.}~\bibnamefont {Choo}},
  \bibinfo {author} {\bibfnamefont {C.}~\bibnamefont {Tauber}}, \bibinfo
  {author} {\bibfnamefont {A.}~\bibnamefont {Tiwari}}, \bibinfo {author}
  {\bibfnamefont {T.}~\bibnamefont {Neupert}},\ and\ \bibinfo {author}
  {\bibfnamefont {R.}~\bibnamefont {Chitra}},\ }\bibfield  {title} {\bibinfo
  {title} {Emergent black hole dynamics in critical floquet systems},\ }\href
  {https://doi.org/10.1103/PhysRevResearch.2.023085} {\bibfield  {journal}
  {\bibinfo  {journal} {Phys. Rev. Res.}\ }\textbf {\bibinfo {volume} {2}},\
  \bibinfo {pages} {023085} (\bibinfo {year} {2020}{\natexlab{a}})}\BibitemShut
  {NoStop}%
\bibitem [{\citenamefont {Fan}\ \emph {et~al.}(2021)\citenamefont {Fan},
  \citenamefont {Gu}, \citenamefont {Vishwanath},\ and\ \citenamefont
  {Wen}}]{2dpapers3}%
  \BibitemOpen
  \bibfield  {author} {\bibinfo {author} {\bibfnamefont {R.}~\bibnamefont
  {Fan}}, \bibinfo {author} {\bibfnamefont {Y.}~\bibnamefont {Gu}}, \bibinfo
  {author} {\bibfnamefont {A.}~\bibnamefont {Vishwanath}},\ and\ \bibinfo
  {author} {\bibfnamefont {X.}~\bibnamefont {Wen}},\ }\bibfield  {title}
  {\bibinfo {title} {Floquet conformal field theories with generally deformed
  hamiltonians},\ }\href {https://doi.org/10.21468/SciPostPhys.10.2.049}
  {\bibfield  {journal} {\bibinfo  {journal} {SciPost Phys.}\ }\textbf
  {\bibinfo {volume} {10}},\ \bibinfo {pages} {049} (\bibinfo {year}
  {2021})}\BibitemShut {NoStop}%
\bibitem [{\citenamefont {Das}\ \emph {et~al.}(2021)\citenamefont {Das},
  \citenamefont {Ghosh},\ and\ \citenamefont {Sengupta}}]{2dpapers4}%
  \BibitemOpen
  \bibfield  {author} {\bibinfo {author} {\bibfnamefont {D.}~\bibnamefont
  {Das}}, \bibinfo {author} {\bibfnamefont {R.}~\bibnamefont {Ghosh}},\ and\
  \bibinfo {author} {\bibfnamefont {K.}~\bibnamefont {Sengupta}},\ }\bibfield
  {title} {\bibinfo {title} {Conformal floquet dynamics with a continuous drive
  protocol},\ }\href {https://doi.org/10.1007/JHEP05(2021)172} {\bibfield
  {journal} {\bibinfo  {journal} {Journal of High Energy Physics}\ }\textbf
  {\bibinfo {volume} {2021}},\ \bibinfo {pages} {172} (\bibinfo {year}
  {2021})}\BibitemShut {NoStop}%
\bibitem [{\citenamefont {Lapierre}\ \emph
  {et~al.}(2020{\natexlab{b}})\citenamefont {Lapierre}, \citenamefont {Choo},
  \citenamefont {Tiwari}, \citenamefont {Tauber}, \citenamefont {Neupert},\
  and\ \citenamefont {Chitra}}]{2dpapers5}%
  \BibitemOpen
  \bibfield  {author} {\bibinfo {author} {\bibfnamefont {B.}~\bibnamefont
  {Lapierre}}, \bibinfo {author} {\bibfnamefont {K.}~\bibnamefont {Choo}},
  \bibinfo {author} {\bibfnamefont {A.}~\bibnamefont {Tiwari}}, \bibinfo
  {author} {\bibfnamefont {C.}~\bibnamefont {Tauber}}, \bibinfo {author}
  {\bibfnamefont {T.}~\bibnamefont {Neupert}},\ and\ \bibinfo {author}
  {\bibfnamefont {R.}~\bibnamefont {Chitra}},\ }\bibfield  {title} {\bibinfo
  {title} {{Fine structure of heating in a quasiperiodically driven critical
  quantum system}},\ }\href {https://doi.org/10.1103/PhysRevResearch.2.033461}
  {\bibfield  {journal} {\bibinfo  {journal} {Phys. Rev. Res.}\ }\textbf
  {\bibinfo {volume} {2}},\ \bibinfo {pages} {033461} (\bibinfo {year}
  {2020}{\natexlab{b}})}\BibitemShut {NoStop}%
\bibitem [{\citenamefont {Han}\ and\ \citenamefont {Wen}(2020)}]{2dpapers6}%
  \BibitemOpen
  \bibfield  {author} {\bibinfo {author} {\bibfnamefont {B.}~\bibnamefont
  {Han}}\ and\ \bibinfo {author} {\bibfnamefont {X.}~\bibnamefont {Wen}},\
  }\bibfield  {title} {\bibinfo {title} {{Classification of $SL_2$ deformed
  Floquet conformal field theories}},\ }\href
  {https://doi.org/10.1103/PhysRevB.102.205125} {\bibfield  {journal} {\bibinfo
   {journal} {Phys. Rev. B}\ }\textbf {\bibinfo {volume} {102}},\ \bibinfo
  {pages} {205125} (\bibinfo {year} {2020})},\ \Eprint
  {https://arxiv.org/abs/2008.01123} {arXiv:2008.01123 [cond-mat.stat-mech]}
  \BibitemShut {NoStop}%
\bibitem [{\citenamefont {Lapierre}\ \emph
  {et~al.}(2025{\natexlab{a}})\citenamefont {Lapierre}, \citenamefont
  {Numasawa}, \citenamefont {Neupert},\ and\ \citenamefont {Ryu}}]{2dpapers7}%
  \BibitemOpen
  \bibfield  {author} {\bibinfo {author} {\bibfnamefont {B.}~\bibnamefont
  {Lapierre}}, \bibinfo {author} {\bibfnamefont {T.}~\bibnamefont {Numasawa}},
  \bibinfo {author} {\bibfnamefont {T.}~\bibnamefont {Neupert}},\ and\ \bibinfo
  {author} {\bibfnamefont {S.}~\bibnamefont {Ryu}},\ }\bibfield  {title}
  {\bibinfo {title} {{Floquet engineered inhomogeneous quantum chaos in
  critical systems}},\ }\href {https://doi.org/10.1103/cn3z-vfgr} {\bibfield
  {journal} {\bibinfo  {journal} {Phys. Rev. B}\ }\textbf {\bibinfo {volume}
  {112}},\ \bibinfo {pages} {104317} (\bibinfo {year} {2025}{\natexlab{a}})},\
  \Eprint {https://arxiv.org/abs/2405.01642} {arXiv:2405.01642
  [cond-mat.str-el]} \BibitemShut {NoStop}%
\bibitem [{\citenamefont {Banerjee}\ \emph {et~al.}(2025)\citenamefont
  {Banerjee}, \citenamefont {Das},\ and\ \citenamefont {Sengupta}}]{2dpapers8}%
  \BibitemOpen
  \bibfield  {author} {\bibinfo {author} {\bibfnamefont {T.}~\bibnamefont
  {Banerjee}}, \bibinfo {author} {\bibfnamefont {S.}~\bibnamefont {Das}},\ and\
  \bibinfo {author} {\bibfnamefont {K.}~\bibnamefont {Sengupta}},\ }\bibfield
  {title} {\bibinfo {title} {{Entanglement asymmetry in periodically driven
  quantum systems}},\ }\href {https://doi.org/10.21468/SciPostPhys.19.2.051}
  {\bibfield  {journal} {\bibinfo  {journal} {SciPost Phys.}\ }\textbf
  {\bibinfo {volume} {19}},\ \bibinfo {pages} {051} (\bibinfo {year} {2025})},\
  \Eprint {https://arxiv.org/abs/2412.03654} {arXiv:2412.03654 [quant-ph]}
  \BibitemShut {NoStop}%
\bibitem [{\citenamefont {Fang}\ \emph {et~al.}(2025)\citenamefont {Fang},
  \citenamefont {Zhou},\ and\ \citenamefont {Wen}}]{2dpapers9}%
  \BibitemOpen
  \bibfield  {author} {\bibinfo {author} {\bibfnamefont {J.}~\bibnamefont
  {Fang}}, \bibinfo {author} {\bibfnamefont {Q.}~\bibnamefont {Zhou}},\ and\
  \bibinfo {author} {\bibfnamefont {X.}~\bibnamefont {Wen}},\ }\bibfield
  {title} {\bibinfo {title} {{Phase transitions in quasiperiodically driven
  quantum critical systems: Analytical results}},\ }\href
  {https://doi.org/10.1103/PhysRevB.111.094304} {\bibfield  {journal} {\bibinfo
   {journal} {Phys. Rev. B}\ }\textbf {\bibinfo {volume} {111}},\ \bibinfo
  {pages} {094304} (\bibinfo {year} {2025})},\ \Eprint
  {https://arxiv.org/abs/2501.04795} {arXiv:2501.04795 [cond-mat.stat-mech]}
  \BibitemShut {NoStop}%
\bibitem [{\citenamefont {Lapierre}\ \emph
  {et~al.}(2025{\natexlab{b}})\citenamefont {Lapierre}, \citenamefont
  {Pelliconi}, \citenamefont {Ryu},\ and\ \citenamefont {Sonner}}]{2dpapers10}%
  \BibitemOpen
  \bibfield  {author} {\bibinfo {author} {\bibfnamefont {B.}~\bibnamefont
  {Lapierre}}, \bibinfo {author} {\bibfnamefont {P.}~\bibnamefont {Pelliconi}},
  \bibinfo {author} {\bibfnamefont {S.}~\bibnamefont {Ryu}},\ and\ \bibinfo
  {author} {\bibfnamefont {J.}~\bibnamefont {Sonner}},\ }\bibfield  {title}
  {\bibinfo {title} {{Driven nonunitary dynamics of quantum critical
  systems}},\ }\href {https://doi.org/10.1103/lwrz-jxrr} {\bibfield  {journal}
  {\bibinfo  {journal} {Phys. Rev. B}\ }\textbf {\bibinfo {volume} {112}},\
  \bibinfo {pages} {104322} (\bibinfo {year} {2025}{\natexlab{b}})},\ \Eprint
  {https://arxiv.org/abs/2505.01508} {arXiv:2505.01508 [cond-mat.str-el]}
  \BibitemShut {NoStop}%
\bibitem [{\citenamefont {Lapierre}\ \emph
  {et~al.}(2025{\natexlab{c}})\citenamefont {Lapierre}, \citenamefont {Mo},\
  and\ \citenamefont {Ryu}}]{2dpapers11}%
  \BibitemOpen
  \bibfield  {author} {\bibinfo {author} {\bibfnamefont {B.}~\bibnamefont
  {Lapierre}}, \bibinfo {author} {\bibfnamefont {L.-H.}\ \bibnamefont {Mo}},\
  and\ \bibinfo {author} {\bibfnamefont {S.}~\bibnamefont {Ryu}},\ }\bibfield
  {title} {\bibinfo {title} {{Entanglement transitions in structured and random
  nonunitary Gaussian circuits}},\ }\href@noop {} {\  (\bibinfo {year}
  {2025}{\natexlab{c}})},\ \Eprint {https://arxiv.org/abs/2507.03768}
  {arXiv:2507.03768 [quant-ph]} \BibitemShut {NoStop}%
\bibitem [{\citenamefont {Wen}\ \emph {et~al.}(2021)\citenamefont {Wen},
  \citenamefont {Fan}, \citenamefont {Vishwanath},\ and\ \citenamefont
  {Gu}}]{dynph2}%
  \BibitemOpen
  \bibfield  {author} {\bibinfo {author} {\bibfnamefont {X.}~\bibnamefont
  {Wen}}, \bibinfo {author} {\bibfnamefont {R.}~\bibnamefont {Fan}}, \bibinfo
  {author} {\bibfnamefont {A.}~\bibnamefont {Vishwanath}},\ and\ \bibinfo
  {author} {\bibfnamefont {Y.}~\bibnamefont {Gu}},\ }\bibfield  {title}
  {\bibinfo {title} {Periodically, quasiperiodically, and randomly driven
  conformal field theories},\ }\href
  {https://doi.org/10.1103/PhysRevResearch.3.023044} {\bibfield  {journal}
  {\bibinfo  {journal} {Phys. Rev. Res.}\ }\textbf {\bibinfo {volume} {3}},\
  \bibinfo {pages} {023044} (\bibinfo {year} {2021})}\BibitemShut {NoStop}%
\bibitem [{\citenamefont {Dey}\ \emph {et~al.}(2026)\citenamefont {Dey},
  \citenamefont {Dutta},\ and\ \citenamefont {Ezhuthachan}}]{2dpapers12}%
  \BibitemOpen
  \bibfield  {author} {\bibinfo {author} {\bibfnamefont {P.}~\bibnamefont
  {Dey}}, \bibinfo {author} {\bibfnamefont {S.}~\bibnamefont {Dutta}},\ and\
  \bibinfo {author} {\bibfnamefont {B.}~\bibnamefont {Ezhuthachan}},\
  }\bibfield  {title} {\bibinfo {title} {{Imprints of dynamical phases in
  semiclassical entanglement entropy in 2D CFT}},\ }\href@noop {} {\  (\bibinfo
  {year} {2026})},\ \Eprint {https://arxiv.org/abs/2606.17625}
  {arXiv:2606.17625 [hep-th]} \BibitemShut {NoStop}%
\bibitem [{\citenamefont {Caputa}\ and\ \citenamefont
  {MacCormack}(2021)}]{2dcurved1}%
  \BibitemOpen
  \bibfield  {author} {\bibinfo {author} {\bibfnamefont {P.}~\bibnamefont
  {Caputa}}\ and\ \bibinfo {author} {\bibfnamefont {I.}~\bibnamefont
  {MacCormack}},\ }\bibfield  {title} {\bibinfo {title} {{Geometry and
  Complexity of Path Integrals in Inhomogeneous CFTs}},\ }\href
  {https://doi.org/10.1007/JHEP01(2021)027} {\bibfield  {journal} {\bibinfo
  {journal} {JHEP}\ }\textbf {\bibinfo {volume} {01}},\ \bibinfo {pages}
  {027}},\ \bibinfo {note} {[Erratum: JHEP 09, 109 (2022)]},\ \Eprint
  {https://arxiv.org/abs/2004.04698} {arXiv:2004.04698 [hep-th]} \BibitemShut
  {NoStop}%
\bibitem [{\citenamefont {de~Boer}\ \emph {et~al.}(2023)\citenamefont
  {de~Boer}, \citenamefont {Godet}, \citenamefont {Kastikainen},\ and\
  \citenamefont {Keski-Vakkuri}}]{2dcurved2}%
  \BibitemOpen
  \bibfield  {author} {\bibinfo {author} {\bibfnamefont {J.}~\bibnamefont
  {de~Boer}}, \bibinfo {author} {\bibfnamefont {V.}~\bibnamefont {Godet}},
  \bibinfo {author} {\bibfnamefont {J.}~\bibnamefont {Kastikainen}},\ and\
  \bibinfo {author} {\bibfnamefont {E.}~\bibnamefont {Keski-Vakkuri}},\
  }\bibfield  {title} {\bibinfo {title} {{Quantum information geometry of
  driven CFTs}},\ }\href {https://doi.org/10.1007/JHEP09(2023)087} {\bibfield
  {journal} {\bibinfo  {journal} {JHEP}\ }\textbf {\bibinfo {volume} {09}},\
  \bibinfo {pages} {087}},\ \Eprint {https://arxiv.org/abs/2306.00099}
  {arXiv:2306.00099 [hep-th]} \BibitemShut {NoStop}%
\bibitem [{\citenamefont {Erdmenger}\ \emph {et~al.}(2026)\citenamefont
  {Erdmenger}, \citenamefont {Kastikainen},\ and\ \citenamefont
  {Schuhmann}}]{2dcurved3}%
  \BibitemOpen
  \bibfield  {author} {\bibinfo {author} {\bibfnamefont {J.}~\bibnamefont
  {Erdmenger}}, \bibinfo {author} {\bibfnamefont {J.}~\bibnamefont
  {Kastikainen}},\ and\ \bibinfo {author} {\bibfnamefont {T.}~\bibnamefont
  {Schuhmann}},\ }\bibfield  {title} {\bibinfo {title} {{Driven inhomogeneous
  CFT as a theory in curved space-time}},\ }\href
  {https://doi.org/10.1007/JHEP02(2026)255} {\bibfield  {journal} {\bibinfo
  {journal} {JHEP}\ }\textbf {\bibinfo {volume} {02}},\ \bibinfo {pages}
  {255}},\ \Eprint {https://arxiv.org/abs/2508.18350} {arXiv:2508.18350
  [hep-th]} \BibitemShut {NoStop}%
\bibitem [{\citenamefont {MacCormack}\ \emph {et~al.}(2019)\citenamefont
  {MacCormack}, \citenamefont {Liu}, \citenamefont {Nozaki},\ and\
  \citenamefont {Ryu}}]{holo1}%
  \BibitemOpen
  \bibfield  {author} {\bibinfo {author} {\bibfnamefont {I.}~\bibnamefont
  {MacCormack}}, \bibinfo {author} {\bibfnamefont {A.}~\bibnamefont {Liu}},
  \bibinfo {author} {\bibfnamefont {M.}~\bibnamefont {Nozaki}},\ and\ \bibinfo
  {author} {\bibfnamefont {S.}~\bibnamefont {Ryu}},\ }\bibfield  {title}
  {\bibinfo {title} {{Holographic Duals of Inhomogeneous Systems: The Rainbow
  Chain and the Sine-Square Deformation Model}},\ }\href
  {https://doi.org/10.1088/1751-8121/ab3944} {\bibfield  {journal} {\bibinfo
  {journal} {J. Phys. A}\ }\textbf {\bibinfo {volume} {52}},\ \bibinfo {pages}
  {505401} (\bibinfo {year} {2019})},\ \Eprint
  {https://arxiv.org/abs/1812.10023} {arXiv:1812.10023 [cond-mat.str-el]}
  \BibitemShut {NoStop}%
\bibitem [{\citenamefont {Das}\ \emph {et~al.}(2023)\citenamefont {Das},
  \citenamefont {Ezhuthachan}, \citenamefont {Kundu}, \citenamefont {Porey},
  \citenamefont {Roy},\ and\ \citenamefont {Sengupta}}]{holo2}%
  \BibitemOpen
  \bibfield  {author} {\bibinfo {author} {\bibfnamefont {S.}~\bibnamefont
  {Das}}, \bibinfo {author} {\bibfnamefont {B.}~\bibnamefont {Ezhuthachan}},
  \bibinfo {author} {\bibfnamefont {A.}~\bibnamefont {Kundu}}, \bibinfo
  {author} {\bibfnamefont {S.}~\bibnamefont {Porey}}, \bibinfo {author}
  {\bibfnamefont {B.}~\bibnamefont {Roy}},\ and\ \bibinfo {author}
  {\bibfnamefont {K.}~\bibnamefont {Sengupta}},\ }\bibfield  {title} {\bibinfo
  {title} {{Brane detectors of a dynamical phase transition in a driven CFT}},\
  }\href {https://doi.org/10.21468/SciPostPhys.15.5.202} {\bibfield  {journal}
  {\bibinfo  {journal} {SciPost Phys.}\ }\textbf {\bibinfo {volume} {15}},\
  \bibinfo {pages} {202} (\bibinfo {year} {2023})},\ \Eprint
  {https://arxiv.org/abs/2212.04201} {arXiv:2212.04201 [hep-th]} \BibitemShut
  {NoStop}%
\bibitem [{\citenamefont {Kudler-Flam}\ \emph {et~al.}(2024)\citenamefont
  {Kudler-Flam}, \citenamefont {Nozaki}, \citenamefont {Numasawa},
  \citenamefont {Ryu},\ and\ \citenamefont {Tan}}]{holo3}%
  \BibitemOpen
  \bibfield  {author} {\bibinfo {author} {\bibfnamefont {J.}~\bibnamefont
  {Kudler-Flam}}, \bibinfo {author} {\bibfnamefont {M.}~\bibnamefont {Nozaki}},
  \bibinfo {author} {\bibfnamefont {T.}~\bibnamefont {Numasawa}}, \bibinfo
  {author} {\bibfnamefont {S.}~\bibnamefont {Ryu}},\ and\ \bibinfo {author}
  {\bibfnamefont {M.~T.}\ \bibnamefont {Tan}},\ }\bibfield  {title} {\bibinfo
  {title} {{Bridging two quantum quench problems {\textemdash} local joining
  quantum quench and M{\"o}bius quench {\textemdash} and their holographic dual
  descriptions}},\ }\href {https://doi.org/10.1007/JHEP08(2024)213} {\bibfield
  {journal} {\bibinfo  {journal} {JHEP}\ }\textbf {\bibinfo {volume} {08}},\
  \bibinfo {pages} {213}},\ \Eprint {https://arxiv.org/abs/2309.04665}
  {arXiv:2309.04665 [hep-th]} \BibitemShut {NoStop}%
\bibitem [{\citenamefont {Jiang}\ and\ \citenamefont {Mezei}(2025)}]{holo4}%
  \BibitemOpen
  \bibfield  {author} {\bibinfo {author} {\bibfnamefont {H.}~\bibnamefont
  {Jiang}}\ and\ \bibinfo {author} {\bibfnamefont {M.}~\bibnamefont {Mezei}},\
  }\bibfield  {title} {\bibinfo {title} {{New horizons for inhomogeneous
  quenches and Floquet CFT}},\ }\href {https://doi.org/10.1007/JHEP04(2025)025}
  {\bibfield  {journal} {\bibinfo  {journal} {JHEP}\ }\textbf {\bibinfo
  {volume} {04}},\ \bibinfo {pages} {025}},\ \Eprint
  {https://arxiv.org/abs/2404.07884} {arXiv:2404.07884 [hep-th]} \BibitemShut
  {NoStop}%
\bibitem [{\citenamefont {Das}\ and\ \citenamefont {Kundu}(2025)}]{holo5}%
  \BibitemOpen
  \bibfield  {author} {\bibinfo {author} {\bibfnamefont {J.}~\bibnamefont
  {Das}}\ and\ \bibinfo {author} {\bibfnamefont {A.}~\bibnamefont {Kundu}},\
  }\bibfield  {title} {\bibinfo {title} {{Flowery horizons {\&} bulk observers:
  sl$^{(q)}$(2,{\,}{\ensuremath{\mathbb{R}}}), drive in 2d holographic CFT}},\
  }\href {https://doi.org/10.1007/JHEP05(2025)035} {\bibfield  {journal}
  {\bibinfo  {journal} {JHEP}\ }\textbf {\bibinfo {volume} {05}},\ \bibinfo
  {pages} {035}},\ \Eprint {https://arxiv.org/abs/2412.18536} {arXiv:2412.18536
  [hep-th]} \BibitemShut {NoStop}%
\bibitem [{\citenamefont {Das}\ \emph {et~al.}(2024)\citenamefont {Das},
  \citenamefont {Das}, \citenamefont {Kundu},\ and\ \citenamefont
  {Sengupta}}]{hd1}%
  \BibitemOpen
  \bibfield  {author} {\bibinfo {author} {\bibfnamefont {D.}~\bibnamefont
  {Das}}, \bibinfo {author} {\bibfnamefont {S.~R.}\ \bibnamefont {Das}},
  \bibinfo {author} {\bibfnamefont {A.}~\bibnamefont {Kundu}},\ and\ \bibinfo
  {author} {\bibfnamefont {K.}~\bibnamefont {Sengupta}},\ }\bibfield  {title}
  {\bibinfo {title} {Exactly solvable floquet dynamics for conformal field
  theories in dimensions greater than two},\ }\href
  {https://doi.org/10.1007/JHEP09(2024)095} {\bibfield  {journal} {\bibinfo
  {journal} {Journal of High Energy Physics}\ }\textbf {\bibinfo {volume}
  {2024}},\ \bibinfo {pages} {95} (\bibinfo {year} {2024})}\BibitemShut
  {NoStop}%
\bibitem [{\citenamefont {Das}\ \emph {et~al.}(2026)\citenamefont {Das},
  \citenamefont {Das}, \citenamefont {Kundu},\ and\ \citenamefont
  {Sengupta}}]{hd2}%
  \BibitemOpen
  \bibfield  {author} {\bibinfo {author} {\bibfnamefont {D.}~\bibnamefont
  {Das}}, \bibinfo {author} {\bibfnamefont {S.~R.}\ \bibnamefont {Das}},
  \bibinfo {author} {\bibfnamefont {A.}~\bibnamefont {Kundu}},\ and\ \bibinfo
  {author} {\bibfnamefont {K.}~\bibnamefont {Sengupta}},\ }\bibfield  {title}
  {\bibinfo {title} {Dynamical phases of higher dimensional floquet cfts},\
  }\href {https://doi.org/10.21468/SciPostPhys.20.2.045} {\bibfield  {journal}
  {\bibinfo  {journal} {SciPost Phys.}\ }\textbf {\bibinfo {volume} {20}},\
  \bibinfo {pages} {045} (\bibinfo {year} {2026})}\BibitemShut {NoStop}%
\bibitem [{\citenamefont {Mo}\ \emph {et~al.}(2026)\citenamefont {Mo},
  \citenamefont {Lapierre},\ and\ \citenamefont {Miao}}]{experiments}%
  \BibitemOpen
  \bibfield  {author} {\bibinfo {author} {\bibfnamefont {L.-H.}\ \bibnamefont
  {Mo}}, \bibinfo {author} {\bibfnamefont {B.}~\bibnamefont {Lapierre}},\ and\
  \bibinfo {author} {\bibfnamefont {Q.}~\bibnamefont {Miao}},\ }\href
  {https://arxiv.org/abs/2605.27530} {\bibinfo {title} {Observing conformal
  floquet dynamics on a digital quantum processor}} (\bibinfo {year} {2026}),\
  \Eprint {https://arxiv.org/abs/2605.27530} {arXiv:2605.27530 [quant-ph]}
  \BibitemShut {NoStop}%
\bibitem [{\citenamefont {Mehen}\ \emph {et~al.}(1999)\citenamefont {Mehen},
  \citenamefont {Stewart},\ and\ \citenamefont {Wise}}]{Mehen}%
  \BibitemOpen
  \bibfield  {author} {\bibinfo {author} {\bibfnamefont {T.}~\bibnamefont
  {Mehen}}, \bibinfo {author} {\bibfnamefont {I.~W.}\ \bibnamefont {Stewart}},\
  and\ \bibinfo {author} {\bibfnamefont {M.~B.}\ \bibnamefont {Wise}},\
  }\bibfield  {title} {\bibinfo {title} {Wigner symmetry in the limit of large
  scattering lengths},\ }\href {https://doi.org/10.1103/PhysRevLett.83.931}
  {\bibfield  {journal} {\bibinfo  {journal} {Phys. Rev. Lett.}\ }\textbf
  {\bibinfo {volume} {83}},\ \bibinfo {pages} {931} (\bibinfo {year}
  {1999})}\BibitemShut {NoStop}%
\bibitem [{\citenamefont {Werner}\ and\ \citenamefont {Castin}(2006)}]{werner}%
  \BibitemOpen
  \bibfield  {author} {\bibinfo {author} {\bibfnamefont {F.}~\bibnamefont
  {Werner}}\ and\ \bibinfo {author} {\bibfnamefont {Y.}~\bibnamefont
  {Castin}},\ }\bibfield  {title} {\bibinfo {title} {Unitary gas in an
  isotropic harmonic trap: Symmetry properties and applications},\ }\href
  {https://doi.org/10.1103/PhysRevA.74.053604} {\bibfield  {journal} {\bibinfo
  {journal} {Phys. Rev. A}\ }\textbf {\bibinfo {volume} {74}},\ \bibinfo
  {pages} {053604} (\bibinfo {year} {2006})}\BibitemShut {NoStop}%
\bibitem [{\citenamefont {Nishida}\ and\ \citenamefont
  {Son}(2007{\natexlab{a}})}]{son1}%
  \BibitemOpen
  \bibfield  {author} {\bibinfo {author} {\bibfnamefont {Y.}~\bibnamefont
  {Nishida}}\ and\ \bibinfo {author} {\bibfnamefont {D.~T.}\ \bibnamefont
  {Son}},\ }\bibfield  {title} {\bibinfo {title} {Fermi gas near unitarity
  around four and two spatial dimensions},\ }\href
  {https://doi.org/10.1103/PhysRevA.75.063617} {\bibfield  {journal} {\bibinfo
  {journal} {Phys. Rev. A}\ }\textbf {\bibinfo {volume} {75}},\ \bibinfo
  {pages} {063617} (\bibinfo {year} {2007}{\natexlab{a}})}\BibitemShut
  {NoStop}%
\bibitem [{\citenamefont {Nishida}\ and\ \citenamefont
  {Son}(2007{\natexlab{b}})}]{son2}%
  \BibitemOpen
  \bibfield  {author} {\bibinfo {author} {\bibfnamefont {Y.}~\bibnamefont
  {Nishida}}\ and\ \bibinfo {author} {\bibfnamefont {D.~T.}\ \bibnamefont
  {Son}},\ }\bibfield  {title} {\bibinfo {title} {Nonrelativistic conformal
  field theories},\ }\href {https://doi.org/10.1103/PhysRevD.76.086004}
  {\bibfield  {journal} {\bibinfo  {journal} {Phys. Rev. D}\ }\textbf {\bibinfo
  {volume} {76}},\ \bibinfo {pages} {086004} (\bibinfo {year}
  {2007}{\natexlab{b}})}\BibitemShut {NoStop}%
\bibitem [{\citenamefont {Chowdhury}\ \emph {et~al.}(2024)\citenamefont
  {Chowdhury}, \citenamefont {Mishra},\ and\ \citenamefont {Son}}]{son3}%
  \BibitemOpen
  \bibfield  {author} {\bibinfo {author} {\bibfnamefont {S.~D.}\ \bibnamefont
  {Chowdhury}}, \bibinfo {author} {\bibfnamefont {R.}~\bibnamefont {Mishra}},\
  and\ \bibinfo {author} {\bibfnamefont {D.~T.}\ \bibnamefont {Son}},\
  }\bibfield  {title} {\bibinfo {title} {Applied nonrelativistic conformal
  field theory: Scattering-length and effective-range corrections to rate of
  production of three neutrons at low relative momenta},\ }\href
  {https://doi.org/10.1103/PhysRevD.109.016001} {\bibfield  {journal} {\bibinfo
   {journal} {Phys. Rev. D}\ }\textbf {\bibinfo {volume} {109}},\ \bibinfo
  {pages} {016001} (\bibinfo {year} {2024})}\BibitemShut {NoStop}%
\bibitem [{\citenamefont {Boisvert}\ \emph {et~al.}(2026)\citenamefont
  {Boisvert}, \citenamefont {Fadda}, \citenamefont {Kulp},\ and\ \citenamefont
  {Yazdi}}]{Boisvert:2025hex}%
  \BibitemOpen
  \bibfield  {author} {\bibinfo {author} {\bibfnamefont {M.}~\bibnamefont
  {Boisvert}}, \bibinfo {author} {\bibfnamefont {S.~H.}\ \bibnamefont {Fadda}},
  \bibinfo {author} {\bibfnamefont {J.}~\bibnamefont {Kulp}},\ and\ \bibinfo
  {author} {\bibfnamefont {R.~M.}\ \bibnamefont {Yazdi}},\ }\href
  {https://arxiv.org/abs/2510.26872} {\bibinfo {title} {Revisiting
  schr\"odinger cfts: Factorization, massless particles, and a path to the
  bootstrap}} (\bibinfo {year} {2026}),\ \Eprint
  {https://arxiv.org/abs/2510.26872} {arXiv:2510.26872 [hep-th]} \BibitemShut
  {NoStop}%
\bibitem [{\citenamefont {Guan}\ \emph {et~al.}(2013)\citenamefont {Guan},
  \citenamefont {Batchelor},\ and\ \citenamefont {Lee}}]{fermirev1}%
  \BibitemOpen
  \bibfield  {author} {\bibinfo {author} {\bibfnamefont {X.-W.}\ \bibnamefont
  {Guan}}, \bibinfo {author} {\bibfnamefont {M.~T.}\ \bibnamefont
  {Batchelor}},\ and\ \bibinfo {author} {\bibfnamefont {C.}~\bibnamefont
  {Lee}},\ }\bibfield  {title} {\bibinfo {title} {Fermi gases in one dimension:
  From bethe ansatz to experiments},\ }\href
  {https://doi.org/10.1103/RevModPhys.85.1633} {\bibfield  {journal} {\bibinfo
  {journal} {Rev. Mod. Phys.}\ }\textbf {\bibinfo {volume} {85}},\ \bibinfo
  {pages} {1633} (\bibinfo {year} {2013})}\BibitemShut {NoStop}%
\bibitem [{\citenamefont {Gubbels}\ and\ \citenamefont
  {Stoof}(2013)}]{fermirev2}%
  \BibitemOpen
  \bibfield  {author} {\bibinfo {author} {\bibfnamefont {K.}~\bibnamefont
  {Gubbels}}\ and\ \bibinfo {author} {\bibfnamefont {H.}~\bibnamefont
  {Stoof}},\ }\bibfield  {title} {\bibinfo {title} {Imbalanced fermi gases at
  unitarity},\ }\href
  {https://doi.org/https://doi.org/10.1016/j.physrep.2012.11.004} {\bibfield
  {journal} {\bibinfo  {journal} {Physics Reports}\ }\textbf {\bibinfo {volume}
  {525}},\ \bibinfo {pages} {255} (\bibinfo {year} {2013})},\ \bibinfo {note}
  {imbalanced Fermi Gases at Unitarity}\BibitemShut {NoStop}%
\bibitem [{\citenamefont {Burovski}\ \emph {et~al.}(2006)\citenamefont
  {Burovski}, \citenamefont {Prokof'ev}, \citenamefont {Svistunov},\ and\
  \citenamefont {Troyer}}]{fermirev3}%
  \BibitemOpen
  \bibfield  {author} {\bibinfo {author} {\bibfnamefont {E.}~\bibnamefont
  {Burovski}}, \bibinfo {author} {\bibfnamefont {N.}~\bibnamefont {Prokof'ev}},
  \bibinfo {author} {\bibfnamefont {B.}~\bibnamefont {Svistunov}},\ and\
  \bibinfo {author} {\bibfnamefont {M.}~\bibnamefont {Troyer}},\ }\bibfield
  {title} {\bibinfo {title} {The fermi–hubbard model at unitarity},\ }\href
  {https://doi.org/10.1088/1367-2630/8/8/153} {\bibfield  {journal} {\bibinfo
  {journal} {New Journal of Physics}\ }\textbf {\bibinfo {volume} {8}},\
  \bibinfo {pages} {153} (\bibinfo {year} {2006})}\BibitemShut {NoStop}%
\bibitem [{\citenamefont {Nishida}\ and\ \citenamefont
  {Son}(2011)}]{fermirev4}%
  \BibitemOpen
  \bibfield  {author} {\bibinfo {author} {\bibfnamefont {Y.}~\bibnamefont
  {Nishida}}\ and\ \bibinfo {author} {\bibfnamefont {D.~T.}\ \bibnamefont
  {Son}},\ }\bibinfo {title} {Unitary fermi gas, $\epsilon$ expansion, and
  nonrelativistic conformal field theories},\ in\ \href
  {https://doi.org/10.1007/978-3-642-21978-8_7} {\emph {\bibinfo {booktitle}
  {The BCS-BEC Crossover and the Unitary Fermi Gas}}}\ (\bibinfo  {publisher}
  {Springer Berlin Heidelberg},\ \bibinfo {year} {2011})\ p.\ \bibinfo {pages}
  {233–275}\BibitemShut {NoStop}%
\bibitem [{\citenamefont {Moroz}(2012)}]{moroz1}%
  \BibitemOpen
  \bibfield  {author} {\bibinfo {author} {\bibfnamefont {S.}~\bibnamefont
  {Moroz}},\ }\bibfield  {title} {\bibinfo {title} {Scale-invariant fermi gas
  in a time-dependent harmonic potential},\ }\href
  {https://doi.org/10.1103/PhysRevA.86.011601} {\bibfield  {journal} {\bibinfo
  {journal} {Phys. Rev. A}\ }\textbf {\bibinfo {volume} {86}},\ \bibinfo
  {pages} {011601(R)} (\bibinfo {year} {2012})}\BibitemShut {NoStop}%
\bibitem [{\citenamefont {Maki}\ and\ \citenamefont {Zhou}(2019)}]{maki2019}%
  \BibitemOpen
  \bibfield  {author} {\bibinfo {author} {\bibfnamefont {J.}~\bibnamefont
  {Maki}}\ and\ \bibinfo {author} {\bibfnamefont {F.}~\bibnamefont {Zhou}},\
  }\bibfield  {title} {\bibinfo {title} {Quantum many-body conformal dynamics:
  Symmetries, geometry, conformal tower states, and entropy production},\
  }\href {https://doi.org/10.1103/PhysRevA.100.023601} {\bibfield  {journal}
  {\bibinfo  {journal} {Phys. Rev. A}\ }\textbf {\bibinfo {volume} {100}},\
  \bibinfo {pages} {023601} (\bibinfo {year} {2019})}\BibitemShut {NoStop}%
\bibitem [{\citenamefont {Maki}\ and\ \citenamefont {Zhou}(2020)}]{maki2020}%
  \BibitemOpen
  \bibfield  {author} {\bibinfo {author} {\bibfnamefont {J.}~\bibnamefont
  {Maki}}\ and\ \bibinfo {author} {\bibfnamefont {F.}~\bibnamefont {Zhou}},\
  }\bibfield  {title} {\bibinfo {title} {Far-away-from-equilibrium
  quantum-critical conformal dynamics: Reversibility, thermalization, and
  hydrodynamics},\ }\href {https://doi.org/10.1103/PhysRevA.102.063319}
  {\bibfield  {journal} {\bibinfo  {journal} {Phys. Rev. A}\ }\textbf {\bibinfo
  {volume} {102}},\ \bibinfo {pages} {063319} (\bibinfo {year}
  {2020})}\BibitemShut {NoStop}%
\bibitem [{\citenamefont {Maki}\ and\ \citenamefont {Zhou}(2024)}]{maki2024}%
  \BibitemOpen
  \bibfield  {author} {\bibinfo {author} {\bibfnamefont {J.}~\bibnamefont
  {Maki}}\ and\ \bibinfo {author} {\bibfnamefont {F.}~\bibnamefont {Zhou}},\
  }\bibfield  {title} {\bibinfo {title} {Signatures of conformal symmetry in
  the dynamics of quantum gases: A cyclic quantum state and entanglement
  entropy},\ }\href {https://doi.org/10.1103/PhysRevA.110.023312} {\bibfield
  {journal} {\bibinfo  {journal} {Phys. Rev. A}\ }\textbf {\bibinfo {volume}
  {110}},\ \bibinfo {pages} {023312} (\bibinfo {year} {2024})}\BibitemShut
  {NoStop}%
\bibitem [{\citenamefont {Deng}\ \emph {et~al.}(2016)\citenamefont {Deng},
  \citenamefont {Shi}, \citenamefont {Diao}, \citenamefont {Yu}, \citenamefont
  {Zhai}, \citenamefont {Qi},\ and\ \citenamefont {Wu}}]{deng1}%
  \BibitemOpen
  \bibfield  {author} {\bibinfo {author} {\bibfnamefont {S.}~\bibnamefont
  {Deng}}, \bibinfo {author} {\bibfnamefont {Z.-Y.}\ \bibnamefont {Shi}},
  \bibinfo {author} {\bibfnamefont {P.}~\bibnamefont {Diao}}, \bibinfo {author}
  {\bibfnamefont {Q.}~\bibnamefont {Yu}}, \bibinfo {author} {\bibfnamefont
  {H.}~\bibnamefont {Zhai}}, \bibinfo {author} {\bibfnamefont {R.}~\bibnamefont
  {Qi}},\ and\ \bibinfo {author} {\bibfnamefont {H.}~\bibnamefont {Wu}},\
  }\bibfield  {title} {\bibinfo {title} {Observation of the efimovian expansion
  in scale-invariant fermi gases},\ }\href
  {https://doi.org/10.1126/science.aaf0666} {\bibfield  {journal} {\bibinfo
  {journal} {Science}\ }\textbf {\bibinfo {volume} {353}},\ \bibinfo {pages}
  {371} (\bibinfo {year} {2016})},\ \Eprint
  {https://arxiv.org/abs/https://www.science.org/doi/pdf/10.1126/science.aaf0666}
  {https://www.science.org/doi/pdf/10.1126/science.aaf0666} \BibitemShut
  {NoStop}%
\bibitem [{\citenamefont {DENG}\ and\ \citenamefont {WU}(2019)}]{deng2}%
  \BibitemOpen
  \bibfield  {author} {\bibinfo {author} {\bibfnamefont {S.}~\bibnamefont
  {DENG}}\ and\ \bibinfo {author} {\bibfnamefont {H.}~\bibnamefont {WU}},\
  }\bibfield  {title} {\bibinfo {title} {An introduction of ultracold strongly
  interacting fermi gases},\ }\href
  {https://doi.org/10.3969/j.issn.0253-9608.2019.01.002} {\bibfield  {journal}
  {\bibinfo  {journal} {Chinese Journal of Nature}\ }\textbf {\bibinfo {volume}
  {41}},\ \bibinfo {pages} {8} (\bibinfo {year} {2019})}\BibitemShut {NoStop}%
\bibitem [{\citenamefont {Chin}\ \emph {et~al.}(2010)\citenamefont {Chin},
  \citenamefont {Grimm}, \citenamefont {Julienne},\ and\ \citenamefont
  {Tiesinga}}]{feshrev}%
  \BibitemOpen
  \bibfield  {author} {\bibinfo {author} {\bibfnamefont {C.}~\bibnamefont
  {Chin}}, \bibinfo {author} {\bibfnamefont {R.}~\bibnamefont {Grimm}},
  \bibinfo {author} {\bibfnamefont {P.}~\bibnamefont {Julienne}},\ and\
  \bibinfo {author} {\bibfnamefont {E.}~\bibnamefont {Tiesinga}},\ }\bibfield
  {title} {\bibinfo {title} {Feshbach resonances in ultracold gases},\ }\href
  {https://doi.org/10.1103/RevModPhys.82.1225} {\bibfield  {journal} {\bibinfo
  {journal} {Rev. Mod. Phys.}\ }\textbf {\bibinfo {volume} {82}},\ \bibinfo
  {pages} {1225} (\bibinfo {year} {2010})}\BibitemShut {NoStop}%
\bibitem [{\citenamefont {Pitaevskii}\ and\ \citenamefont
  {Rosch}(1997)}]{pitaevski1}%
  \BibitemOpen
  \bibfield  {author} {\bibinfo {author} {\bibfnamefont {L.~P.}\ \bibnamefont
  {Pitaevskii}}\ and\ \bibinfo {author} {\bibfnamefont {A.}~\bibnamefont
  {Rosch}},\ }\bibfield  {title} {\bibinfo {title} {Breathing modes and hidden
  symmetry of trapped atoms in two dimensions},\ }\href
  {https://doi.org/10.1103/PhysRevA.55.R853} {\bibfield  {journal} {\bibinfo
  {journal} {Phys. Rev. A}\ }\textbf {\bibinfo {volume} {55}},\ \bibinfo
  {pages} {R853(R)} (\bibinfo {year} {1997})}\BibitemShut {NoStop}%
\bibitem [{\citenamefont {Saint-Jalm}\ \emph {et~al.}(2019)\citenamefont
  {Saint-Jalm}, \citenamefont {Castilho}, \citenamefont {Le~Cerf},
  \citenamefont {Bakkali-Hassani}, \citenamefont {Ville}, \citenamefont
  {Nascimbene}, \citenamefont {Beugnon},\ and\ \citenamefont
  {Dalibard}}]{dalibard1}%
  \BibitemOpen
  \bibfield  {author} {\bibinfo {author} {\bibfnamefont {R.}~\bibnamefont
  {Saint-Jalm}}, \bibinfo {author} {\bibfnamefont {P.~C.~M.}\ \bibnamefont
  {Castilho}}, \bibinfo {author} {\bibfnamefont {E.}~\bibnamefont {Le~Cerf}},
  \bibinfo {author} {\bibfnamefont {B.}~\bibnamefont {Bakkali-Hassani}},
  \bibinfo {author} {\bibfnamefont {J.-L.}\ \bibnamefont {Ville}}, \bibinfo
  {author} {\bibfnamefont {S.}~\bibnamefont {Nascimbene}}, \bibinfo {author}
  {\bibfnamefont {J.}~\bibnamefont {Beugnon}},\ and\ \bibinfo {author}
  {\bibfnamefont {J.}~\bibnamefont {Dalibard}},\ }\bibfield  {title} {\bibinfo
  {title} {Dynamical symmetry and breathers in a two-dimensional bose gas},\
  }\href {https://doi.org/10.1103/PhysRevX.9.021035} {\bibfield  {journal}
  {\bibinfo  {journal} {Phys. Rev. X}\ }\textbf {\bibinfo {volume} {9}},\
  \bibinfo {pages} {021035} (\bibinfo {year} {2019})}\BibitemShut {NoStop}%
\bibitem [{\citenamefont {Minguzzi}\ and\ \citenamefont
  {Gangardt}(2005)}]{tonks1}%
  \BibitemOpen
  \bibfield  {author} {\bibinfo {author} {\bibfnamefont {A.}~\bibnamefont
  {Minguzzi}}\ and\ \bibinfo {author} {\bibfnamefont {D.~M.}\ \bibnamefont
  {Gangardt}},\ }\bibfield  {title} {\bibinfo {title} {Exact coherent states of
  a harmonically confined tonks-girardeau gas},\ }\href
  {https://doi.org/10.1103/PhysRevLett.94.240404} {\bibfield  {journal}
  {\bibinfo  {journal} {Phys. Rev. Lett.}\ }\textbf {\bibinfo {volume} {94}},\
  \bibinfo {pages} {240404} (\bibinfo {year} {2005})}\BibitemShut {NoStop}%
\bibitem [{\citenamefont {Polychronakos}(2006)}]{calogero1}%
  \BibitemOpen
  \bibfield  {author} {\bibinfo {author} {\bibfnamefont {A.~P.}\ \bibnamefont
  {Polychronakos}},\ }\bibfield  {title} {\bibinfo {title} {{Physics and
  Mathematics of Calogero particles}},\ }\href
  {https://doi.org/10.1088/0305-4470/39/41/S07} {\bibfield  {journal} {\bibinfo
   {journal} {J. Phys. A}\ }\textbf {\bibinfo {volume} {39}},\ \bibinfo {pages}
  {12793} (\bibinfo {year} {2006})},\ \Eprint
  {https://arxiv.org/abs/hep-th/0607033} {arXiv:hep-th/0607033} \BibitemShut
  {NoStop}%
\bibitem [{\citenamefont {Manuel}\ and\ \citenamefont
  {Tarrach}(1993)}]{manuel}%
  \BibitemOpen
  \bibfield  {author} {\bibinfo {author} {\bibfnamefont {C.}~\bibnamefont
  {Manuel}}\ and\ \bibinfo {author} {\bibfnamefont {R.}~\bibnamefont
  {Tarrach}},\ }\bibfield  {title} {\bibinfo {title} {Contact interactions and
  dirac anyons},\ }\href
  {https://doi.org/https://doi.org/10.1016/0370-2693(93)90723-U} {\bibfield
  {journal} {\bibinfo  {journal} {Physics Letters B}\ }\textbf {\bibinfo
  {volume} {301}},\ \bibinfo {pages} {72} (\bibinfo {year} {1993})}\BibitemShut
  {NoStop}%
\bibitem [{\citenamefont {Chou}(1991)}]{chou}%
  \BibitemOpen
  \bibfield  {author} {\bibinfo {author} {\bibfnamefont {C.}~\bibnamefont
  {Chou}},\ }\bibfield  {title} {\bibinfo {title} {{The Multi - anyon spectra
  and wave functions}},\ }\href {https://doi.org/10.1103/PhysRevD.44.2533}
  {\bibfield  {journal} {\bibinfo  {journal} {Phys. Rev. D}\ }\textbf {\bibinfo
  {volume} {44}},\ \bibinfo {pages} {2533} (\bibinfo {year} {1991})},\ \bibinfo
  {note} {[Erratum: Phys.Rev.D 45, 1433 (1992)]}\BibitemShut {NoStop}%
\bibitem [{\citenamefont {Bergman}\ and\ \citenamefont
  {Lozano}(1994)}]{bergman}%
  \BibitemOpen
  \bibfield  {author} {\bibinfo {author} {\bibfnamefont {O.}~\bibnamefont
  {Bergman}}\ and\ \bibinfo {author} {\bibfnamefont {G.}~\bibnamefont
  {Lozano}},\ }\bibfield  {title} {\bibinfo {title} {Aharonov-bohm scattering,
  contact interactions, and scale invariance},\ }\href
  {https://doi.org/https://doi.org/10.1006/aphy.1994.1013} {\bibfield
  {journal} {\bibinfo  {journal} {Annals of Physics}\ }\textbf {\bibinfo
  {volume} {229}},\ \bibinfo {pages} {416} (\bibinfo {year}
  {1994})}\BibitemShut {NoStop}%
\bibitem [{\citenamefont {Hammer}\ and\ \citenamefont {Son}(2021)}]{nuc1}%
  \BibitemOpen
  \bibfield  {author} {\bibinfo {author} {\bibfnamefont {H.-W.}\ \bibnamefont
  {Hammer}}\ and\ \bibinfo {author} {\bibfnamefont {D.~T.}\ \bibnamefont
  {Son}},\ }\bibfield  {title} {\bibinfo {title} {Unnuclear physics: Conformal
  symmetry in nuclear reactions},\ }\href
  {https://doi.org/10.1073/pnas.2108716118} {\bibfield  {journal} {\bibinfo
  {journal} {Proceedings of the National Academy of Sciences}\ }\textbf
  {\bibinfo {volume} {118}},\ \bibinfo {pages} {e2108716118} (\bibinfo {year}
  {2021})},\ \Eprint
  {https://arxiv.org/abs/https://www.pnas.org/doi/pdf/10.1073/pnas.2108716118}
  {https://www.pnas.org/doi/pdf/10.1073/pnas.2108716118} \BibitemShut {NoStop}%
\bibitem [{\citenamefont {Balasubramanian}\ and\ \citenamefont
  {McGreevy}(2008)}]{dual1}%
  \BibitemOpen
  \bibfield  {author} {\bibinfo {author} {\bibfnamefont {K.}~\bibnamefont
  {Balasubramanian}}\ and\ \bibinfo {author} {\bibfnamefont {J.}~\bibnamefont
  {McGreevy}},\ }\bibfield  {title} {\bibinfo {title} {Gravity duals for
  nonrelativistic conformal field theories},\ }\href
  {https://doi.org/10.1103/PhysRevLett.101.061601} {\bibfield  {journal}
  {\bibinfo  {journal} {Phys. Rev. Lett.}\ }\textbf {\bibinfo {volume} {101}},\
  \bibinfo {pages} {061601} (\bibinfo {year} {2008})}\BibitemShut {NoStop}%
\bibitem [{\citenamefont {Son}(2008)}]{dual2}%
  \BibitemOpen
  \bibfield  {author} {\bibinfo {author} {\bibfnamefont {D.~T.}\ \bibnamefont
  {Son}},\ }\bibfield  {title} {\bibinfo {title} {Toward an ads/cold atoms
  correspondence: A geometric realization of the schr\"odinger symmetry},\
  }\href {https://doi.org/10.1103/PhysRevD.78.046003} {\bibfield  {journal}
  {\bibinfo  {journal} {Phys. Rev. D}\ }\textbf {\bibinfo {volume} {78}},\
  \bibinfo {pages} {046003} (\bibinfo {year} {2008})}\BibitemShut {NoStop}%
\bibitem [{\citenamefont {Scopa}\ \emph {et~al.}(2018)\citenamefont {Scopa},
  \citenamefont {Unterberger},\ and\ \citenamefont {Karevski}}]{scopa1}%
  \BibitemOpen
  \bibfield  {author} {\bibinfo {author} {\bibfnamefont {S.}~\bibnamefont
  {Scopa}}, \bibinfo {author} {\bibfnamefont {J.}~\bibnamefont {Unterberger}},\
  and\ \bibinfo {author} {\bibfnamefont {D.}~\bibnamefont {Karevski}},\
  }\bibfield  {title} {\bibinfo {title} {Exact dynamics of a one dimensional
  bose gas in a periodic time-dependent harmonic trap},\ }\href
  {https://doi.org/10.1088/1751-8121/aab8a5} {\bibfield  {journal} {\bibinfo
  {journal} {Journal of Physics A: Mathematical and Theoretical}\ }\textbf
  {\bibinfo {volume} {51}},\ \bibinfo {pages} {185001} (\bibinfo {year}
  {2018})}\BibitemShut {NoStop}%
\bibitem [{\citenamefont {Sun}\ \emph {et~al.}(2025)\citenamefont {Sun},
  \citenamefont {Min}, \citenamefont {Yan}, \citenamefont {Wang}, \citenamefont
  {Xie}, \citenamefont {Wu}, \citenamefont {Maki}, \citenamefont {Zhang},
  \citenamefont {Peng}, \citenamefont {Zhan},\ and\ \citenamefont
  {Jiang}}]{pra1}%
  \BibitemOpen
  \bibfield  {author} {\bibinfo {author} {\bibfnamefont {D.}~\bibnamefont
  {Sun}}, \bibinfo {author} {\bibfnamefont {J.}~\bibnamefont {Min}}, \bibinfo
  {author} {\bibfnamefont {X.}~\bibnamefont {Yan}}, \bibinfo {author}
  {\bibfnamefont {L.}~\bibnamefont {Wang}}, \bibinfo {author} {\bibfnamefont
  {X.}~\bibnamefont {Xie}}, \bibinfo {author} {\bibfnamefont {X.}~\bibnamefont
  {Wu}}, \bibinfo {author} {\bibfnamefont {J.}~\bibnamefont {Maki}}, \bibinfo
  {author} {\bibfnamefont {S.}~\bibnamefont {Zhang}}, \bibinfo {author}
  {\bibfnamefont {S.-G.}\ \bibnamefont {Peng}}, \bibinfo {author}
  {\bibfnamefont {M.}~\bibnamefont {Zhan}},\ and\ \bibinfo {author}
  {\bibfnamefont {K.}~\bibnamefont {Jiang}},\ }\bibfield  {title} {\bibinfo
  {title} {Persistent breather and dynamical symmetry in a unitary fermi gas},\
  }\href {https://doi.org/10.1103/PhysRevA.111.053317} {\bibfield  {journal}
  {\bibinfo  {journal} {Phys. Rev. A}\ }\textbf {\bibinfo {volume} {111}},\
  \bibinfo {pages} {053317} (\bibinfo {year} {2025})}\BibitemShut {NoStop}%
\bibitem [{\citenamefont {Yan}\ \emph {et~al.}(2026)\citenamefont {Yan},
  \citenamefont {Min}, \citenamefont {Sun}, \citenamefont {Peng}, \citenamefont
  {Xie}, \citenamefont {Wu}, \citenamefont {Li}, \citenamefont {Yang},\ and\
  \citenamefont {Jiang}}]{pra2}%
  \BibitemOpen
  \bibfield  {author} {\bibinfo {author} {\bibfnamefont {X.}~\bibnamefont
  {Yan}}, \bibinfo {author} {\bibfnamefont {J.}~\bibnamefont {Min}}, \bibinfo
  {author} {\bibfnamefont {D.}~\bibnamefont {Sun}}, \bibinfo {author}
  {\bibfnamefont {S.-G.}\ \bibnamefont {Peng}}, \bibinfo {author}
  {\bibfnamefont {X.}~\bibnamefont {Xie}}, \bibinfo {author} {\bibfnamefont
  {X.}~\bibnamefont {Wu}}, \bibinfo {author} {\bibfnamefont {J.}~\bibnamefont
  {Li}}, \bibinfo {author} {\bibfnamefont {W.}~\bibnamefont {Yang}},\ and\
  \bibinfo {author} {\bibfnamefont {K.}~\bibnamefont {Jiang}},\ }\bibfield
  {title} {\bibinfo {title} {Energy dynamics of a nonequilibrium unitary fermi
  gas},\ }\href {https://doi.org/10.1103/1jfx-yntk} {\bibfield  {journal}
  {\bibinfo  {journal} {Phys. Rev. A}\ }\textbf {\bibinfo {volume} {113}},\
  \bibinfo {pages} {053312} (\bibinfo {year} {2026})}\BibitemShut {NoStop}%
\bibitem [{\citenamefont {Pinney}(1950)}]{Pinney1950}%
  \BibitemOpen
  \bibfield  {author} {\bibinfo {author} {\bibfnamefont {E.}~\bibnamefont
  {Pinney}},\ }\bibfield  {title} {\bibinfo {title} {The nonlinear differential
  equation $y''+p(x)y+cy^{-3}=0$},\ }\href
  {https://doi.org/10.1090/S0002-9939-1950-0037979-4} {\bibfield  {journal}
  {\bibinfo  {journal} {Proceedings of the American Mathematical Society}\
  }\textbf {\bibinfo {volume} {1}},\ \bibinfo {pages} {681} (\bibinfo {year}
  {1950})}\BibitemShut {NoStop}%
\bibitem [{\citenamefont {Lewis}(1968)}]{Lewis1968}%
  \BibitemOpen
  \bibfield  {author} {\bibinfo {author} {\bibfnamefont {H.~R.}\ \bibnamefont
  {Lewis}},\ }\bibfield  {title} {\bibinfo {title} {Class of exact invariants
  for classical and quantum time-dependent harmonic oscillators},\ }\href
  {https://doi.org/10.1063/1.1664532} {\bibfield  {journal} {\bibinfo
  {journal} {Journal of Mathematical Physics}\ }\textbf {\bibinfo {volume}
  {9}},\ \bibinfo {pages} {1976} (\bibinfo {year} {1968})}\BibitemShut
  {NoStop}%
\bibitem [{sic()}]{sicit}%
  \BibitemOpen
  \href@noop {} {\bibinfo {title} {See {S}upplemental {M}aterial for
  details}}\BibitemShut {NoStop}%
\bibitem [{\citenamefont {Martínez-Tibaduiza}\ \emph
  {et~al.}(2020)\citenamefont {Martínez-Tibaduiza}, \citenamefont {Aragão},
  \citenamefont {Farina},\ and\ \citenamefont {Zarro}}]{bch1}%
  \BibitemOpen
  \bibfield  {author} {\bibinfo {author} {\bibfnamefont {D.}~\bibnamefont
  {Martínez-Tibaduiza}}, \bibinfo {author} {\bibfnamefont {A.}~\bibnamefont
  {Aragão}}, \bibinfo {author} {\bibfnamefont {C.}~\bibnamefont {Farina}},\
  and\ \bibinfo {author} {\bibfnamefont {C.}~\bibnamefont {Zarro}},\ }\bibfield
   {title} {\bibinfo {title} {New bch-like relations of the su(1,1), su(2) and
  so(2,1) lie algebras},\ }\href
  {https://doi.org/https://doi.org/10.1016/j.physleta.2020.126937} {\bibfield
  {journal} {\bibinfo  {journal} {Physics Letters A}\ }\textbf {\bibinfo
  {volume} {384}},\ \bibinfo {pages} {126937} (\bibinfo {year}
  {2020})}\BibitemShut {NoStop}%
\bibitem [{\citenamefont {Emerson}\ \emph {et~al.}(2002)\citenamefont
  {Emerson}, \citenamefont {Weinstein}, \citenamefont {Lloyd},\ and\
  \citenamefont {Cory}}]{Emerson:2002lgj}%
  \BibitemOpen
  \bibfield  {author} {\bibinfo {author} {\bibfnamefont {J.}~\bibnamefont
  {Emerson}}, \bibinfo {author} {\bibfnamefont {Y.~S.}\ \bibnamefont
  {Weinstein}}, \bibinfo {author} {\bibfnamefont {S.}~\bibnamefont {Lloyd}},\
  and\ \bibinfo {author} {\bibfnamefont {D.~G.}\ \bibnamefont {Cory}},\
  }\bibfield  {title} {\bibinfo {title} {{Fidelity Decay as an Efficient
  Indicator of Quantum Chaos}},\ }\href
  {https://doi.org/10.1103/PhysRevLett.89.284102} {\bibfield  {journal}
  {\bibinfo  {journal} {Phys. Rev. Lett.}\ }\textbf {\bibinfo {volume} {89}},\
  \bibinfo {pages} {284102} (\bibinfo {year} {2002})},\ \Eprint
  {https://arxiv.org/abs/quant-ph/0207099} {arXiv:quant-ph/0207099}
  \BibitemShut {NoStop}%
\bibitem [{\citenamefont {Adams}\ \emph {et~al.}(2008)\citenamefont {Adams},
  \citenamefont {Balasubramanian},\ and\ \citenamefont
  {McGreevy}}]{Adams_2008}%
  \BibitemOpen
  \bibfield  {author} {\bibinfo {author} {\bibfnamefont {A.}~\bibnamefont
  {Adams}}, \bibinfo {author} {\bibfnamefont {K.}~\bibnamefont
  {Balasubramanian}},\ and\ \bibinfo {author} {\bibfnamefont {J.}~\bibnamefont
  {McGreevy}},\ }\bibfield  {title} {\bibinfo {title} {Hot spacetimes for cold
  atoms},\ }\href {https://doi.org/10.1088/1126-6708/2008/11/059} {\bibfield
  {journal} {\bibinfo  {journal} {Journal of High Energy Physics}\ }\textbf
  {\bibinfo {volume} {2008}},\ \bibinfo {pages} {059–059} (\bibinfo {year}
  {2008})}\BibitemShut {NoStop}%
\bibitem [{\citenamefont {Herzog}\ \emph {et~al.}(2008)\citenamefont {Herzog},
  \citenamefont {Rangamani},\ and\ \citenamefont {Ross}}]{Herzog:2008wg}%
  \BibitemOpen
  \bibfield  {author} {\bibinfo {author} {\bibfnamefont {C.~P.}\ \bibnamefont
  {Herzog}}, \bibinfo {author} {\bibfnamefont {M.}~\bibnamefont {Rangamani}},\
  and\ \bibinfo {author} {\bibfnamefont {S.~F.}\ \bibnamefont {Ross}},\
  }\bibfield  {title} {\bibinfo {title} {{Heating up Galilean holography}},\
  }\href {https://doi.org/10.1088/1126-6708/2008/11/080} {\bibfield  {journal}
  {\bibinfo  {journal} {JHEP}\ }\textbf {\bibinfo {volume} {11}},\ \bibinfo
  {pages} {080}},\ \Eprint {https://arxiv.org/abs/0807.1099} {arXiv:0807.1099
  [hep-th]} \BibitemShut {NoStop}%
\bibitem [{\citenamefont {Maldacena}\ \emph {et~al.}(2008)\citenamefont
  {Maldacena}, \citenamefont {Martelli},\ and\ \citenamefont
  {Tachikawa}}]{Maldacena:2008wh}%
  \BibitemOpen
  \bibfield  {author} {\bibinfo {author} {\bibfnamefont {J.}~\bibnamefont
  {Maldacena}}, \bibinfo {author} {\bibfnamefont {D.}~\bibnamefont
  {Martelli}},\ and\ \bibinfo {author} {\bibfnamefont {Y.}~\bibnamefont
  {Tachikawa}},\ }\bibfield  {title} {\bibinfo {title} {{Comments on string
  theory backgrounds with non-relativistic conformal symmetry}},\ }\href
  {https://doi.org/10.1088/1126-6708/2008/10/072} {\bibfield  {journal}
  {\bibinfo  {journal} {JHEP}\ }\textbf {\bibinfo {volume} {10}},\ \bibinfo
  {pages} {072}},\ \Eprint {https://arxiv.org/abs/0807.1100} {arXiv:0807.1100
  [hep-th]} \BibitemShut {NoStop}%
\bibitem [{\citenamefont {Rangamani}\ \emph {et~al.}(2015)\citenamefont
  {Rangamani}, \citenamefont {Rozali},\ and\ \citenamefont
  {Wong}}]{Rangamani2015}%
  \BibitemOpen
  \bibfield  {author} {\bibinfo {author} {\bibfnamefont {M.}~\bibnamefont
  {Rangamani}}, \bibinfo {author} {\bibfnamefont {M.}~\bibnamefont {Rozali}},\
  and\ \bibinfo {author} {\bibfnamefont {A.}~\bibnamefont {Wong}},\ }\bibfield
  {title} {\bibinfo {title} {Driven holographic cfts},\ }\href
  {https://doi.org/10.1007/JHEP04(2015)093} {\bibfield  {journal} {\bibinfo
  {journal} {Journal of High Energy Physics}\ }\textbf {\bibinfo {volume}
  {2015}},\ \bibinfo {pages} {93} (\bibinfo {year} {2015})}\BibitemShut
  {NoStop}%
\bibitem [{\citenamefont {Hashimoto}\ \emph {et~al.}(2017)\citenamefont
  {Hashimoto}, \citenamefont {Kinoshita}, \citenamefont {Murata},\ and\
  \citenamefont {Oka}}]{Hashimoto2016}%
  \BibitemOpen
  \bibfield  {author} {\bibinfo {author} {\bibfnamefont {K.}~\bibnamefont
  {Hashimoto}}, \bibinfo {author} {\bibfnamefont {S.}~\bibnamefont
  {Kinoshita}}, \bibinfo {author} {\bibfnamefont {K.}~\bibnamefont {Murata}},\
  and\ \bibinfo {author} {\bibfnamefont {T.}~\bibnamefont {Oka}},\ }\bibfield
  {title} {\bibinfo {title} {{Holographic Floquet states I: a strongly coupled
  Weyl semimetal}},\ }\href {https://doi.org/10.1007/JHEP05(2017)127}
  {\bibfield  {journal} {\bibinfo  {journal} {JHEP}\ }\textbf {\bibinfo
  {volume} {05}},\ \bibinfo {pages} {127}},\ \Eprint
  {https://arxiv.org/abs/1611.03702} {arXiv:1611.03702 [hep-th]} \BibitemShut
  {NoStop}%
\bibitem [{\citenamefont {Ishii}\ and\ \citenamefont {Murata}(2018)}]{murata}%
  \BibitemOpen
  \bibfield  {author} {\bibinfo {author} {\bibfnamefont {T.}~\bibnamefont
  {Ishii}}\ and\ \bibinfo {author} {\bibfnamefont {K.}~\bibnamefont {Murata}},\
  }\bibfield  {title} {\bibinfo {title} {{Floquet superconductor in
  holography}},\ }\href {https://doi.org/10.1103/PhysRevD.98.126005} {\bibfield
   {journal} {\bibinfo  {journal} {Phys. Rev. D}\ }\textbf {\bibinfo {volume}
  {98}},\ \bibinfo {pages} {126005} (\bibinfo {year} {2018})},\ \Eprint
  {https://arxiv.org/abs/1804.06785} {arXiv:1804.06785 [hep-th]} \BibitemShut
  {NoStop}%
\bibitem [{\citenamefont {Macr\`{\i}}\ \emph {et~al.}(2016)\citenamefont
  {Macr\`{\i}}, \citenamefont {Smerzi},\ and\ \citenamefont {Pezz\`e}}]{echo1}%
  \BibitemOpen
  \bibfield  {author} {\bibinfo {author} {\bibfnamefont {T.}~\bibnamefont
  {Macr\`{\i}}}, \bibinfo {author} {\bibfnamefont {A.}~\bibnamefont {Smerzi}},\
  and\ \bibinfo {author} {\bibfnamefont {L.}~\bibnamefont {Pezz\`e}},\
  }\bibfield  {title} {\bibinfo {title} {Loschmidt echo for quantum
  metrology},\ }\href {https://doi.org/10.1103/PhysRevA.94.010102} {\bibfield
  {journal} {\bibinfo  {journal} {Phys. Rev. A}\ }\textbf {\bibinfo {volume}
  {94}},\ \bibinfo {pages} {010102(R)} (\bibinfo {year} {2016})}\BibitemShut
  {NoStop}%
\bibitem [{\citenamefont {Goussev}\ \emph {et~al.}(2012)\citenamefont
  {Goussev}, \citenamefont {Jalabert}, \citenamefont {Pastawski},\ and\
  \citenamefont {Wisniacki}}]{echo2}%
  \BibitemOpen
  \bibfield  {author} {\bibinfo {author} {\bibfnamefont {A.}~\bibnamefont
  {Goussev}}, \bibinfo {author} {\bibfnamefont {R.~A.}\ \bibnamefont
  {Jalabert}}, \bibinfo {author} {\bibfnamefont {H.~M.}\ \bibnamefont
  {Pastawski}},\ and\ \bibinfo {author} {\bibfnamefont {D.}~\bibnamefont
  {Wisniacki}},\ }\bibfield  {title} {\bibinfo {title} {{Loschmidt echo}},\
  }\href {https://doi.org/10.4249/scholarpedia.11687} {\bibfield  {journal}
  {\bibinfo  {journal} {Scholarpedia}\ }\textbf {\bibinfo {volume} {7}},\
  \bibinfo {pages} {11687} (\bibinfo {year} {2012})},\ \Eprint
  {https://arxiv.org/abs/1206.6348} {arXiv:1206.6348 [nlin.CD]} \BibitemShut
  {NoStop}%
\bibitem [{\citenamefont {Shkedrov}\ \emph {et~al.}(2022)\citenamefont
  {Shkedrov}, \citenamefont {Menashes}, \citenamefont {Ness}, \citenamefont
  {Vainbaum}, \citenamefont {Altman},\ and\ \citenamefont {Sagi}}]{heatref1}%
  \BibitemOpen
  \bibfield  {author} {\bibinfo {author} {\bibfnamefont {C.}~\bibnamefont
  {Shkedrov}}, \bibinfo {author} {\bibfnamefont {M.}~\bibnamefont {Menashes}},
  \bibinfo {author} {\bibfnamefont {G.}~\bibnamefont {Ness}}, \bibinfo {author}
  {\bibfnamefont {A.}~\bibnamefont {Vainbaum}}, \bibinfo {author}
  {\bibfnamefont {E.}~\bibnamefont {Altman}},\ and\ \bibinfo {author}
  {\bibfnamefont {Y.}~\bibnamefont {Sagi}},\ }\bibfield  {title} {\bibinfo
  {title} {Absence of heating in a uniform fermi gas created by periodic
  driving},\ }\href {https://doi.org/10.1103/PhysRevX.12.011041} {\bibfield
  {journal} {\bibinfo  {journal} {Phys. Rev. X}\ }\textbf {\bibinfo {volume}
  {12}},\ \bibinfo {pages} {011041} (\bibinfo {year} {2022})}\BibitemShut
  {NoStop}%
\end{thebibliography}%


\begin{thebibliography}{9}%
\makeatletter
\providecommand \@ifxundefined [1]{%
 \@ifx{#1\undefined}
}%
\providecommand \@ifnum [1]{%
 \ifnum #1\expandafter \@firstoftwo
 \else \expandafter \@secondoftwo
 \fi
}%
\providecommand \@ifx [1]{%
 \ifx #1\expandafter \@firstoftwo
 \else \expandafter \@secondoftwo
 \fi
}%
\providecommand \natexlab [1]{#1}%
\providecommand \enquote  [1]{``#1''}%
\providecommand \bibnamefont  [1]{#1}%
\providecommand \bibfnamefont [1]{#1}%
\providecommand \citenamefont [1]{#1}%
\providecommand \href@noop [0]{\@secondoftwo}%
\providecommand \href [0]{\begingroup \@sanitize@url \@href}%
\providecommand \@href[1]{\@@startlink{#1}\@@href}%
\providecommand \@@href[1]{\endgroup#1\@@endlink}%
\providecommand \@sanitize@url [0]{\catcode `\\12\catcode `\$12\catcode
  `\&12\catcode `\#12\catcode `\^12\catcode `\_12\catcode `\%12\relax}%
\providecommand \@@startlink[1]{}%
\providecommand \@@endlink[0]{}%
\providecommand \url  [0]{\begingroup\@sanitize@url \@url }%
\providecommand \@url [1]{\endgroup\@href {#1}{\urlprefix }}%
\providecommand \urlprefix  [0]{URL }%
\providecommand \Eprint [0]{\href }%
\providecommand \doibase [0]{https://doi.org/}%
\providecommand \selectlanguage [0]{\@gobble}%
\providecommand \bibinfo  [0]{\@secondoftwo}%
\providecommand \bibfield  [0]{\@secondoftwo}%
\providecommand \translation [1]{[#1]}%
\providecommand \BibitemOpen [0]{}%
\providecommand \bibitemStop [0]{}%
\providecommand \bibitemNoStop [0]{.\EOS\space}%
\providecommand \EOS [0]{\spacefactor3000\relax}%
\providecommand \BibitemShut  [1]{\csname bibitem#1\endcsname}%
\let\auto@bib@innerbib\@empty
\bibitem [{\citenamefont {Yan}\ \emph {et~al.}(2026)\citenamefont {Yan},
  \citenamefont {Min}, \citenamefont {Sun}, \citenamefont {Peng}, \citenamefont
  {Xie}, \citenamefont {Wu}, \citenamefont {Li}, \citenamefont {Yang},\ and\
  \citenamefont {Jiang}}]{pra2}%
  \BibitemOpen
  \bibfield  {author} {\bibinfo {author} {\bibfnamefont {X.}~\bibnamefont
  {Yan}}, \bibinfo {author} {\bibfnamefont {J.}~\bibnamefont {Min}}, \bibinfo
  {author} {\bibfnamefont {D.}~\bibnamefont {Sun}}, \bibinfo {author}
  {\bibfnamefont {S.-G.}\ \bibnamefont {Peng}}, \bibinfo {author}
  {\bibfnamefont {X.}~\bibnamefont {Xie}}, \bibinfo {author} {\bibfnamefont
  {X.}~\bibnamefont {Wu}}, \bibinfo {author} {\bibfnamefont {J.}~\bibnamefont
  {Li}}, \bibinfo {author} {\bibfnamefont {W.}~\bibnamefont {Yang}},\ and\
  \bibinfo {author} {\bibfnamefont {K.}~\bibnamefont {Jiang}},\ }\bibfield
  {title} {\bibinfo {title} {Energy dynamics of a nonequilibrium unitary fermi
  gas},\ }\href {https://doi.org/10.1103/1jfx-yntk} {\bibfield  {journal}
  {\bibinfo  {journal} {Phys. Rev. A}\ }\textbf {\bibinfo {volume} {113}},\
  \bibinfo {pages} {053312} (\bibinfo {year} {2026})}\BibitemShut {NoStop}%
\bibitem [{\citenamefont {Das}\ \emph {et~al.}(2026{\natexlab{a}})\citenamefont
  {Das}, \citenamefont {Das}, \citenamefont {Kundu},\ and\ \citenamefont
  {Sengupta}}]{NRFloquet}%
  \BibitemOpen
  \bibfield  {author} {\bibinfo {author} {\bibfnamefont {D.}~\bibnamefont
  {Das}}, \bibinfo {author} {\bibfnamefont {S.~R.}\ \bibnamefont {Das}},
  \bibinfo {author} {\bibfnamefont {A.}~\bibnamefont {Kundu}},\ and\ \bibinfo
  {author} {\bibfnamefont {K.}~\bibnamefont {Sengupta}},\ }\bibfield  {title}
  {\bibinfo {title} {Non-relativistic floquet conformal field theory and
  fermions at unitarity},\ }\href@noop {} {\  (\bibinfo {year}
  {2026}{\natexlab{a}})},\ \bibinfo {note} {main text}\BibitemShut {NoStop}%
\bibitem [{\citenamefont {Pinney}(1950)}]{Pinney1950}%
  \BibitemOpen
  \bibfield  {author} {\bibinfo {author} {\bibfnamefont {E.}~\bibnamefont
  {Pinney}},\ }\bibfield  {title} {\bibinfo {title} {The nonlinear differential
  equation $y''+p(x)y+cy^{-3}=0$},\ }\href
  {https://doi.org/10.1090/S0002-9939-1950-0037979-4} {\bibfield  {journal}
  {\bibinfo  {journal} {Proceedings of the American Mathematical Society}\
  }\textbf {\bibinfo {volume} {1}},\ \bibinfo {pages} {681} (\bibinfo {year}
  {1950})}\BibitemShut {NoStop}%
\bibitem [{\citenamefont {Lewis}(1968)}]{Lewis1968}%
  \BibitemOpen
  \bibfield  {author} {\bibinfo {author} {\bibfnamefont {H.~R.}\ \bibnamefont
  {Lewis}},\ }\bibfield  {title} {\bibinfo {title} {Class of exact invariants
  for classical and quantum time-dependent harmonic oscillators},\ }\href
  {https://doi.org/10.1063/1.1664532} {\bibfield  {journal} {\bibinfo
  {journal} {Journal of Mathematical Physics}\ }\textbf {\bibinfo {volume}
  {9}},\ \bibinfo {pages} {1976} (\bibinfo {year} {1968})}\BibitemShut
  {NoStop}%
\bibitem [{\citenamefont {Lewis}\ and\ \citenamefont
  {Riesenfeld}(1969)}]{LewisRiesenfeld1969}%
  \BibitemOpen
  \bibfield  {author} {\bibinfo {author} {\bibfnamefont {H.~R.}\ \bibnamefont
  {Lewis}}\ and\ \bibinfo {author} {\bibfnamefont {W.~B.}\ \bibnamefont
  {Riesenfeld}},\ }\bibfield  {title} {\bibinfo {title} {An exact quantum
  theory of the time-dependent harmonic oscillator and of a charged particle in
  a time-dependent electromagnetic field},\ }\href
  {https://doi.org/10.1063/1.1664991} {\bibfield  {journal} {\bibinfo
  {journal} {Journal of Mathematical Physics}\ }\textbf {\bibinfo {volume}
  {10}},\ \bibinfo {pages} {1458} (\bibinfo {year} {1969})}\BibitemShut
  {NoStop}%
\bibitem [{\citenamefont {Castin}(2004)}]{Castin2004}%
  \BibitemOpen
  \bibfield  {author} {\bibinfo {author} {\bibfnamefont {Y.}~\bibnamefont
  {Castin}},\ }\bibfield  {title} {\bibinfo {title} {Exact scaling transform
  for a unitary quantum gas in a time-dependent harmonic potential},\ }\href
  {https://doi.org/10.1016/j.crhy.2004.03.017} {\bibfield  {journal} {\bibinfo
  {journal} {Comptes Rendus Physique}\ }\textbf {\bibinfo {volume} {5}},\
  \bibinfo {pages} {407} (\bibinfo {year} {2004})},\ \Eprint
  {https://arxiv.org/abs/cond-mat/0406020} {arXiv:cond-mat/0406020 [cond-mat]}
  \BibitemShut {NoStop}%
\bibitem [{\citenamefont {Werner}\ and\ \citenamefont {Castin}(2006)}]{werner}%
  \BibitemOpen
  \bibfield  {author} {\bibinfo {author} {\bibfnamefont {F.}~\bibnamefont
  {Werner}}\ and\ \bibinfo {author} {\bibfnamefont {Y.}~\bibnamefont
  {Castin}},\ }\bibfield  {title} {\bibinfo {title} {Unitary gas in an
  isotropic harmonic trap: Symmetry properties and applications},\ }\href
  {https://doi.org/10.1103/PhysRevA.74.053604} {\bibfield  {journal} {\bibinfo
  {journal} {Phys. Rev. A}\ }\textbf {\bibinfo {volume} {74}},\ \bibinfo
  {pages} {053604} (\bibinfo {year} {2006})}\BibitemShut {NoStop}%
\bibitem [{\citenamefont {Blau}\ \emph {et~al.}(2009)\citenamefont {Blau},
  \citenamefont {Hartong},\ and\ \citenamefont {Rollier}}]{Blau_2009}%
  \BibitemOpen
  \bibfield  {author} {\bibinfo {author} {\bibfnamefont {M.}~\bibnamefont
  {Blau}}, \bibinfo {author} {\bibfnamefont {J.}~\bibnamefont {Hartong}},\ and\
  \bibinfo {author} {\bibfnamefont {B.}~\bibnamefont {Rollier}},\ }\bibfield
  {title} {\bibinfo {title} {Geometry of schrödinger space-times, global
  coordinates, and harmonic trapping},\ }\href
  {https://doi.org/10.1088/1126-6708/2009/07/027} {\bibfield  {journal}
  {\bibinfo  {journal} {Journal of High Energy Physics}\ }\textbf {\bibinfo
  {volume} {2009}},\ \bibinfo {pages} {027–027} (\bibinfo {year}
  {2009})}\BibitemShut {NoStop}%
\bibitem [{\citenamefont {Das}\ \emph {et~al.}(2026{\natexlab{b}})\citenamefont
  {Das}, \citenamefont {Das}, \citenamefont {Kundu},\ and\ \citenamefont
  {Sengupta}}]{hd2}%
  \BibitemOpen
  \bibfield  {author} {\bibinfo {author} {\bibfnamefont {D.}~\bibnamefont
  {Das}}, \bibinfo {author} {\bibfnamefont {S.~R.}\ \bibnamefont {Das}},
  \bibinfo {author} {\bibfnamefont {A.}~\bibnamefont {Kundu}},\ and\ \bibinfo
  {author} {\bibfnamefont {K.}~\bibnamefont {Sengupta}},\ }\bibfield  {title}
  {\bibinfo {title} {Dynamical phases of higher dimensional floquet cfts},\
  }\href {https://doi.org/10.21468/SciPostPhys.20.2.045} {\bibfield  {journal}
  {\bibinfo  {journal} {SciPost Phys.}\ }\textbf {\bibinfo {volume} {20}},\
  \bibinfo {pages} {045} (\bibinfo {year} {2026}{\natexlab{b}})}\BibitemShut
  {NoStop}%
\end{thebibliography}%

\appendix

\section{End Matter} 
\label{endmat}

\begin{figure}
\rotatebox{0}{\includegraphics*[width= 0.98 \linewidth]{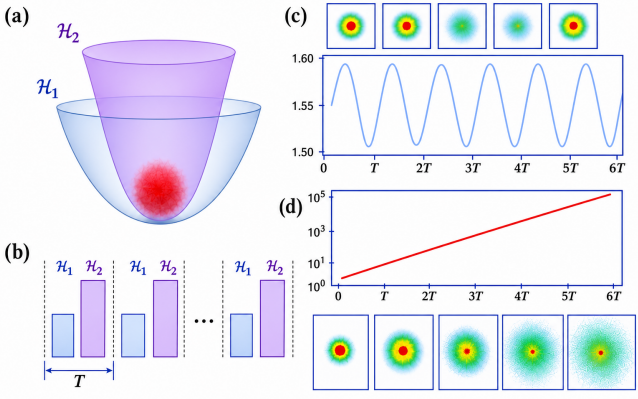}}
\caption{(a) Schematic representation of (a) ultracold fermions in a trap and (b) the square-pulse drive protocol with period $T$. Here $H_1$ and $H_2$ correspond to Hamiltonians of trapped fermions with trap frequencies $\mu_1$ and $\mu_2$ respectively. (c) and (d) Fermion clouds in different dynamical phases. (c) In the elliptic phase, the cloud oscillates around a mean position while (d) in the hyperbolic phase, its radius expands. See text for details.   }  \label{figem0}
\end{figure}

In the End Matter we sketch possible experiments that can validate the theoretical predictions of our work. To this end, we consider a system of trapped fermions at unitarity as schematically shown in Fig.\ \ref{figem0}(a). For such a system, the conformal generators $H$, $C$ and $D$ have the representation (in the Schrodinger picture)
\bea
H & = & \int d^dx \left[\frac{1}{2} |\nabla \psi_\sigma|^2 - g^2 \psi^\dagger_{\uparrow}\psi^\dagger_{\downarrow} \psi_\uparrow \psi_\downarrow \right]\nonumber \\
C & = & \int d^d x \frac{1}{2}x^2 \psi^\dagger_\sigma \psi_\sigma,  \nonumber \\
D & = & -\frac{i}{2} \int d^dx~ x^i \left[ \psi^\dagger_\sigma \nabla_i \psi_\sigma - ({\rm h.c.}) \right],
\label{2-2}
\eea
where $\psi_\sigma$ are fermion fields in $d$ dimensions, $\sigma = \uparrow,\downarrow$, and $g$ denotes interaction strength of the fermions. Here and in rest of the End Matter, we set $m=\hbar=1$, where $m$ is the fermion mass. In an expansion in $\epsilon = d-2$, the fixed point is at $g^2 = 2\pi\epsilon$ \cite{son2}. The Hamiltonian for critical fermions in a harmonic trap with frequency $\mu$  is then given by ${\cal H} = H + \mu^2 C$.  This allows us to compute  one-point and autocorrelation functions for these fermions using a square-pulse protocol (Fig.\ \ref{figem0}(b)).  The fermionic ground state with $n$ particles, which serves as the initial state for such computation, is a primary state of the CFT and is denoted by $|\Delta_0\rangle$. 

In the context of fermions at unitarity in a harmonic trap, the quantity 
\ben
_{\mu_1}\langle \Delta_0| C(\ell) |\Delta_0\rangle_{\mu_1} = \frac{1}{2} \sum_{i=1}^n~ _{\mu_1}\langle
\Delta_0| 
 (\vx_i)^2|\Delta_0\rangle_{\mu_1} (\ell T)  \label{2-19}
\een
where $\vx_i$ denote the coordinates of the $n$ particles inside the trap, is of particular interest since it yields information about the spatial profile of the fermionic cloud inside the trap. The three dynamical phases we have identified correspond to the particle cloud oscillating in (stroboscopic) time, or increasing in time. In the latter two phases, the number density $n(\vx)$ becomes large near the boundary of the system at late times, i.e. the cloud is pushed to the boundaries. The three phases can therefore be identified experimentally by simply looking at the cloud profile. This is schematically sketched in Fig.\ \ref{figem0}(c) and (d).

\begin{figure}
\rotatebox{0}{\includegraphics*[width= 0.98 \linewidth]{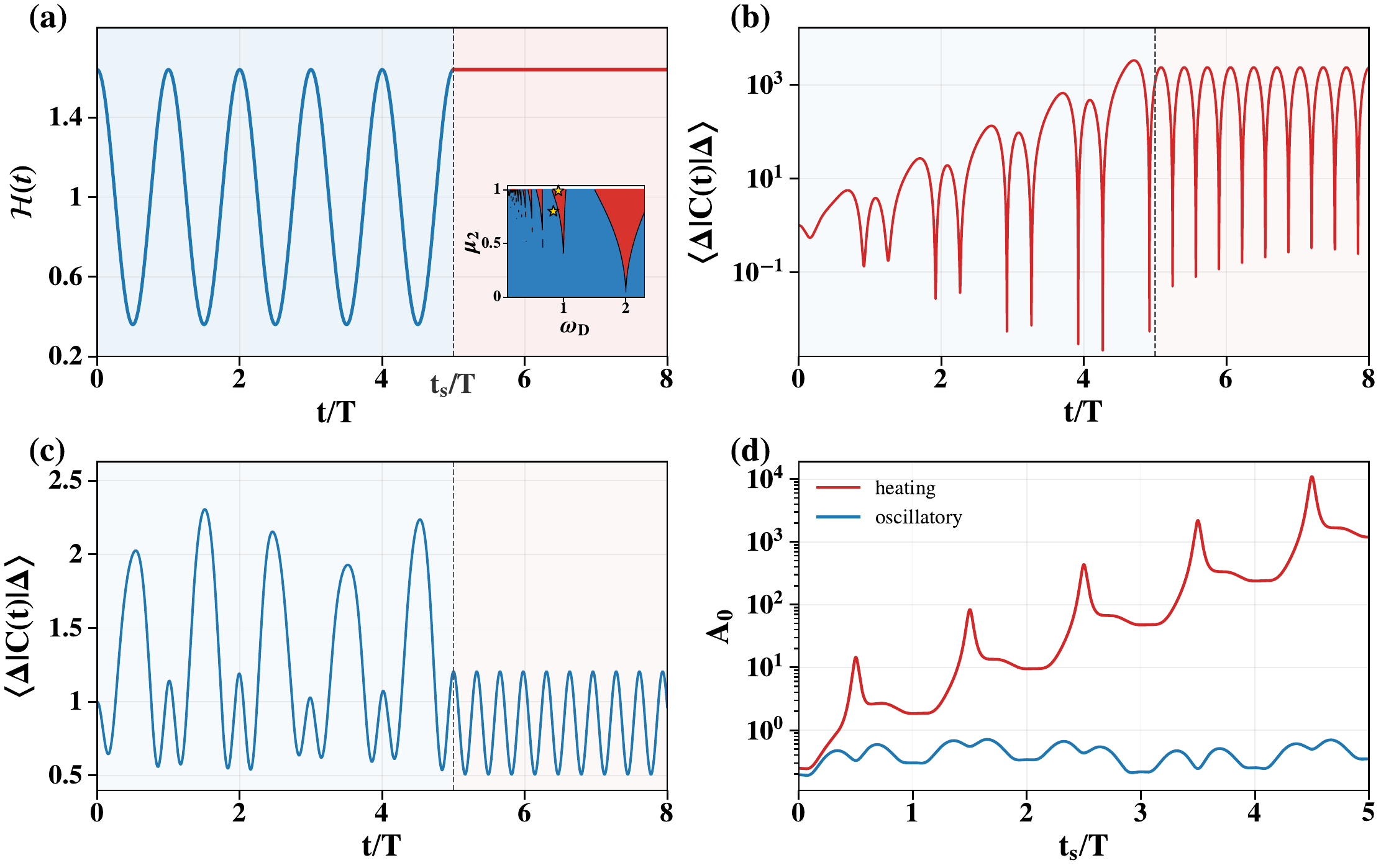}}
\caption{(a) Schematic representation of an experimentally relevant protocol where energy is pumped using a continuous cosine drive ($\mu^2(t)= \mu_1^2 + \mu_2^2 \cos \omega_D t$) till $t_s=5T$ (vertical dashed line) followed by a Hamiltonian evolution with ${\mathcal H}= H+ (\mu_1^2 +\mu_2^2) C$ for $8T\geq t> t_s$. The inset indicates the phase diagram in the $\mu_2-\omega_D$ plane showing hyperbolic (red) and elliptic (blue) phases. (b) Plot of $C(t)$ for $\mu_2=0.99$, $\omega_D=0.915$ showing large amplitude oscillations whose amplitude grows with the driving time. This is  a signature of the hyperbolic phase and (c) Same as (b) but for $\mu_2=0.8$ and  $\omega_D=0.835$ showing smaller amplitude oscillations signifying the elliptic phase. (d) Plot of oscillation amplitude $A_0$ for $t>t_s$ in the two phases showing a growing envelope in the hyperbolic phase and a bounded behavior in the elliptic phase. $\mu_1=1$ for all plots. See text for details. }   \label{figem1}
\end{figure}

The three dynamical phases can be simply understood in terms of the equivalent time independent Hamiltonian $\calh_F^\prime$ discussed in the main text. This transformed Hamiltonian represents the particles in a harmonic oscillator potential in the elliptic phase ( $\kappa^2 > 0$ ) and in an inverted harmonic oscillator potential in the hyperbolic phase ($\kappa^2 < 0$). The growth of the cloud in the hyperbolic phase is a consequence of the instability of the inverted oscillator.

However, we note that the square-pulse protocol is difficult to generate in realistic experiments. A more realistic protocol consists of a periodic drive with some finite number of periods, followed by evolution by $\calh_1$ for a further time $t$. The evolution operator for the latter protocol is $U(\ell,t) = e^{-it(H +\mu_1^2 C)}U(\ell T)$. The resulting expectation value 
$C(\ell,t)  \equiv  _{\mu_1}\langle \Delta_0| U^{-1} (\ell,t) C U(\ell,t)  |\Delta_0\rangle_{\mu_1}$ is given by
\bea
C(\ell,t) & = & \frac{\Delta}{4 \mu_1^3} [ B_0 + B_1 \cos(2 \mu_1 t) + B_2 \sin (2\mu_1 t) ] \nonumber\\
B_0 &=& (\mu_1^4 b_\ell^2 + \mu_1^2 a_\ell^2 + \mu_1^2 d_\ell^2 + c_\ell^2) \nonumber \\
B_1 &= & (\mu_1^4 b_\ell^2 + \mu_1^2 a_\ell^2 - \mu_1^2 d_\ell^2 - c_\ell^2) \nonumber \\
B_2 &=& 2\mu_1 (\mu_1^2 b_\ell d_\ell + a_\ell c_\ell ) \label{proteq}
\eea
where $a_\ell \cdots d_\ell$ are given in (\ref{2-5}) for the square-pulse protocol and numerically computed for the cosine protocol in the SM \cite{sicit}. We note that replacing a square-pulse protocol with cosine drive (Fig.\ \ref{figem1}) does not change the behavior of $C(\ell,t)$ qualitatively since the dynamical phases discussed in the main text also exist for a cosine protocol (see SM \cite{sicit} for details).  The system displays persistent oscillations with frequency $2\mu_1$ once the drive is removed, reflecting the breathing mode. The oscillation amplitude given by 
\begin{eqnarray} 
A_0 &=&  \frac{\Delta }{ 4 \mu_1^3} \sqrt{B_1^2 + B_2^2} \label{ampeq} 
\end{eqnarray} 
grows exponentially with the number of time periods of the drive in the hyperbolic phase, grows linearly in the parabolic phase and oscillates with frequency $\rho$ in the elliptic phase. 

Experiments that can check the validity of the results constitute realization of a drive protocol sketched in Fig.\ \ref{figem1}(a) \cite{pra1,pra2} (see SM \cite{sicit} for a dictionary between our analysis and the experimental setup \cite{pra1}). A continuous periodic drive $\mu^2(t)= \mu_1^2 +\mu_2^2 \cos \omega_D t$ is used for energy injection in this protocol; the corresponding dynamical phases are sketched in the SM \cite{sicit}. The corresponding phase diagram in the $\mu_2-\omega_D$ plane for $\mu_1=1$ is shown in the inset of Fig.\ref{figem1}(a). For this protocol (Fig.\ \ref{figem1}(a)), we numerically calculate $U(t)$ (see SM \cite{sicit} for details) followed by a calculation of $C(t)= \langle \Delta|U^{-1} C U |\Delta\rangle$ similar to that discussed in Eq.\ \ref{proteq}. 

The time variation of $C(t)$ in the hyperbolic phase is shown in Fig.\ \ref{figem1}(b); this constitutes oscillations with a large amplitude. The amplitude of these oscillations during Hamiltonian evolution for $t>t_s$ ($t_s$ is the time at which the drive is stopped) depends on the energy absorbed by the system during the periodic drive. The corresponding behavior in the elliptic phase (Fig.\ \ref{figem1}(c)) exhibits smaller amplitude oscillations. The amplitude $A_0$ (Eq.\ \ref{ampeq}) of these oscillations for $t>t_s$ is plotted in Fig.\ \ref{figem1}(d), as a function of $t_s$. $A_0$ exhibits an overall growing envelope in the hyperbolic phase and a bounded behavior in the elliptic phase; thus, it provides a sharp distinction between the two dynamical phases. 

We note that Ref.\ \onlinecite{pra2} has reported a growth of absorbed energy for the protocol given in Fig.\ \ref{figem1}(a) (signature of the hyperbolic phase). This stems from the particular choice of  experimental parameters (see SM \cite{sicit} for details). Bounded oscillations of fermions (signature of the elliptic phase) have not been experimentally observed so far; our work provides a concrete prediction regarding the existence of such an oscillatory phase. Moreover, it also predicts the presence of  two distinct dynamical phases in the above-mentioned experimentally realizable drive protocol from a universal analysis based solely on CFT. We note here that typical experimental systems are not exactly at unitarity but close to it \cite{heatref1}. In such situations, our analysis is expected to hold until a finite number of drive cycles making the protocol in Fig.\ \ref{figem1}a particularly relevant for testing FCFT predictions \cite{heatref1,pra1}.

Before concluding this section, we note that for resonant anyons described by particles in the presence of a $U(1)$ Chern Simons field one has 
\bea
H & = & \int d^2x \left[ |{\vec D}\Phi|^2 + \frac{v}{4} |\phi|^4 \right] ~~{\vec D} ={ \vec \nabla }
- ie {\vec A} \nonumber \\
{\vec \nabla} \times {\vec A} & = &-\frac{e}{\kappa} |\phi|^2
\eea
The renormalized coupling $v$ is related to the Chern-Simons coupling $\kappa$ by $v = \frac{2 e^2}{|\kappa|}$. The generators $C,D$ are the same as in (\ref{2-2}) with $\psi$ replaced by $\phi$ . As for fermions at unitarity a term proportional to $C$ in the Hamiltonian corresponds to a harmonic trap. The considerations of this paper extend to this system in a straightforward way and would lead to similar predictions if such a system can be realized in experiments.

\end{document}


\title{Supplemental Material for\\
\textit{Non-relativistic Floquet Conformal Field Theory}}

\author{Diptarka Das$^{1}$, Sumit R. Das,$^{2}$ Arnab Kundu$^{3,4}$ and Krishnendu Sengupta $^5$}
\affiliation{$^1$ Department of Physics, Indian Institute of Technology, Kanpur, UP 208016, INDIA.}
\affiliation{$^2$Department of Physics and Astronomy, University of Kentucky, Lexington, KY 40506, U.S.A.}
\affiliation{$^3$Saha Institute of Nuclear Physics, 1/AF Bidhannagar, Kolkata 700064, INDIA.}
\affiliation{$^4$Homi Bhabha National Institute, Training School Complex, Anushaktinagar, Mumbai 400094, INDIA.}
\affiliation{$^5$School of Physical Sciences, Indian Association for the Cultivation of Science, Jadavpur, Kolkata 700032, INDIA.}

\maketitle

\tableofcontents 


\section{Transformation of the primary operators}
\label{primtrans} 

In this section, we will provide the explicit derivation of the transformation of a primary operator under a finite transformation by the ${\rm SL}(2, {\mathbb R}) \equiv {\rm SO}(2,1)$ group. Let us begin with the ${\rm so}(2,1)$ algebra:
\begin{eqnarray}
[D, H]= 2 i H \ , ~ [D,C] = - 2 i C \ , ~ [H, C] = - i D \ ,
\end{eqnarray}
where the generators can be written in terms of the following differential representation:
\begin{eqnarray}
&& D\equiv  - i \left( 2t \partial_t + x\partial_x \right) \ ,  ~~ H \equiv - i \partial_t \ ,  \\
&&  C \equiv -i (t^2 \partial_t + t x \partial_x) \ .  
\end{eqnarray}
Everything will trivially generalize for a $d$-dimensional spacelike vector, ${\vec x}_d$, and to avoid clutter, we will omit the vector symbols from now on. On the other hand, the adjoint action of the algebra generators on a local operator with dimension $\Delta$ and particle number $N$ satisfies the following relations:
\begin{eqnarray}
    && [D, {\cal O}] = - i \left( 2t \partial_t + x\partial_x+\Delta \right){\cal O} \ , \label{Oact1}  \\
    && [N, {\cal O}] = N{\cal O} \ ,  \label{Oact2} \\
    && [H, {\cal O}] = -i \partial_t {\cal O}  \ ,\label{Oact3} \\
    && [C, {\cal O}] = -i \left( t^2 \partial_t + tx \partial_x + t \Delta \right) {\cal O} + \frac{x^2}{2} N {\cal O}. \label{Oact4}
\end{eqnarray}
Note that the adjoint action of the special conformal generator, $C$, is valid only for a primary operator, while the other actions are defined for an arbitrary local operator. Henceforth, we will focus only on primary operators. 

Consider a general group element $U=e^{i s X} \in {\rm SO}(2,1)$, where $X = \gamma D + \alpha_- C + \alpha_+ H$, with $\{\alpha_{\pm},\gamma\} \in {\mathbb R}$. Expressing the generator $X = X^\mu(s) \partial_\mu$ on a local coordinate basis that spans the tangent space of the group manifold at identity, we can write:
\begin{eqnarray}
X_H &=& i X = v(t) \partial_t + w(t, x) \partial_x , \quad 
 v(t) = \alpha_- t^2 + 2 \gamma t + \alpha_+, \quad w(t,x) = x (\alpha_- t+ \gamma) \ . \label{vandw}
\end{eqnarray}

Let us first consider the finite transformation of the coordinates. This is obtained by solving the integral curve equations generated by $X_H$. Therefore, along the flow:
\begin{eqnarray}
    \frac{dt}{ds} = v(t) \ , \quad \frac{dx}{ds} = w(t, x) \ . \label{intcureqn}
\end{eqnarray}
 The full solution of the above coupled differential equation is given by
\begin{eqnarray}
    t(s) = \frac{a(s) t_0 + b(s) }{c(s) t_0 + d(s)} \ , \quad x(s) = \frac{x_0 }{c(s) t_0 + d(s)} \ , \label{txs}
\end{eqnarray}
where
\begin{eqnarray}
&& a(s) = \cos\left( s \delta\right) +\frac{\gamma}{\delta} \sin\left( s\delta\right) \ , \label{as1} \\
&& b(s) = \frac{\alpha_+}{\delta} \sin(s \delta)  ,  c(s) = - \frac{\alpha_-}{\delta} \sin(s\delta)  , \\
&& d(s) =  \cos\left( s \delta\right) - \frac{\gamma}{\delta} \sin\left( s \delta\right) \ , \\
&& \delta^2 = - \gamma^2 + \alpha_+ \alpha_- \ . \label{ds1}
\end{eqnarray}
The simplest way to obtain this is to substitute an ansatz of the form in (\ref{txs}) into the integral curve equations in (\ref{intcureqn}) and then solve for the functions. This substitution yields:
\begin{eqnarray}
    \frac{d}{ds} \begin{pmatrix}
a & b \\
c & d
\end{pmatrix} = \begin{pmatrix}
\gamma & \alpha_+ \\
-\alpha_- & -\gamma
\end{pmatrix} \begin{pmatrix}
a & b \\
c & d
\end{pmatrix} = M \begin{pmatrix}
a & b \\
c & d
\end{pmatrix} \ ,
\end{eqnarray}
where $M^2 = - \delta^2 \mathds{1}$.
%
%
Using the above, it can easily be established that:
\begin{eqnarray}
    e^{sM} = \cos\left( s \delta\right) \mathds{1} + \frac{\sin\left(s \delta \right)}{\delta} M \ . 
\end{eqnarray}
This finally yields the solutions for $\{a(s), b(s), c(s), d(s)\}$.

Let us now consider the finite transformation of a primary operator. Given an element of the algebra $X$, the group element can be written as: $U^{-1} = e^{i s X}$. The transformation of an operator is given by
\begin{eqnarray}
    {\cal O}(s) = e^{i s X} {\cal O} e^{- i s X} \implies \frac{\partial{\cal O}}{\partial s} = i \left[X, {\cal O} \right]  \ . \label{opevolve}
\end{eqnarray}
Using the adjoint actions in (\ref{Oact1})-(\ref{Oact4}), we obtain:
\begin{eqnarray}
    \frac{\partial {\cal O}}{\partial s} = i [X, {\cal O}] &=& v(t) \partial_t {\cal O} + w(t, x) \partial_x {\cal O} + \left( \Delta (\gamma + \alpha_- t) + \frac{i\alpha_-}{2}N x^2\right) {\cal O} \ . 
\end{eqnarray}
Let us now define an ``inverse coordinate flow" as follows:
\begin{eqnarray}
    \frac{dt}{ds} = - v(t(s)) \ , \quad \frac{dx}{ds} = - w(t(s), x(s)), \label{coordflow}
\end{eqnarray}
along which the operator evolution takes the form: 
\begin{eqnarray}
 && \frac{\partial {\cal O}}{\partial s} + \frac{dt}{ds} \partial_t {\cal O} + \frac{dx}{ds} \partial_x {\cal O} = \Big ( \Delta (\gamma + \alpha_- t(s))  + \frac{i\alpha_-}{2}N x^2\Big) {\cal O},  \label{opevolvefinal} \\
 && \implies \quad \frac{d}{ds}{\cal O} = \left( \Delta (\gamma + \alpha_- t(s)) + \frac{i\alpha_-}{2}N x^2\right) {\cal O}. \nonumber
\end{eqnarray}
Our task now is to solve the equation in (\ref{opevolvefinal}). The formal solution can be written down easily as follows:  
\begin{eqnarray}
{\cal O}(t, x) &=&  {\rm exp} \Big[ \int_0^s \Big( \Delta (\gamma + \alpha_- t(\sigma) ) + \frac{i\alpha_-}{2} N x(\sigma)^2\Big) d\sigma \Big] {\cal O}(t_0, x_0). \label{Otx}
\end{eqnarray}
Here: $t_0 = t(s=0), x_0 = x(s=0)$. Now, we need to evaluate the integrals in the exponential.  

Note that in writing the operator flow equation above, we have chosen a coordinate flow in (\ref{coordflow}), which is related to the coordinate flow equation in (\ref{intcureqn}) by sending $s\to - s$.  The coordinate transformations are given by the same formulae as in (\ref{txs}).

To evaluate the integral in the exponential, let us first note that:
\begin{eqnarray}
&& D(s) = c(s) t_0 + d(s) \ ,  \\
&&    \frac{d}{ds}D(s) = \left( \gamma + \alpha_- t(s) \right) D(s)\ . \label{eqnD(s)}  
\end{eqnarray}
In deriving the above, we have used the coordinate flow equation in (\ref{coordflow}), together with the substitution: $a(s) t_0 + b(s) = t(s) D(s)$. Therefore, we obtain:
\begin{eqnarray}
    {\rm exp}\left[ \Delta \int_0^s \left( \gamma + \alpha_- t(\sigma)\right) d\sigma\right]  = (D(s) )^\Delta   \ ,
\end{eqnarray}
by directly substituting the above solution into the equation in (\ref{eqnD(s)}). This takes care of the first term in (\ref{Otx}). 

To evaluate the second term, let us first begin with the following coordinate flow equation: 
\begin{eqnarray}
\frac{d}{ds} x(s) &=& - (\gamma + \alpha_- t(s)) x(s) \ . \\
 \implies \frac{d}{d\sigma} x(\sigma)^2 &=& - 2 \left( \gamma + \alpha_- t(\sigma)\right) x(\sigma)^2 = - \frac{d}{dt}v(t) x(\sigma)^2 \ , \nonumber\\
\end{eqnarray}
where we have used the definition of $v(t)$ from (\ref{vandw}). Using this expression, we can now show that:
\begin{eqnarray}
    \frac{d}{d\sigma}\left( \frac{x(\sigma)^2}{v(t(\sigma ))}\right) = 0 \ ,
\end{eqnarray}
where we have also used 
\begin{eqnarray} 
\frac{d}{d\sigma}v(t(\sigma)) &=& - \frac{d}{dt}v(t) \frac{d}{d\sigma}t(\sigma).
\end{eqnarray} 
 The upshot is that we have now identified a conserved quantity along the coordinate flow, using which we can write:
\begin{eqnarray}
    \frac{x(\sigma)^2}{v(t(\sigma))} = \frac{x_0^2}{v(t_0)} \ . 
\end{eqnarray}
This yields:
\begin{eqnarray}
    \int_0^s x(\sigma)^2 d\sigma = \frac{x_0^2}{v(t_0)} \left(t_0 - t(s) \right) = \frac{c(s)}{\alpha_- D(s)} x_0^2 \ ,
\end{eqnarray}
where the last line is obtained using the relation between $t(s)$ and $t_0$, as well as using the explicit solution of the flow parameters in (\ref{as1})-(\ref{ds1}) with a substitution: $s \to - s$. Therefore, the total phase is obtained to be:
\begin{eqnarray}
    {\cal O}(t(s), x(s)) = D(s)^\Delta {\rm exp}\left[ \frac{i N}{2} \frac{c(s) x_0^2}{D(s)}\right] {\cal O}(t_0, x_0) \ . 
\end{eqnarray}
Note that the above transformation holds along the inverse coordinate flow and therefore a primary operator transformation can be written as:
\begin{eqnarray}
 &&   e^{i s X}{\cal O}(t, x) e^{- i s X}  \nonumber\\ 
 && = \left(c(s) t + d(s) \right)^{- \Delta} {\rm exp}\left[ - \frac{i N}{2} \frac{c(s) x^2}{c(s) t + d(s)}\right] \nonumber\\
 && {\cal O} \left(\frac{a(s) t + b(s)}{c(s) t + d(s)}, \frac{x}{c(s) t + d(s) } \right) \ ,
\end{eqnarray}
where $\{a(s), b(s), c(s), d(s)\}$ are given by (\ref{as1})-(\ref{ds1}). We have used this transformation rule to calculate correlation functions in this article.

Before leaving this section, let us offer some comments regarding the coordinate transformation under the full Schr\"{o}dinger transformation. In a $(d+1)$-dimensional system, with $d$ spatial directions, the full Schr\"{o}dinger transformation can be obtained ({\it e.g.}~following the treatment above) to be:
\begin{eqnarray}
t' = \frac{a t + b}{c t + d}  \ , \quad \vec{x}' = \frac{\vec{x} + \vec{v} t + \vec{A}}{c t + d} \ , \label{schrgen}
\end{eqnarray}
where $\vec{v}$ is the $d$-dimensional velocity vector and $\vec{A}$ is a $d$-dimensional vector that accounts for the shift of the coordinate system. Given an initial point $(t_0, \vec{x}_0)$, a repeated use of this coordinate transformation in (\ref{schrgen}) will send the point towards the fixed point of the transformation in (\ref{schrgen}), provided a fixed point exists. Clearly, the corresponding physics in the asymptotic limit of this sequence of coordinate transformations depends crucially on the nature of the fixed point. 

It is easy to see that a fixed point solution obeying $t' = t$ and $\vec{x}' = \vec{x}$ obeys the following features: (i) The fixed points in time, denoted by $t_*$, are independent of the existence of a fixed point in $\vec{x}$. Therefore, the asymptotic stroboscopic dynamics is decided purely by the nature of $t_*$. (ii) It is straightforward to check that $t_*$ has two real solutions in the hyperbolic phase. One attractive and one repulsive. In the parabolic phase, there is a unique $t_*$ which is neutral and the elliptic class does not admit a real solution for $t_*$. Correspondingly, setting $\vec{v}=0=\vec{A}$, it is clear that $x_*=0$ irrespective of the nature of $t_*$. This is a clear distinction compared to the fixed point structure of a relativistic CFT, where the basic $sl(2, {\mathbb R})$ algebra still governs the dynamics. At the same time, it is also evident that turning $\vec{v}$ and $\vec{A}$ to non-vanishing values, generically, the fixed point structure of $x_*$ does not affect the time-dynamical behaviour although the fixed point data is enriched. Therefore, for a large class of correlation functions, the underlying $sl(2, {\mathbb R})$ conjugacy classes determine the asymptotic stroboscopic dynamics, even when the full Schr\"{o}dinger algebra is considered.

\section{Two-time driven correlator}
\label{corrsec}
\noindent
The primary correlator in a non-relativistic CFT (NRCFT) at equilibrium is given by 
\begin{equation}
\left\langle O(t,\vec{x})\,O^\dagger(0,\vec{0})\right\rangle
\sim
\frac{1}{t^\Delta}
\exp\!\left(
\frac{i n_O \vec{x}^{\,2}}{2t}
\right),
\label{eq:nrcft-corr}
\end{equation}
where $n_O$ is the eigenvalue of the number operator $N$. We want to determine the driven two time correlator of primaries, i.e., $\left\langle O(\ell_1 T,\vec{x}_1)\,O^\dagger(\ell_2 T,\vec{x}_2)\right\rangle$ . A primary field ${\cal O}(\vx,t)$, under a generic ${\rm SO}(2,1)$ Floquet drive transforms after $\ell$ cycles to:  
\ben
 \left(U^{\dagger}\right)^{\ell} {\cal O}(t,\vx)U^\ell = \frac{e^{\frac{in_O c_\ell \vx^2}{2(c_\ell t+d_\ell)}}}{
(c_\ell t+d_\ell)^{\Delta}}{\cal O} \left(  \frac{a_\ell t+b_\ell}{c_\ell t+d_\ell},\frac{\vx}{c_\ell t+d_\ell } \right).
\label{2-4}
\een
The ${\rm SO}(2,1)$ transformation is parametrized using the group element
\begin{equation}
	U_{\ell_i} = g_{\ell_i}
	=
	\begin{pmatrix}
		a_{\ell_i} & b_{\ell_i}\\
		c_{\ell_i} & d_{\ell_i}
	\end{pmatrix},
	\qquad
	a_{\ell_i}  d_{\ell_i} -b_{\ell_i}  c_{\ell_i} =1.  \label{upar1}
\end{equation}
Therefore for an autocorrelator,  we can express
\begin{align}
	\Gamma_{\ell_1,\ell_2} &=\left\langle O(\ell_1 T , \vec{x}_1) O^\dagger(\ell_2 T , \vec{x}_2) \right\rangle= 
		\left\langle O(0 , \vec{x}_1)U^\dagger_{\ell_2-\ell_1} O^{\dagger}(0 , \vec{x}_2) \right\rangle \nonumber\\
		&= \left\langle O(0 , \vec{x}_1)U^\dagger_{\ell_2-\ell_1} O^\dagger(0 , \vec{x}_2)U_{\ell_2-\ell_1} \right\rangle, \label{a1-15}
\end{align}
where we have used the fact that the action of $U_\ell$ on vacuum is trivial. The unitary element that appears above is $U^\dagger_{\ell_2-\ell_1}  = U_{\ell_1} U^\dagger_{\ell_2}$, which therefore corresponds to the ${\rm SO}(2,1)$ group element
\begin{equation}
	U_{\ell_2 - \ell_1 } = g_{\ell_1}^{-1}g_{\ell_2}
	  = 		\begin{pmatrix}
	d_{\ell_1} & -b_{\ell_1}\\
	-c_{\ell_1} & a_{\ell_1}
	\end{pmatrix}.\begin{pmatrix}
	a_{\ell_2} & b_{\ell_2}\\
	c_{\ell_2} & d_{\ell_2}
	\end{pmatrix}  = 	\begin{pmatrix}
	A & B\\
	C & D
	\end{pmatrix}
\end{equation}
whose entries explicitly are
\begin{align}
	A&=d_{\ell_1}a_{\ell_2}-b_{\ell_1}c_{\ell_2},
	&
	B&=d_{\ell_1}b_{\ell_2}-b_{\ell_1}d_{\ell_2},
	\nonumber\\
	C&=-c_{\ell_1}a_{\ell_2}+a_{\ell_1}c_{\ell_2},
	&
	D&=-c_{\ell_1}b_{\ell_2}+a_{\ell_1}d_{\ell_2}.
	\label{eq-abcd}
\end{align}
Therefore using the transformation \eqref{2-4} and the expression for the NRCFT Greens function \eqref{eq:nrcft-corr} we get
\begin{align}
	\Gamma_{\ell_1,\ell_2} &= \frac{e^{-\frac{i n_O}{2} C\vec x_2^{\,2}/D}}{D^\Delta}
	\left\langle
	O(0,\vec x_1)
	O^\dagger\!\left(\frac{B}{D},\frac{\vec x_2}{D}\right)
	\right\rangle =
	\frac{\exp\!\left[
		\frac{i n_O}{2} \bigg( -\frac{C}{D}\vec x_2^{\,2}
		+\frac{\left(\vec x_1-\vec x_2/D\right)^2}{-B/D}  \bigg)
		\right]}{(-B)^\Delta}. \label{grfn1} 
	\end{align}
This completes our derivation of the driven Greens function for NRCFT.

\section{Computation of Autocorrelations}
\label{autosec} 

\noindent
In this section, we discuss the properties of the autocorrelations of the driven NRCFT, for the case of two step drive considered in Ref.\ \cite{NRFloquet} 

\begin{figure}[h]
	\centering
	\rotatebox{0}{\includegraphics*[width= 0.48 \linewidth]{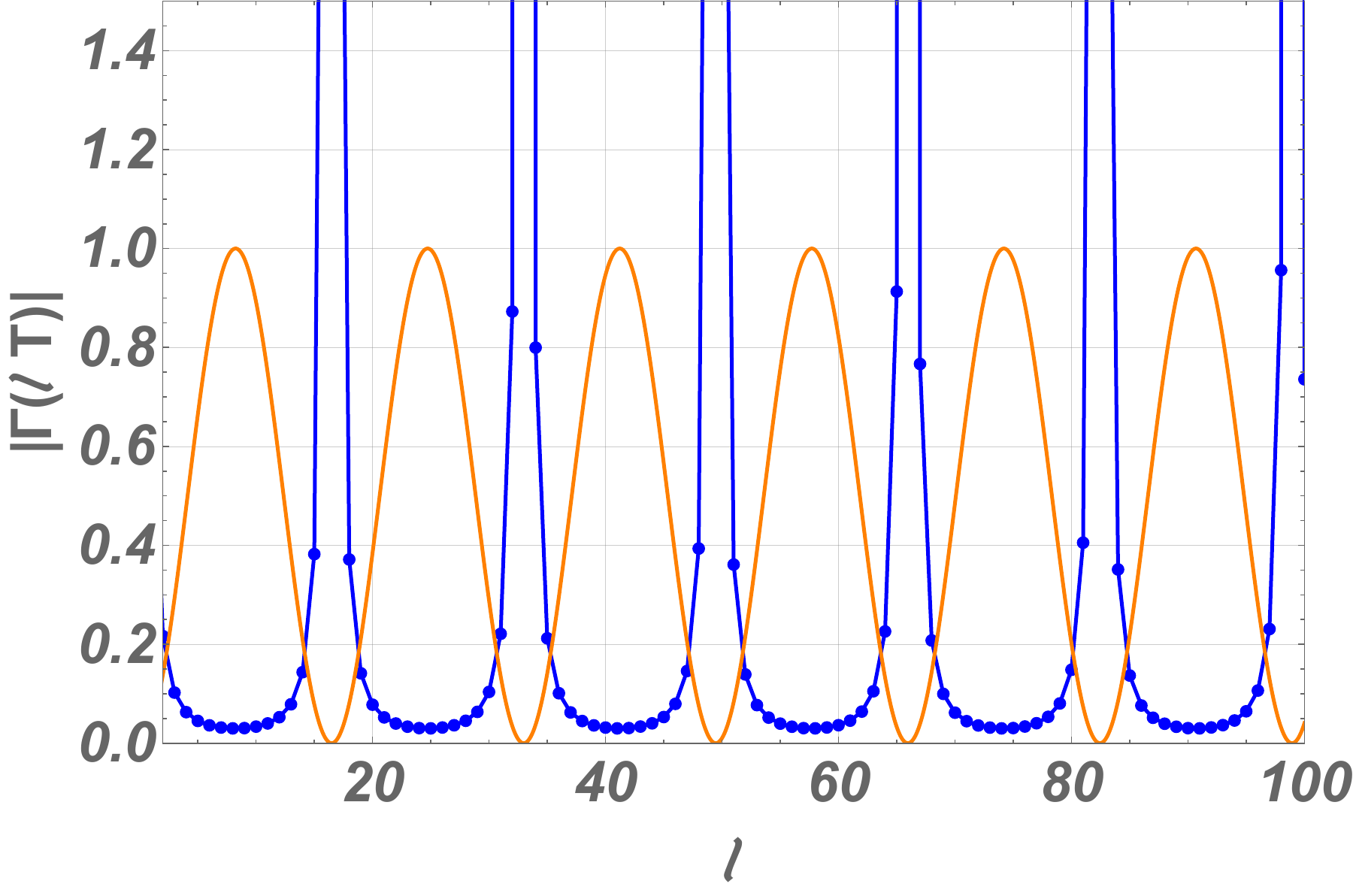}}
	\rotatebox{0}{\includegraphics*[width= 0.48\linewidth]{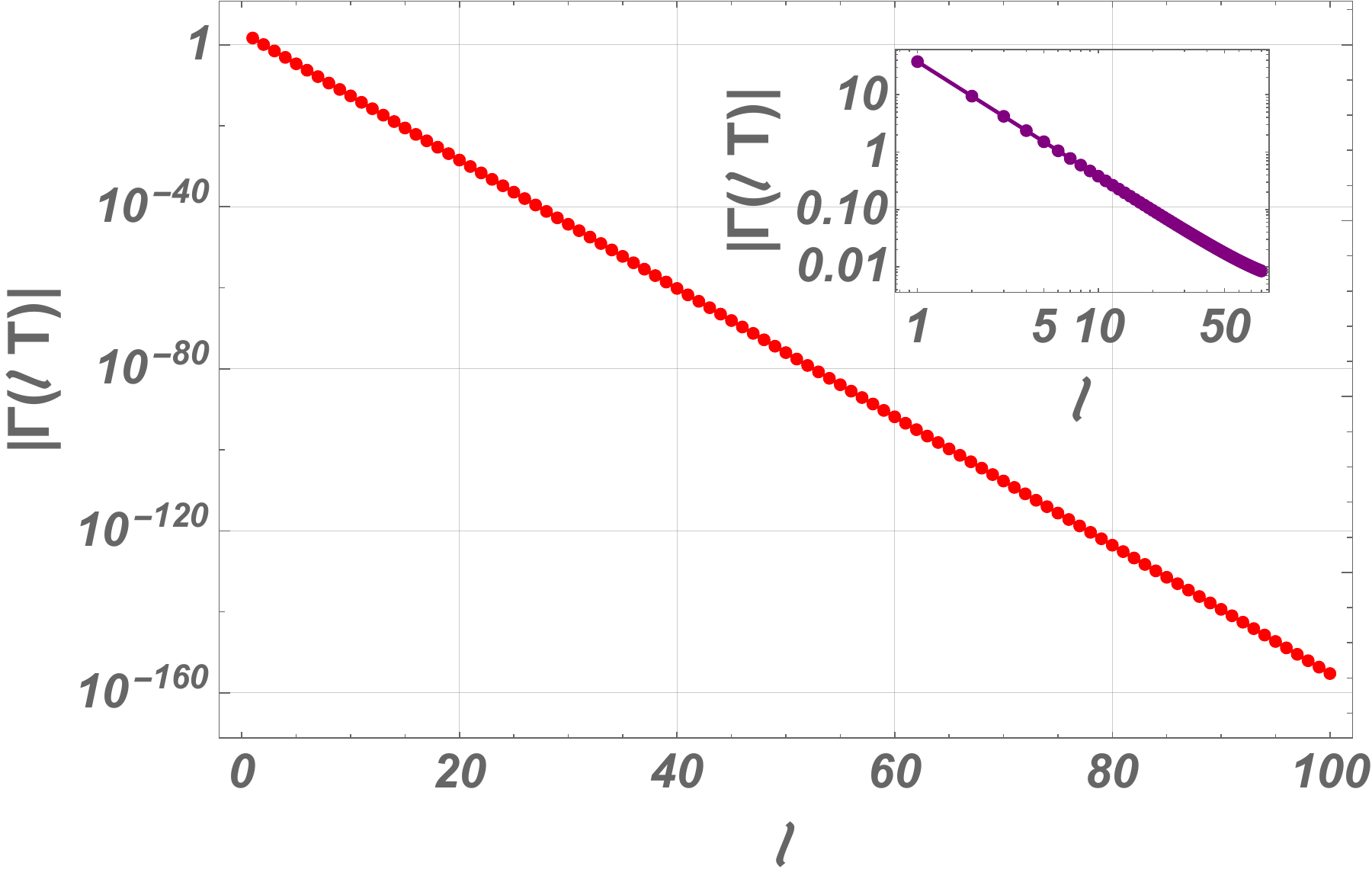}}
	\caption{Plots of absolute value of $\Gamma(t)$ in different steady states. We choose $n_O=1$ and $\Delta =2$. The left panel corresponds to $\mu_1=0.6$, $\mu_2=0.7$, $T_1=0.2$ and $T=0.3$, blue curve indicating the elliptic stroboscopic response, while the orange curve plots scaled $\sin^2( \ell \rho)$. The red points in the log plot of the right panel corresponds to exponential decay with $\mu_1=0.1$, $\mu_2=1.3$, $T_1=5$ and $T=9$.  The inset (log-log plot) is chosen with $\mu_1= 4.833$, $\mu_2 = 15.011, T_1 = 1, T = 2$, shows the parabolic phase. See text for details.} \label{figend1}
\end{figure}

{\it Vacuum Primary}: Firstly, we evaluate the vacuum primary autocorrelator $\Gamma_{\ell_1,\ell_2}$ by setting  $\vec{x}_2$ and $\ell_2$ to zero for simplicity: 
\begin{align}
	\Gamma(\ell T) & =  \frac{1}{b_\ell^\Delta} \exp \bigg\{  \frac{i}{2} \frac{n_O \vec{x}^2}{ d_\ell} \left( c_\ell  + \frac{1}{b_\ell } \right) \bigg\},
\end{align}
where we have written $\ell_1=\ell$ and $\vec x_1= \vec x$.  A plot of $|\Gamma(\ell T)|$ as a function of $\ell$, shown in Fig.\ \ref{figend1}, clearly shows the different behavior in the ellptic (left panel) and hyperbolic  phases (right panel). The inset of the right panel of Fig.\ \ref{figend1} indicates the behavior of $|\Gamma(\ell)|$ near the transition line $(\cos \rho \sim 0.9999)$. For the hyperbolic phase, $|\Gamma|$ shows an exponential decay with $\ell$ while near the transition line, the decay reduces to almost linear. For the elliptic phase the left panel exhibits oscillations interspersed with divergences at points where $b_\ell$ goes to zero. This happens whenever $\sin (\rho \ell )/\sin \rho= 0$ as the explicit formulae in \cite{NRFloquet} shows. These are exactly the positions of the zeros of $\sin^2 (\ell \rho)$ leaving out the first zero as the overlaying orange curves show in the figure.
Thus the vacuum autocorrelation function of primary operators can distinguish between the dynamic phases. 

\noindent
\paragraph{Conformal generators in trapped primary eigenstates}
In the context of fermions at unitarity, it is practically simpler to access the primary states $|\Delta\rangle $ corresponding to a fixed particle number $n$. We therefore consider the autocorrelators of $C$ and ${\mathcal E}$ starting from the state $|\Delta\rangle_{\mu_1}$. We note at the outset that these operators, in the context of trapped fermions, are unbounded allowing for the possibility of unbounded growth of autocorrelators. We shall see that this is indeed the case for the hyperbolic phase of the driven CFT. These autocorrelators are defined as 
\begin{align} 
{\mathcal A}_C &= {_{\mu_1}}\langle \Delta| (U_{T}^{-1})^{\ell} C U_T^{\ell} C|\Delta\rangle_{\mu_1} , \,\,\, 
{\mathcal A}_{\mathcal E} = {_{\mu_1}}\langle \Delta| (U_{T}^{-1})^{\ell} {\mathcal E} U_T^{\ell} {\mathcal E}|\Delta\rangle_{\mu_1}
\label{autodef1}
\end{align} 
To compute these we note that one can write $C_{\ell} C= (U_{T}^{-1})^{\ell} C U_T^{\ell} C$ and ${\mathcal E}_{\ell} {\mathcal E}$ as
\begin{eqnarray} 
C_{\ell} C &=& S_1=   a_{\ell}^2 C^2 +   b_{\ell}^2 HC +   b_{\ell}   a_{\ell} D C,  \nonumber\\ 
H_{\ell} H &=& S_2 =   d_{\ell}^2 H^2 +   c_{\ell}^2 C H +   d_{\ell}   c_{\ell} D H \nonumber\\
C_{\ell} H &=& S_3=  a_{\ell}^2 C H +   b_{\ell}^2 H^2 +  b_{\ell}   a_{\ell} D H, \nonumber\\
H_{\ell} C &=& S_4 =   d_{\ell}^2 H C +   c_{\ell}^2 C^2 +   d_{\ell}   c_{\ell} D C, \nonumber\\
{\mathcal E}_{\ell} {\mathcal E} &=& \mu_1^4 S_1 +\mu_1^2 (S_3+ S_4) + S_2.
\label{autodef2} 
\end{eqnarray} 

The computation of the autocorrelation therefore reduces computation of the expectation values of $C^2$, $H^2$, $HC$, $CH$, $CD$ and $HD$ for the trapped primary state $|\Delta\rangle_{\mu_1}$. For this, we note that from the definition of $L_0$ and $L_{\pm}$ [ see Eq.\ (3) of \cite{NRFloquet}] we can write 
\begin{eqnarray} 
H &=& \mu_1 [L_0 + \frac{1}{2} ( L_+ + L_-)], \quad D = -i(L_+-L_-), \quad
C = \frac{1}{\mu_1}[ L_0 - \frac{1}{2} (L_+ + L_-)]  \label{autodef3} 
\end{eqnarray} 
Using the properties ${_{\mu_1}}\langle \Delta| L_0|\Delta\rangle_{\mu_1} = \Delta/2$ and ${_{\mu_1}}\langle \Delta|L_+= L_- |\Delta\rangle_{\mu_1}=0$, we find that 
\begin{eqnarray}
{_{\mu_1}}\langle \Delta| H^2|\Delta\rangle_{\mu_1} &=& \mu_1^4 \,{_{\mu_1}}\langle \Delta| C^2|\Delta\rangle_{\mu_1} =\frac{\Delta(\Delta+1)}{4} \mu_1^2, \nonumber\\
{_{\mu_1}}\langle \Delta| H C|\Delta\rangle_{\mu_1} &=& {_{\mu_1}}\langle \Delta| C H |\Delta\rangle_{\mu_1} =\frac{\Delta(\Delta-1)}{4}, \nonumber\\
{_{\mu_1}}\langle \Delta| D H|\Delta\rangle_{\mu_1} &=& -\mu_1^2\, {_{\mu_1}}\langle \Delta| D C|\Delta\rangle_{\mu_1} = i\frac{\Delta \mu_1}{2}. \label{autodef4} 
\end{eqnarray} 
Substituting Eq.\ \ref{autodef4} in Eq.\ \ref{autodef1} and using Eq.\ \ref{autodef2}, we find after some algebra that ${\mathcal A}_C= {\mathcal F}_1 = {_{\mu_1}}\langle \Delta| S_1 |\Delta\rangle_{\mu_1}$ is given by 
\begin{eqnarray} 
{\mathcal A}_C  &=& \frac{\Delta}{2} \left( \frac{(\Delta+1)  a_{\ell}^2}{2 \mu_1^2} + \frac{(\Delta-1)  b_{\ell}^2}{2} 
- \frac{i}{\mu_1}   b_{\ell}   a_{\ell} \right).  \label{autodef5} 
\end{eqnarray} 

\begin{figure}[h]
	\centering
	\rotatebox{0}{\includegraphics*[width= 0.48 \linewidth]{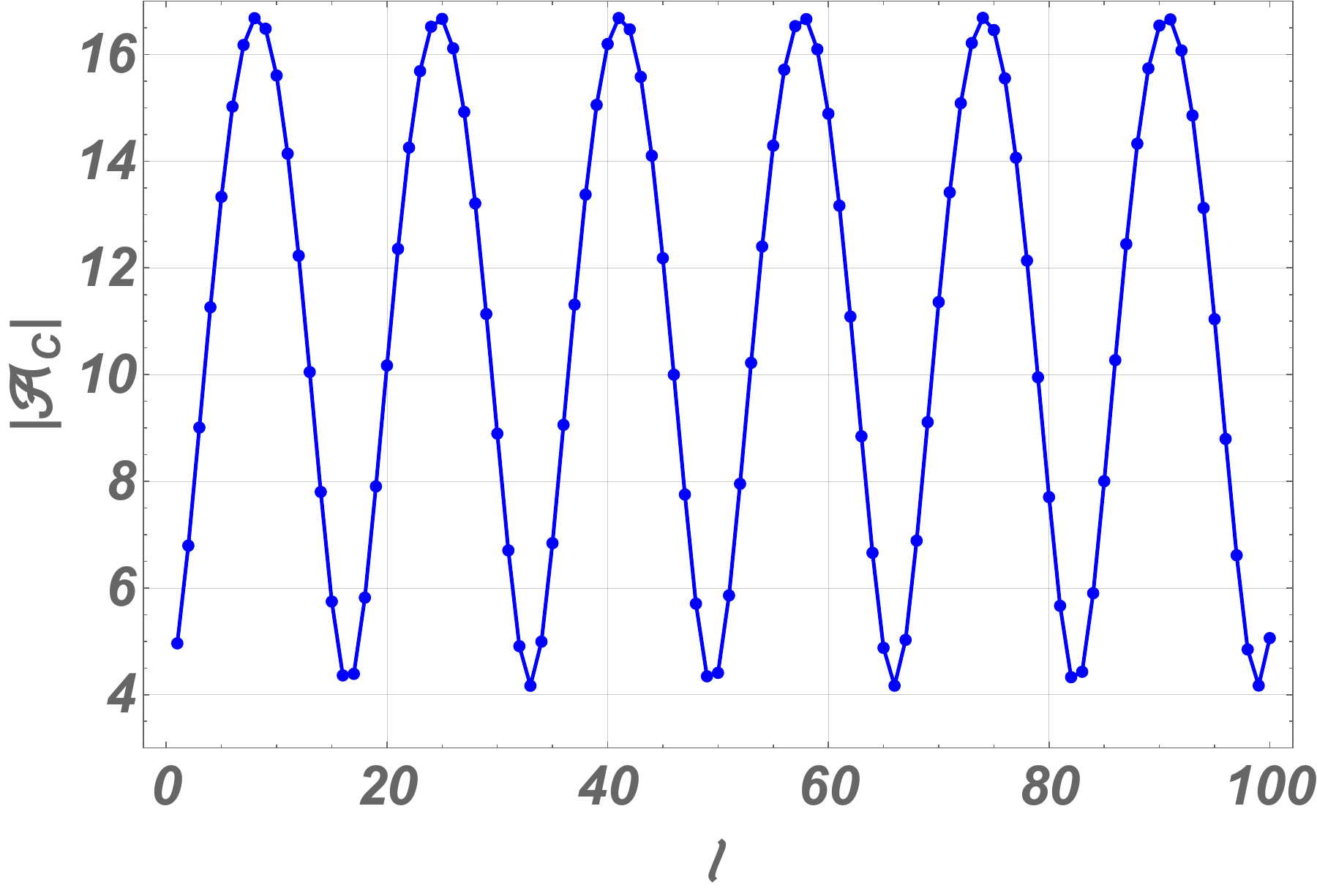}}
	\rotatebox{0}{\includegraphics*[width= 0.51\linewidth]{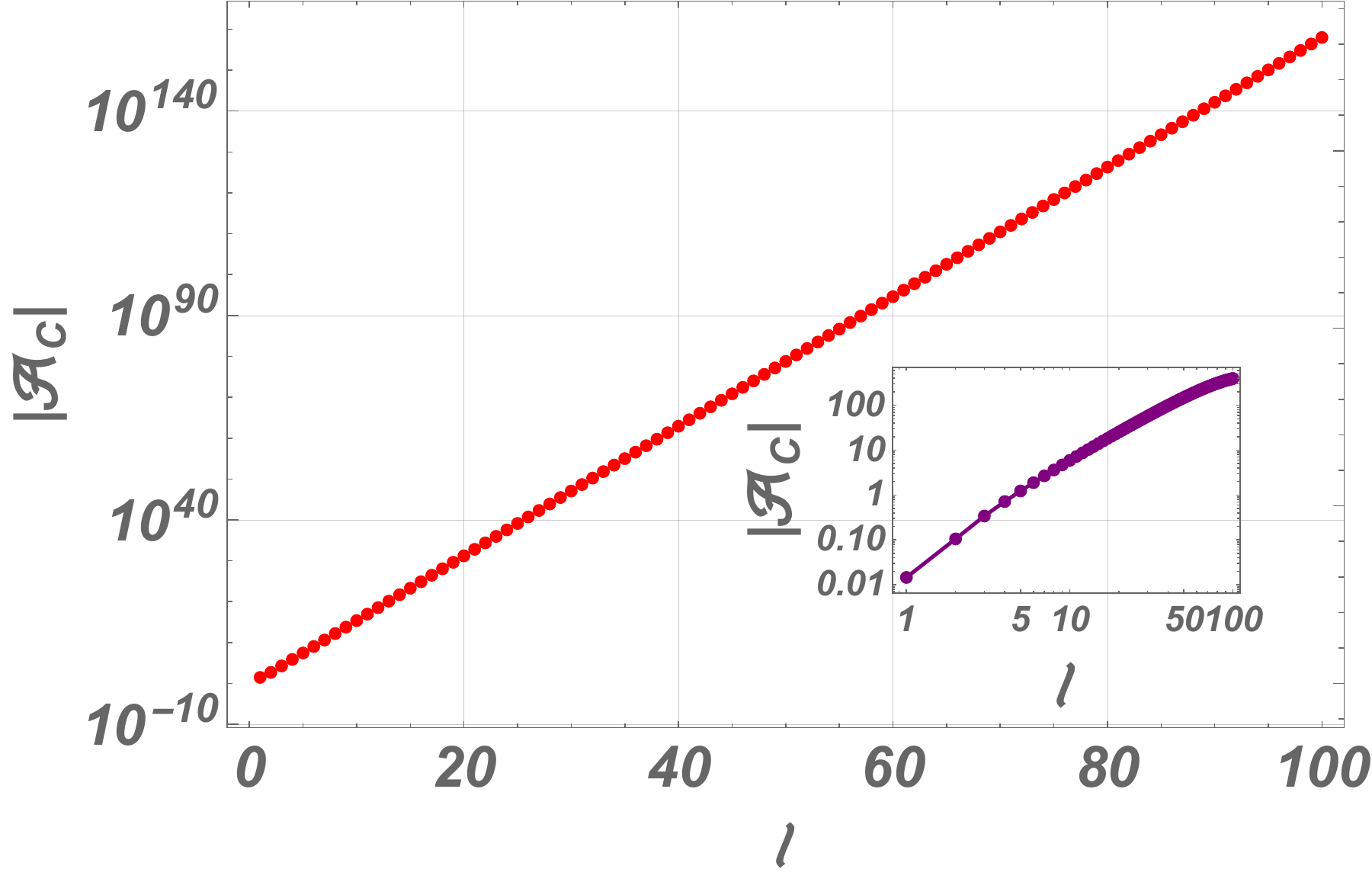}}
    \rotatebox{0}{\includegraphics*[width= 0.48 \linewidth]{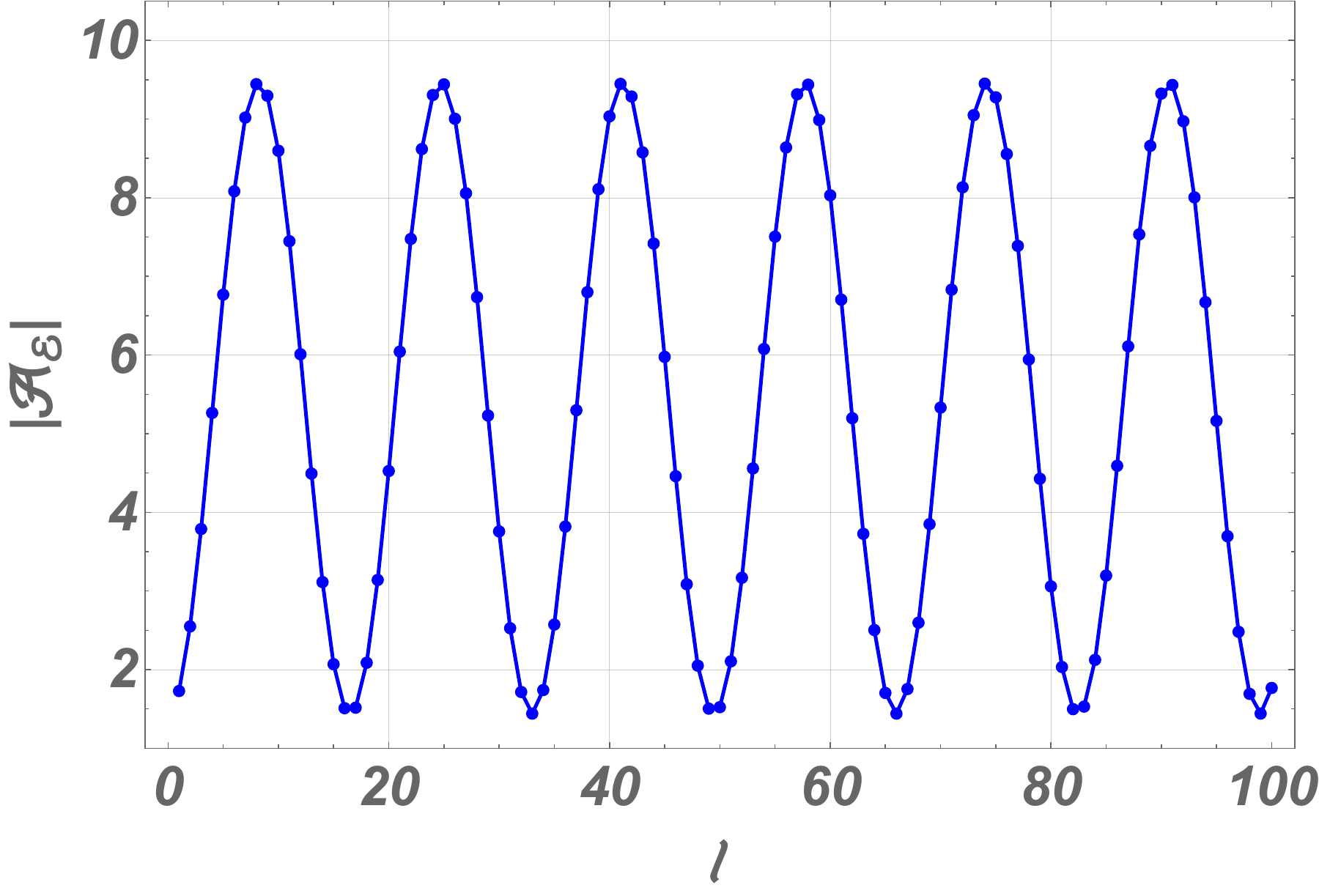}}
	\rotatebox{0}{\includegraphics*[width= 0.51\linewidth]{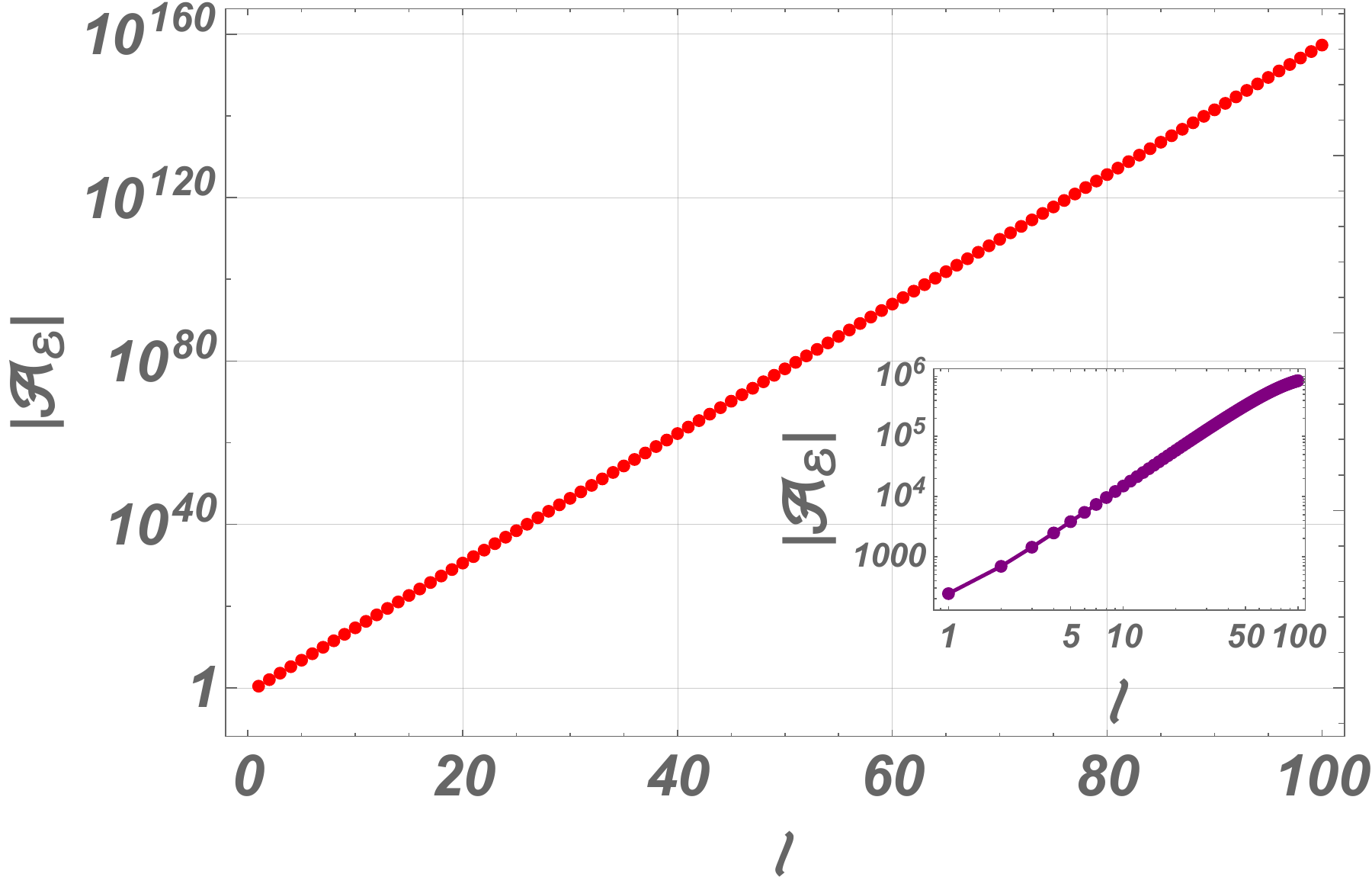}}
	\caption{Plots of absolute value of ${\mathcal A}_{C}$ as a function of $\ell$ in (a) ellptic and (b) hyperbolic phases. The inset shows its behavior near the transition line. (c) and (d) Same as in (a) and (b)
respectively, but for absolute value of ${\mathcal A}_{{\mathcal E}}$. The parameters for all plots are same as those in Fig.\ \ref{figend1}. See text for details.}
 \label{figend2} 
\end{figure}

To compute ${\mathcal A}_{{\mathcal E}}$, we need the expectation values of the operators $S_2$, $S_3$ and $S_4$. These are obtained easily using Eqs.\  \ref{autodef2} and \ref{autodef4}. 
Defining ${\mathcal F} _{2,3,4} = {_{\mu_1}}\langle \Delta| S_{2,3,4} |\Delta\rangle_{\mu_1}$, we find 
\begin{eqnarray} 
{\mathcal F}_2  &=& \frac{\Delta}{2} \left( \frac{(\Delta+1)\mu_1^2   d_{\ell}^2}{2} + \frac{(\Delta-1)  c_{\ell}^2}{2} + i \mu_1   c_{\ell}   d_{\ell} \right), \nonumber\\
{\mathcal F}_3  &=& \frac{\Delta}{4} \left( (\Delta-1)  a_{\ell}^2 + (\Delta+1) \mu_1^2   b_{\ell}^2 + 2 i \mu_1   b_{\ell}   a_{\ell} \right), \nonumber\\
{\mathcal F}_4  &=& \frac{\Delta}{4} \left( (\Delta-1)\mu_1^2   d_{\ell}^2  + \frac{(\Delta+1)  c_{\ell}^2}{\mu_1^2} - \frac{2 i}{ \mu_1}   c_{\ell}   d_{\ell} \right). \label{autodef6}
\end{eqnarray} 
Using Eq.\ \ref{autodef6} one finds 
\begin{eqnarray} 
{\mathcal A}_{\mathcal E} &=&  \mu_1^4 {\mathcal F}_1 + \mu_1^2 ( {\mathcal F}_3 + {\mathcal F}_4 ) + {\mathcal F}_2. \label{autodef7} 
\end{eqnarray} 

The behavior of these autocorrelators for $t= \ell T$ can be numerically computed by substituting values of $  a_{\ell}$, $  b_{\ell}$, $  c_{\ell}$ and $  d_{\ell}$ given in Eqs.\ \ref{autodef5} and \ref{autodef7}. The resulting behavior is shown in Fig.\ \ref{figend2}. The top panels shows the behavior for ${\mathcal A}_C$ in the elliptic ( left panel) and hyperbolic (right panel). The former shows oscillatory behavior while the latter indicates an exponential decay with $\ell$. The inset shows behavior of ${\mathcal A}_C$ near the transition line ($\cos \rho \sim 0.9999$); here ${\mathcal A}_c$ exhibits an almost linear decay.

The bottom panels indicates the behavior of ${\mathcal A}_{{\mathcal E}}$. We find that its behavior is similar ot that of ${\mathcal A}_{C}$ in both the phases; this indicates that the instantaneous energy autocorrelations can also distinguish between the dynamical phases discussed in this work.

\begin{figure}
\rotatebox{0}{\includegraphics*[width= 0.48 \linewidth]{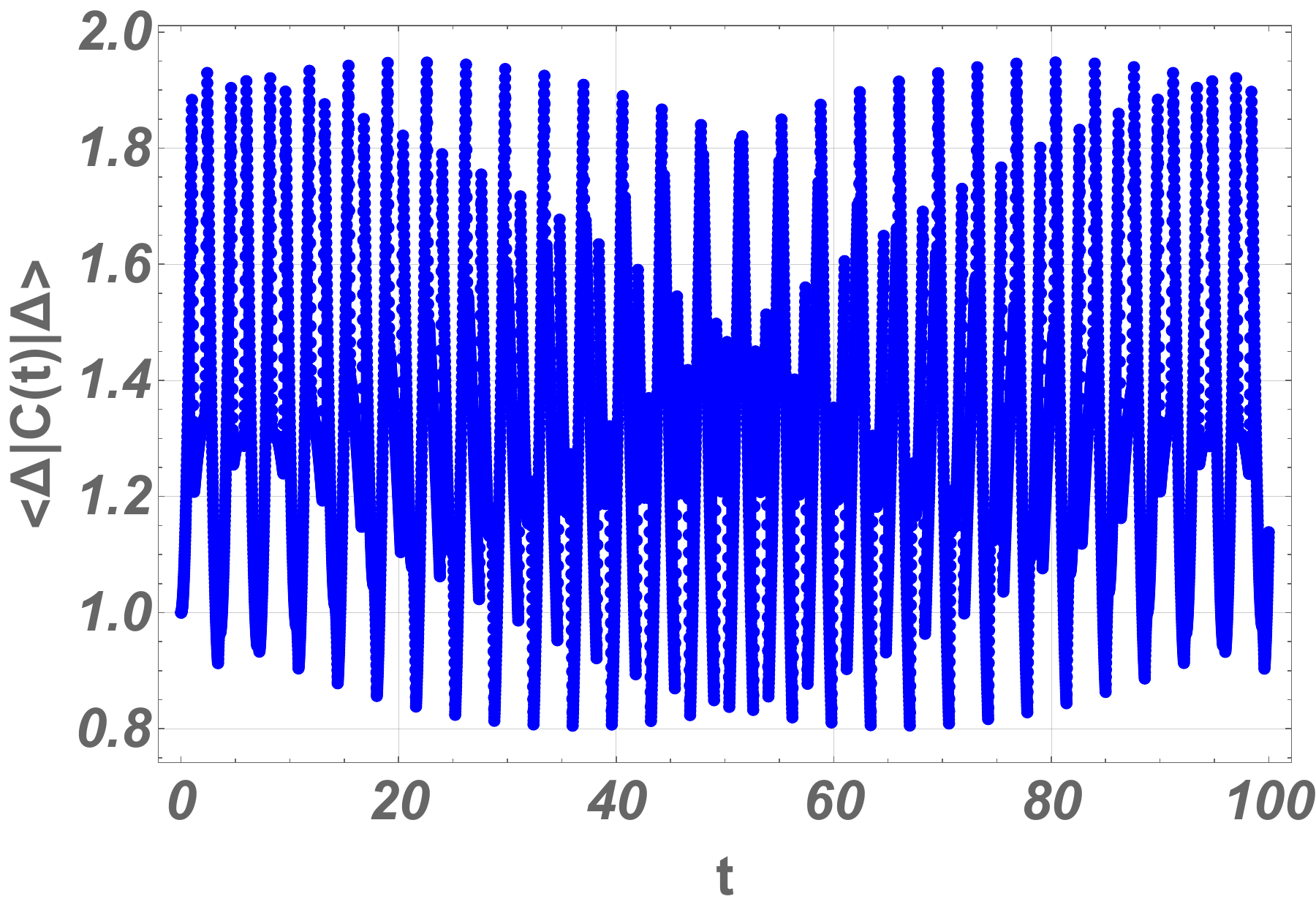}}
\rotatebox{0}{\includegraphics*[width= 0.48 \linewidth]{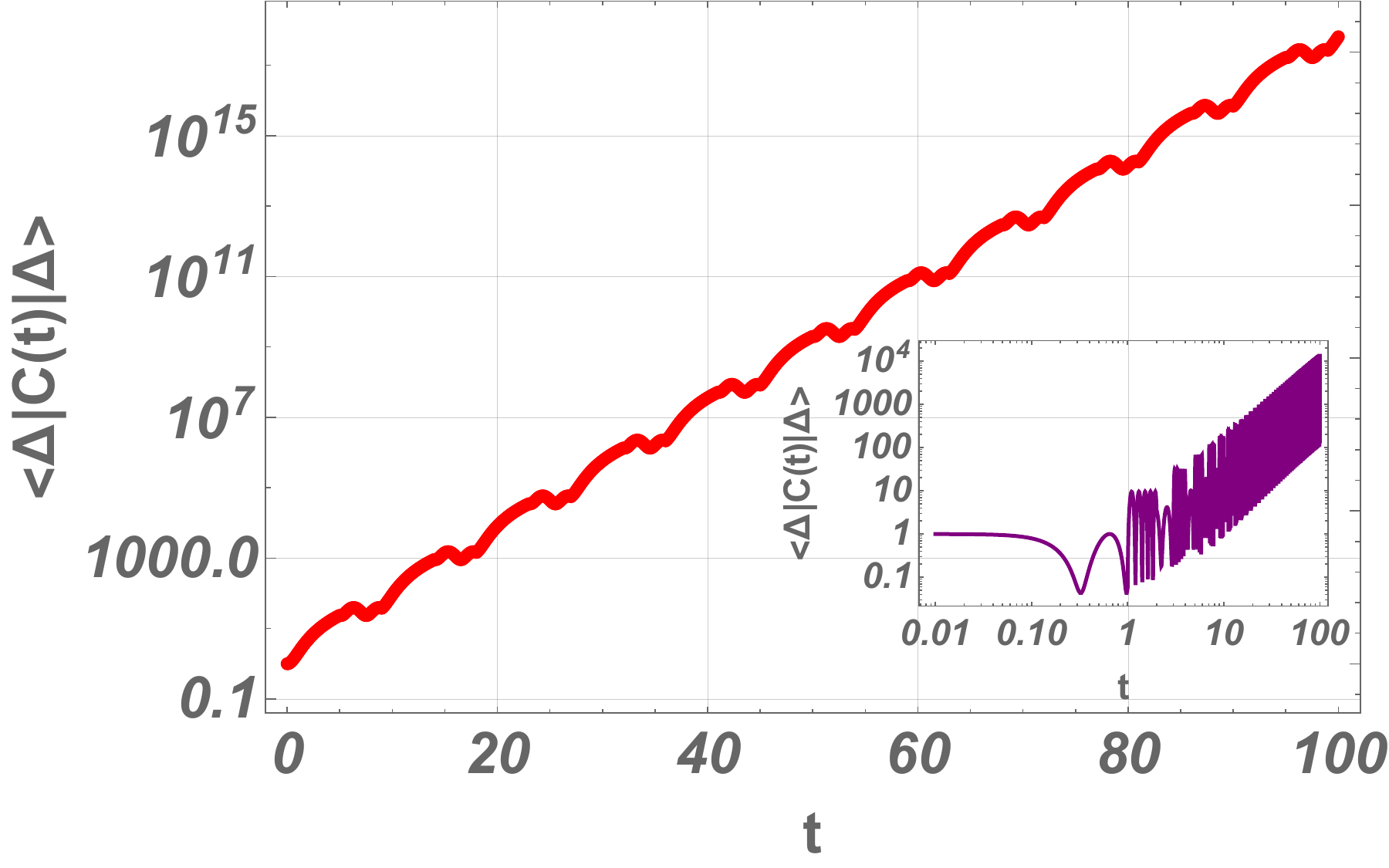}}
\caption{Plot of  expectation value of $C(t)$ as a function of $t= \ell T +\delta t$. The left panel corresponds to behavior in the elliptic phase while the right panel corresponds to the hyperbolic phase.  
The inset shows the near-linear initial growth followed by rapid oscillations of $C$ near the transition line. {The parameters for the oscillatory phase are $\mu_1 = 0.3, \mu_2 = 2, T_1 = 1$ and $T = 1.2$. The parameters for the right panel are the same as in Fig.\ \ref{figend1}. }See text for details}  \label{figns1}
\end{figure}

\section{Non-stroboscopic response}
\label{nssec} 
In this section, we study the non-stroboscopic response of the system. This is motivated by the fact that the information about micromotion of a periodically driven system requires probing it arbitrary times $t=\ell T+\delta t$. In the present problem, the exact evolution naturally factorizes as $U(t)=U_{\rm res}(\delta t)U_T^\ell$, where $U_{\rm res}(\delta t)$ is the residual evolution within 
the current driving cycle.
\begin{equation}
	U_{\rm res}(\delta t)=
	\begin{cases}
		e^{-i {\cal H}_1 \,\delta t},
		&0\leq \delta t\leq T_1,\\[6pt]
			e^{-i {\cal H}_2 \,(\delta t-T_1)}e^{-i {\cal H}_1 \,T_1},
		&T_1<\delta t<T.
	\end{cases}
	\label{eq:Mres-anytime}
\end{equation}
The non-stroboscopic correlator therefore probes both the Floquet Hamiltonian and the micromotion operator. The phase of the drive is still controlled by the reality of the $\rho$ parameter [see Eq.(7) in \cite{NRFloquet}]. The non-stroboscopic correlator inherits these behaviors, dressed by the intra-cycle remainder matrix as seen in Fig. \ref{figns1}. The left panel of Fig.\ \ref{figns1} shows behavior in the elliptic phase; here the presence of the micromotion do not alter the nature of the oscillation of $C$. This can be seen comparing the behavior of $C(t)$ to its stroboscopic 
counterpart in \cite{NRFloquet}. The right panel shows the behavior of $C(t)$ in the hyperbolic phase; we find that the exponential growth is modified by small oscillations. These oscillations do not arrest the exponential growth and consequently the nature of $C(t)$ is similar to that in the stroboscopic case. The inset of the Fig.\ \ref{figns1}(b) exhibits the behavior of $C(t)$ near the transition line in the elliptic phase. Here we find oscillatory features at late times along with the linear growth; these oscillations were absent when $C$ was measured only in stroboscopic times. Thus we find that the effect of the micromotion manifests itself near the transition line; the properties of the dynamical phase do not change due to its presence.    
\noindent
\subsection{Numerical details}
For the numerical plots, we sample the correlator at uniformly spaced times
\begin{equation}
	t_j=j\Delta t,
	\qquad
	j=0,1,\ldots,N-1,
	\qquad
	T_{\rm obs}=N\Delta t.
\end{equation}
The point $t=0$ is avoided because the correlator has the expected short-time
singularity.  At each time we decompose
$t_j=\ell_jT+\delta t_j$, construct $U(t_j)$ using
Eq.~\eqref{eq:Mres-anytime}, and $U(t) = U_{{\rm res}}(\delta) U_T^\ell$. 
Writing
\begin{equation}
	U(t)=
	\begin{pmatrix}
		a(t)&b(t)\\
		c(t)&d(t)
	\end{pmatrix},
	\label{eq:Mt-components}
\end{equation}
Points where $b(t_j)$ or $d(t_j)$ becomes numerically too small are treated as singular points and are omitted or regularized in the plotting routine.

\section{ Cosine drive protocol}

In \cite{NRFloquet} we discuss the dynamical phases generated by periodically
driving the trap.  Here we spell out the numerical procedure used for the
continuous cosine protocol
\begin{equation}
{\cal H}(t)=H+\mu^2(t) C,\qquad
\mu^2(t)=\mu_1^2+\mu_2^2\cos(\omega_D t),
\label{eq:cosine-drive-hamiltonian}
\end{equation}
where $H,C,D$ are the generators of the $SO(2,1)$ algebra.  The drive period is
$T=2\pi/\omega_D$.  The phase diagram is obtained from the one-period evolution
matrix and its conjugacy class.

Since ${\cal H}(t)$ is always a linear combination of the $SO(2,1)$ generators,
the evolution may be computed in any faithful representation of this algebra.
We use the two-dimensional representation $D=i \sigma_z$, $C=-i \sigma_-$ and $H= i \sigma_+$ 
as given in \cite{NRFloquet}. In this representation
\begin{equation}
-i{\cal H}(t)=
\begin{pmatrix}
0&1\\
-\kappa(t)&0
\end{pmatrix},
\qquad
\kappa(t)=\mu_1^2+\mu_2^2\cos(\omega_D t).
\label{eq:matrix-generator}
\end{equation}

\begin{figure}[t]
\centering
\includegraphics[width=0.58\linewidth]{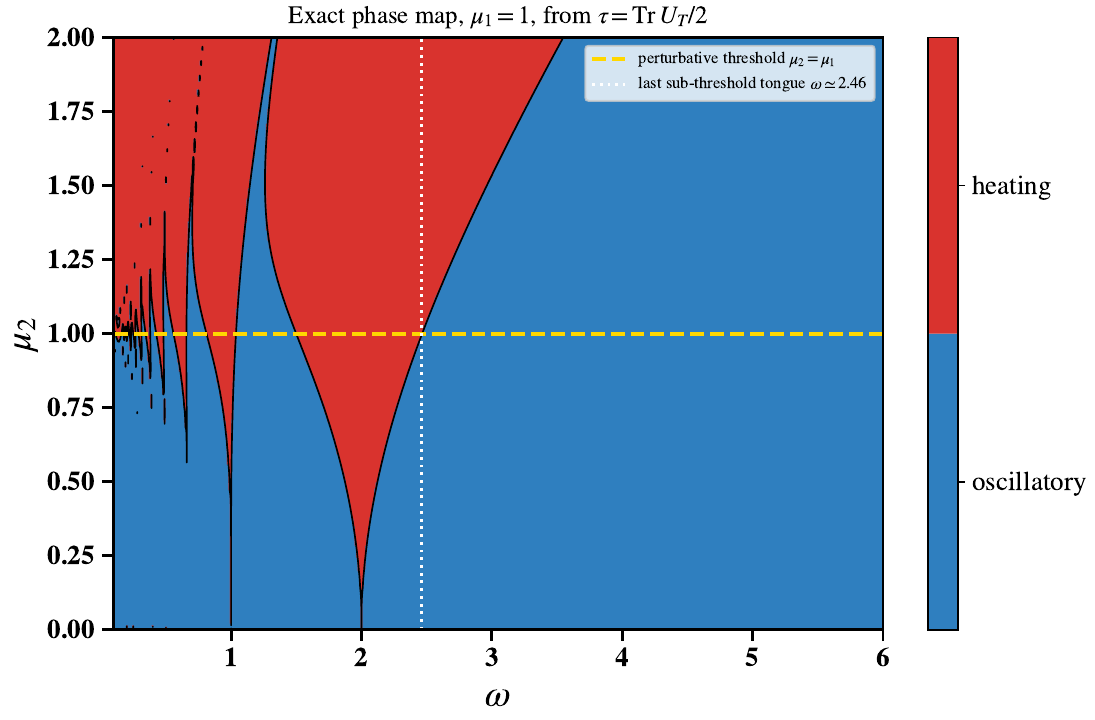}
\caption{Cosine-drive phase diagram for $\mu_1=1$.  The blue region denotes
the elliptic/oscillatory phase, $|\calt|\leq1$, and the red region denotes the
hyperbolic/heating phase, $|\calt|>1$.  The black curve indicates the
transition $|\calt|=1$. The region below the yellow line is the experimentally relevant $\mu_2 < \mu_1$ region, and the dotted white line indicates the upper bound on $\omega_D \simeq 2.46$, in order to realize the exponential phase in this relevant region.} 
\label{fig:cosine-phase-map}
\end{figure}

For the Suzuki-Trotter decomposition of the evolution operator $U$, we divide a drive period of length $T$ into $N_0$ equal time steps.
To this end, we define 
\begin{equation}
\delta t=\frac{T}{N_0},\qquad t_j=j\,\delta t,\qquad j=0,1,\ldots,N_0.
\label{eq:N0-definition}
\end{equation}
Here $N_0 \gg 1$ is chosen to accurately approximate the time-ordered exponential over one full drive period.  The numerical results
below use a right-endpoint discretization on each interval.  We denote the
evolution over the interval $[t_j,t_{j+1}]$ by $U(t_j,t_{j+1})$:
\begin{equation}
U(t_j,t_{j+1})
\simeq \exp[-i{\cal H}(t_{j+1})\delta t].
\label{eq:interval-evolution}
\end{equation}
Writing $s_{j+1}=\sqrt{-\kappa(t_{j+1})}$,  this interval matrix is given by 
\begin{equation}
U(t_j,t_{j+1})=
\begin{pmatrix}
\cosh(s_{j+1}\delta t) &
\sinh(s_{j+1}\delta t)/s_{j+1}\\
s_{j+1}\sinh(s_{j+1}\delta t) &
\cosh(s_{j+1}\delta t)
\end{pmatrix}.
\label{eq:interval-matrix}
\end{equation}
At points where $s_{j+1}=0$, the matrix element
$\sinh(s_{j+1}\delta t)/s_{j+1}$ is evaluated by its smooth limit
$\delta t \rightarrow 0$.
The one-period evolution is the chronological product of these interval
matrices,
\begin{equation}
U(T,0)\simeq
U(t_{N_0-1},t_{N_0})\,
U(t_{N_0-2},t_{N_0-1})\cdots
U(t_0,t_1).
\label{eq:one-period-product}
\end{equation}
This product is an ordinary matrix product, and the dynamical phase is diagnosed by
\begin{equation}
\calt=\frac{1}{2}\tr\,U(T,0).
\label{eq:trace-diagnostic}
\end{equation}
The evolution is elliptic, and the corresponding response is oscillatory, when
$|\calt|\leq 1$.  It is hyperbolic, corresponding to heating in the driven
trap, when $|\calt|>1$.  The boundary $|\calt|=1$ is the parabolic transition. Fig.\ \ref{fig:cosine-phase-map} shows the phase diagram for $\mu_1=1$ in the
$\omega_D$--$\mu_2$ plane.  The scan uses $0\leq \mu_2\leq 2,\quad 0.1\leq\omega_D\leq 6$, grid size: $
401\times591$  and Trotter steps $N_0=2048$. The $ \max|\mathcal T|=3.5922\times10^{14}$ and $
\quad 26 \%$ of the phase space is in the exponential phase. See the next subsection for an investigation of convergence.

The high-frequency region of the phase diagram can be understood using a Magnus expansion 
which provides a semi-analytic, albeit approximate, understanding of the heating and non-heating phases. 
To this end, we first write the Floquet Hamiltonian within the Magnus expansion and up to third-order in $T$. 
This is given by  $H_F = \sum_{n=1,,\infty} H_F^{(n)}$ where the first three terms of the expansion can be written as 
\begin{eqnarray}
H_F^{(1)} &=& \frac{1}{T} \int_0^T dt_1 {\mathcal H}(t_1), \quad   H_F^{(2)} = \frac{1}{2T^2} \int_0^T dt_1 \int_0^{t_1} dt_2 [ {\mathcal H}(t_1), {\mathcal H}(t_2)], \nonumber\\
H_F^{(3)} &=& \frac{1}{6T^3} \int_0^T dt_1 \int_0^{t_1} dt_2 \int_0^{t_2} dt_3 ( [ {\mathcal H}(t_1), [ {\mathcal H}(t_2), {\mathcal H}(t_3)]] + [ {\mathcal H}(t_3), [ {\mathcal H}(t_2), {\mathcal H}(t_1)]], \label{mageq1}
\end{eqnarray} 
where $ {\mathcal H}(t)$ is given by Eq.\ \ref{eq:cosine-drive-hamiltonian}. A straightforward evaluation of these integrals yield
\begin{eqnarray}
H_F^{(1)} &=& H + \mu_1^2 C, \quad   H_F^{(2)} = 0, \quad H_F^{(3)} = k_1 H + k_2 C, \label{mageq3}
\end{eqnarray}  
where $k_1= - 3\mu_2^2/(68 \pi^2)$, $k_2=3\mu_2^2(4\mu_0^2-\mu_1^2)/(272 \pi^2)$ and we have used $\omega_D T=2 \pi$. 

Using the Pauli representation for $H$ and $C$, it is straightforward to obtain the evolution operator $U_{\rm Magnus} = \exp[-i (H_F^{(1)} + H_F^{(3)})T/\hbar]$. After some algebra, we find 
\begin{eqnarray} 
{\mathcal T}_{\rm Magnus} &=& \frac{1}{2} {\rm Tr} U_{\rm Magnus} = \cosh [N_m/(68 \pi \omega_D)], \nonumber\\
N_m &=& \sqrt{ 204 \mu_2^4 \pi^2 - 9 \mu_2^6 + 4 \mu_1^2( 9 \mu_2^4- 4624 \pi^4)}. \label{mageq4}
\end{eqnarray} 
We note that the ellptic and the hyperbolic phases corresponds to $N_m^2<0$ and $N_m^2>0$ respectively, while the transition occurs at $N_m=0$ leading to a critical value of $\mu_2=\mu_2^c$ given by 
\begin{eqnarray}
\mu_2^c &=& \sqrt{2 \mu_1 + \sqrt{ \mu_1^2 + 68 \mu_1 \pi^2/3}}. \label{mageq5} 
\end{eqnarray} 
We note that $\mu_2^c >\mu_1$; consequently, $N_m^2 >0$ for $\mu_2 >\mu_1$ in the high frequency regime. This indicates that for experimentally relevant trap parameters for which $\mu_1 >\mu_2$, 
the cosine drive protocol  can only yield elliptic phase in the high-frequency regime where the third-order Magnus yield accurate results. This feature can be numerically confirmed from Fig.\ \ref{fig:cosine-phase-map} at large $\omega_D$ where only non-heating phase is present in the experimentally relevant region. 

\subsection{Convergence with the number of Trotter steps}
\label{sec:convergence}

The convergence of the Trotter calculation is
summarized below, with all errors measured relative to the \(N_0=2048\)
calculation on the same grid.  In the scan space \(|\mathcal T|\) itself becomes very large.  The most useful convergence 
diagnostics for the phase diagram are the class-flip fraction and the heating
fraction.  The class-flip fraction denotes the fraction of grid points whose phase
classification, heating for \(|\mathcal T|>1\) and oscillatory for
\(|\mathcal T|<1\), differs from the \(N_0=2048\) reference calculation.
Thus it directly measures the stability of the plotted phase diagram rather
than the pointwise numerical error in \(\mathcal T\). The phase classification is stable on the plotted grid by
\(N_0=240\), and the heating fraction converges to \(0.260170\). For \(N_0\geq 240\), no grid point changes its heating/oscillatory classification relative to the \(N_0=2048\) reference calculation on the
\(401\times591\) sampled grid. 

\begin{center}
	\begin{tabular}{c c c}
		\hline
		\(N_0\) 
		& class-flip fraction & heating fraction \\
		\hline
		30   
		& \(4.35\times10^{-4}\) & \(0.259913\) \\
		60   
		& \(1.01\times10^{-4}\) & \(0.260120\) \\
		120 
		& \(1.27\times10^{-5}\) & \(0.260166\) \\
		240  
		& \(0\) & \(0.260170\) \\
		480 
		& \(0\) & \(0.260170\) \\
		960 
		& \(0\) & \(0.260170\) \\
		1024 
		& \(0\) & \(0.260170\) \\
		2048  
		& \(0\) & \(0.260170\) \\
		\hline
	\end{tabular}
\end{center}
\section{Computation resembling the experimental protocol in \cite{pra2}}

\noindent
The protocol discussed in Ref.\ \onlinecite{pra2} results only in the heating phase, which is observed as the oscillations with growing amplitudes in their Fig. 1. Below, we explain the reason for this. 
We note that for the driven CFT, one should compute the one-period ${\rm SO}(2,1)$  monodromy
matrix given by
\begin{eqnarray} 
 U_T={\cal P}\exp\left[-i\int_0^T dt\,
 \bigl(H+\mu^2(t)C\bigr)\right],
\end{eqnarray} 
and then compute $ {\cal T}\equiv \frac12 {\rm Tr}\,U_T$ to determine the phase. The conjugacy class of this matrix determines the stroboscopic phase:
\begin{eqnarray} 
 |{\cal T}|<1
 \quad\Rightarrow\quad
 \text{elliptic / oscillatory},
\end{eqnarray} 
\begin{eqnarray} 
 |{\cal T}|=1
 \quad\Rightarrow\quad
 \text{parabolic / critical},
\end{eqnarray} 
\begin{eqnarray} 
 |{\cal T}|>1
 \quad\Rightarrow\quad
 \text{hyperbolic / exponentially growing}.
\end{eqnarray} 
Equivalently, in the notation of \cite{NRFloquet}, 
\begin{eqnarray} 
 {\cal T}=\frac12{\rm Tr}\,U_T=\cos\rho,
\end{eqnarray} 
so that real \(\rho\) gives the elliptic phase, imaginary \(\rho\) gives the
hyperbolic phase, and \(\rho=0\) gives the parabolic transition.  This is the
classification underlying the elliptic, hyperbolic and parabolic regimes
discussed in Ref.~\cite{NRFloquet}.

For the drive protocol of Ref.\ \onlinecite{pra2}, the relevant protocol is the resonant modulation
\begin{eqnarray} 
 \mu^2(t)=\mu_0^2\left[1+\beta\sin(2\mu_0t)\right].
\end{eqnarray} 
After preparing the cloud in equilibrium, the trapping potential is modulated at frequency
\(2\mu_0\), with \(\mu_0=2\pi\times720\,{\rm Hz}\), and then the trap is
held fixed.  The same paragraph gives Eq.~(5).  The caption of
Fig.~1(c) of Ref.~\cite{pra2} specifies the modulation amplitude
\(\beta=0.04\).  The caption of Fig.~2(b) specifies the stroboscopic
modulation times \(t_1=2T,4T,6T\), where \(T=\pi/\mu_0\) is the modulation
period.  The larger-amplitude anharmonicity data in Fig.~5 use
\(\beta=0.1\), as stated in the caption of that figure.

\begin{figure}[t]
\centering
\includegraphics[width=0.58\linewidth]{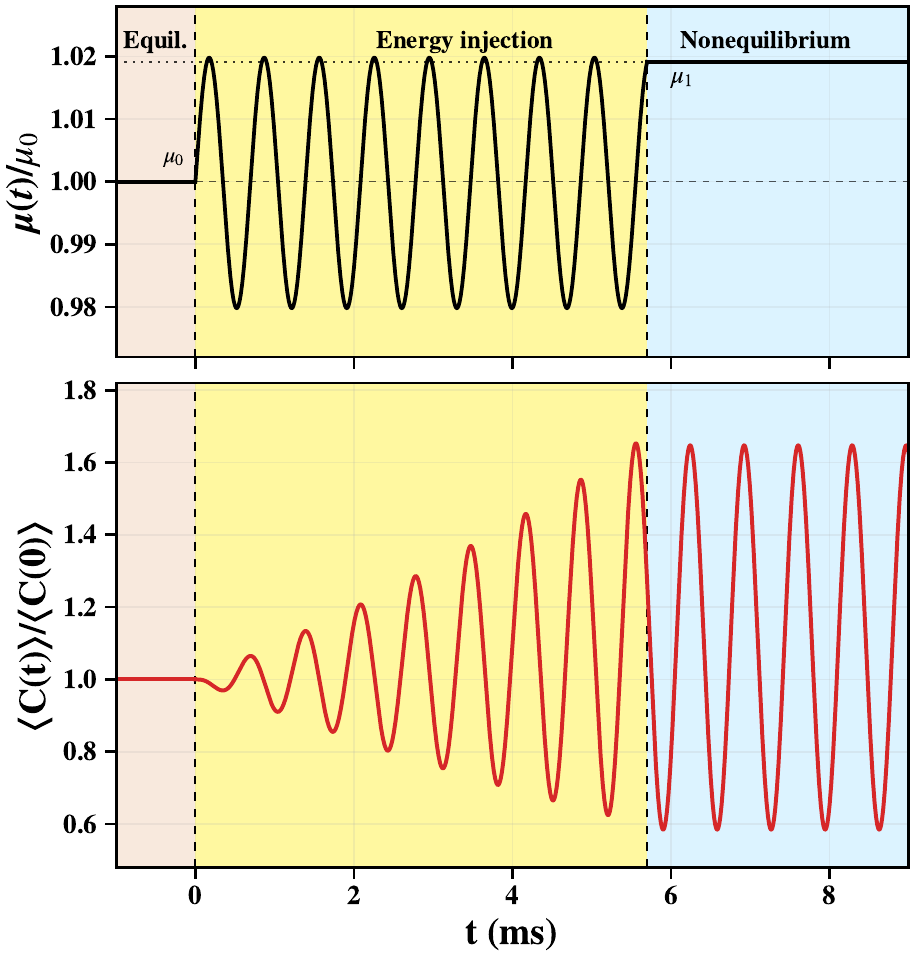}
\caption{Protocol-matched NRCFT computation for the trap modulation used in Ref.~\cite{pra2}.
The upper panel shows the imposed trap protocol, with resonant modulation
\(\mu^2(t)=\mu_0^2[1+\beta\sin(2\mu_0 t)]\) followed by evolution at
\(\mu_1^2=\mu_0^2[1+\beta\sin(2\mu_0 t_1)]\).
The lower panel shows the computed conformal response
\(\langle C(t)\rangle/\langle C(0)\rangle=A(t)^2+B(t)^2\), obtained directly from
the unfitted ${\rm SO}(2,1)$ evolution matrix
\(U(t)=\begin{pmatrix}A&B\\ C&D\end{pmatrix}\).
The parameters are \(\mu_0/(2\pi)=720\,{\rm Hz}\), \(\beta=0.04\),
\(T=\pi/\mu_0=0.694\,{\rm ms}\), \(t_1=5.70\,{\rm ms}\), and
\(\mu_1/\mu_0=1.0191\).}
\label{fig:expt}
\end{figure}

It is useful to make the dimensionless time variable
\begin{eqnarray} 
 s=\mu_0 t.
\end{eqnarray} 
Then one modulation period is \(s\in[0,\pi]\), and the two-dimensional
${\rm SO}(2,1)$ evolution obeys
\begin{eqnarray} 
 \frac{dU}{ds}
 =
 \begin{pmatrix}
 0&1\\
 -[1+\beta\sin(2s)]&0
 \end{pmatrix}U,
 \qquad U(0)={\bf 1}.
\end{eqnarray} 
The Floquet diagnostic is therefore
\begin{eqnarray} 
 {\cal T}_{2}(\beta)
 &=& \frac12{\rm Tr}\,U(\pi).
\end{eqnarray} 
For \(\beta=0\), \(U(\pi)=-{\bf 1}\), so \({\cal T}_{2}=-1\). The
unmodulated system therefore lies exactly at the parabolic point for the modulation
period \(T=\pi/\mu_0\).  Turning on the resonant modulation moves the
system into the hyperbolic phase.  Perturbatively one finds
\begin{eqnarray} 
 {\cal T}_{2}(\beta)
 &=& -1-\frac{\pi^2}{32}\beta^2+O(\beta^4),
\end{eqnarray} 
so every sufficiently small nonzero \(\beta\) on this resonant cut has
\(|{\cal T}_2|>1\).

The exact numerical values, obtained by Trotterizing Eq.~(6) with exact
constant-\(\mu^2\) evolution on each time step, are shown in the table
below.  The determinant of the computed monodromy stays
equal to one up to numerical roundoff.  In particular, for the 
amplitude in Ref.\ \onlinecite{pra2}, \(\beta=0.04\), and we find 
\begin{eqnarray} 
 {\cal T}_{2}=-1.000493474543,
 \qquad
 |{\cal T}_{2}|=1.000493474543>1,
\end{eqnarray} 
and for the larger-amplitude run for which $\beta=0.1$,
\begin{eqnarray} 
 {\cal T}_{2}=-1.003084029633,
 \qquad
 |{\cal T}_{2}|=1.003084029633>1.
\end{eqnarray} 
Thus the experiment in Ref.\ \onlinecite{pra2} follows a resonant cut through the principal
heating tongue.  It changes the amplitude and the number of modulation
periods, but does not detune the drive frequency across the elliptic regions
of the full Floquet phase diagram.  This is why the elliptic/oscillatory
Floquet phase is not visible in that protocol. This is further elaborated in the table below. 

\begin{center}
\begin{tabular}{cccccc}
\toprule
\(\beta\) &
\({\cal T}_2=\frac12{\rm Tr}\,U(\pi)\) &
\(|{\cal T}_2|\) &
\(\frac12{\rm Tr}\,U^2\) &
\(\frac12{\rm Tr}\,U^4\) &
Floquet class\\
\midrule
0      & \(-0.999999999994\) & \(0.999999999994\) & \(1.000000000\) & \(1.000000000\) & parabolic\\
0.01   & \(-1.000030842492\) & \(1.000030842492\) & \(1.000123372\) & \(1.000493518\) & hyperbolic\\
0.02   & \(-1.000123369700\) & \(1.000123369700\) & \(1.000493509\) & \(1.001974524\) & hyperbolic\\
0.04   & \(-1.000493474543\) & \(1.000493474543\) & \(1.001974385\) & \(1.007905337\) & hyperbolic\\
0.06   & \(-1.001110301757\) & \(1.001110301757\) & \(1.004443673\) & \(1.017814183\) & hyperbolic\\
0.08   & \(-1.001973830054\) & \(1.001973830054\) & \(1.007903112\) & \(1.031737367\) & hyperbolic\\
0.10   & \(-1.003084029633\) & \(1.003084029633\) & \(1.012355141\) & \(1.049725863\) & hyperbolic\\
\bottomrule
\end{tabular}
\par\smallskip
\begin{minipage}{0.93\textwidth}
\small
\noindent{\bf Table 1.}
Floquet half-trace for the resonant protocol in Ref.\ \onlinecite{pra2},
\(\mu^2(t)=\mu_0^2[1+\beta\sin(2\mu_0t)]\), over one modulation
period \(T=\pi/\mu_0\).  The columns \(\frac12{\rm Tr}\,U^2\) and
\(\frac12{\rm Tr}\,U^4\) correspond to two and four modulation periods,
respectively; Ref.~\cite{pra2}, Fig.~2(b), uses \(t_1=2T,4T,6T\).  The
experimental periodic-drive amplitude in Figs.~1--3 is \(\beta=0.04\), while
Fig.~5 uses \(\beta=0.1\).  Except for the unmodulated parabolic point
\(\beta=0\), the listed values all satisfy \(|{\cal T}_2|>1\).
\end{minipage}
\end{center}

\section{Ermakov--Pinney representation of the ${\rm SO}(2,1)$ evolution}
\label{sec:SI-ermakov}

We explain here how an Ermakov--Pinney description follows from a
general ${\rm SO}(2,1)$ evolution and how the corresponding scaling
variable determines the evolved wavefunction.

\noindent
Consider the most general time-dependent Hamiltonian constructed from
the ${\rm SO}(2,1)$ generators,
\begin{equation}
	{\cal H}(t)=u(t)H+v(t)D+w(t)C ,
	\label{eq:SI-general-A}
\end{equation}
For $u(t) > 0$, if we define \begin{equation}
	V_\eta(t)=e^{\ii\beta(t)C}e^{-\ii\ell(t)D},
	\qquad
	\ell=\ln(\sqrt u\,\eta),
	\qquad
	\beta=\frac{1}{u}\left(\frac{\dot\eta}{\eta}-\delta_u\right), \qquad 
	\delta_u(t) \equiv v(t) - \frac{\dot{u}(t)}{2u(t)}, 
	\label{eq:Vr}
\end{equation}
then in the frame $| \widetilde{\psi} (t) \rangle = V_\eta(t)^{-1} | \psi(t) \rangle$ the transformed Hamiltonian becomes: 
\begin{align}
	{\cal H}'(t) &= V_\eta(t)^{-1} {\cal H}(t) V_\eta(t) - i \, V_\eta(t)^{-1} \dot{V}_\eta(t)  \nonumber \\
	&=\frac{1}{\eta^2(t)}H
	+\eta(t)\left[\ddot\eta(t)+\Omega^2(t)\eta(t)\right]C, 
	\label{eq:Atilderho}
\end{align}
where $\Omega_u^2(t)
=
u(t)w(t)-\delta_u^2(t)-\dot\delta_u(t)$. Therefore, if we solve $\eta(t)$ such that 
\begin{align}
		\ddot{\eta}+\Omega^2(t)\eta=\frac{W^2}{\eta^3} \label{eq:EP}
\end{align}
then the transformed Hamiltonian becomes: 
\begin{align}
	{\cal H}'(t) &= \eta(t)^{-2} \left( H + W^2 C \right)
\end{align}
so that if we introduce the Ermakov time
\begin{equation}
	\tau(t)-\tau(t_0)=\int_{t_0}^{t}\frac{\dd s}{\eta^2(s)}
	\label{eq:tau}
\end{equation}
then the Schr\"odinger equation simply becomes
\begin{equation}
	\ii\frac{\partial}{\partial\tau}|\widetilde\psi\rangle
	=\left(H+W^2\, C\right)|\widetilde\psi\rangle.
	\label{eq:refeq}
\end{equation}
Note that in the time-independent frame, $W^2$ is exactly the Casimir invariant. Therefore from Eq.\eqref{eq:EP} the nature of the Casimir characterizes the nature of the EP variable $\eta(t)$ and the hyperbolic, elliptic and parabolic classes translate to unstable, stable and marginal solutions.  The equation \eqref{eq:EP} is called the Ermakov-Pinney equation  \cite{Pinney1950}. 
Now if we transform the time-independent piece back using the same unitary, i.e., 
\begin{align}
H + W^2 \, C &\rightarrow	V_\eta(t) \left( H + W^2 \, C \right) V_\eta(t)^{-1} \equiv I(t)
\end{align}
and then $I(t)$ turns out to be the Lewis invariant \cite{Lewis1968,LewisRiesenfeld1969} of the original frame as long as Eq.\eqref{eq:EP} is satisfied. In the Ermakov frame $H+ W^2 \,C$ itself is the Lewis invariant. 
\noindent

Next, we specialize to the case of the time-dependent trap $H+\mu^2(t)C$ which is relevant for cold-atom experiments discussed in the End Matter of the main text.  Let $\Psi_0(\{\mathbf r_i\})$ be an eigenstate of the reference trapped Hamiltonian $H+\mu_0^2 C$ with energy $E$.  The evolution operator for $\psi(t)$ now is:
\begin{align}
U(t) &= e^{ i \frac{\dot {\eta}}{\eta } C } e^{ -i (\ln \eta) D} e^{ - i \tau(t) (H + \mu_0^2 \, C ) }. 
\end{align}	
where $\tau(t)$ is the EP time of Eq.\eqref{eq:tau}. When generators have coordinate representations of the form as in Eq. (20) of \cite{NRFloquet} then the exact evolved $N$-body wavefunction in $d$ spatial dimensions \cite{Castin2004,werner} is
therefore
\begin{equation}
	\Psi(\{\mathbf r_i\},t)
	=
	\frac{1}{\eta^{Nd/2}(t)}
	\exp\!\left[
	\frac{i \dot\eta(t)}{2\eta(t)}
	\sum_{i=1}^{N}\mathbf r_i^2
	-iE\,\tau(t)
	\right]
	\Psi_0\!\left(
	\left\{\frac{\mathbf r_i}{\eta(t)}\right\}
	\right).
	\label{eq:SI-manybody-wavefunction}
\end{equation}
Thus, once the solution $\eta(t)$ of the Ermakov--Pinney equation is known, 
the complete many-body wavefunction is determined in terms of the initial wavefunction.

\section{Geometries with Schr\"{o}dinger Isometry}
\label{sec:SI-holographic}

In this section we study geometries with Schr\"{o}dinger Isometry. To this end, we first note that an infinite family of Schr\"{o}dinger geometries can also be obtained by solving such ten or eleven-dimensional supergravity equations of motion. The advantage of this construction is that, we can identify the precise dual quantum field theory, since the description is rooted in a UV-complete picture in string theory. The simplest example of this class is perhaps the discrete light-cone quantization of the dipole-deformed ${\cal N}=4$ super Yang-Mills (SYM) theory, in which the ordinary products of quantum fields are lifted to a star-product structure and a compactification on a null direction is performed. The full conformal symmetry of the ${\cal N}=4$ SYM theory is broken down to the $(3+1)$-dimensional Schr\"{o}dinger symmetry under the dipole deformation and the null-reduction provides us with a conserved charge, which is the central element ({\it i.e.}~the total particle number) of the Schr\"{o}dinger algebra. Other such examples fall under a similar identification structure: deformation of a well-understood relativistic conformal theory.

Strong coupled non-relativistic systems with a large number of degrees of freedom are conjectured to be holographically dual to asymptotically Schr\"{o}dinger symmetric geometries, which can be obtained as solutions of both bottom-up Einstein-gravitional theory with an appropriate matter field, or from a ten/eleven-dimensional supergravity theory. A $(d+1)$-dimensional Schr\"{o}dinger-invariant theory is dual to a $(d+3)$-dimensional bulk geometry of the form:
\begin{eqnarray}
    ds^2 = L^2 \left( - \beta^2 \frac{dt^2}{r^{2z}} + \frac{2 dt d\xi + dr^2 + d\vec{x}^2}{r^2}\right) \ , \label{schrmet}
\end{eqnarray}
where $r$ is the holographic radial coordinate, $\vec{x} \in {\mathbb R}^d$ is the spatial $d$-dimensional vector, $\xi$ is a null coordinate which is related to the central extension of the Schr\"{o}dinger algebra. The length-scale, $L$, corresponds to the curvature of the geometry, and the dynamical exponent, $z=2$. 

The conformal boundary, which is parametrized by $\{ t, \vec{x}, \xi\}$, is located at $r \to 0$. The boundary is defined in terms of the Newton-Cartan data consisting of a non-degenerate one-form, which is physically equivalent to a clock; a non-degenerate spatial metric, which defines the spatial manifold on which the Schr\"{o}dinger-invariant system is defined and a U$(1)$-connection which physically corresponds to the particle number symmetry.  

The corresponding Killing vectors of the Schr\"{o}dinger geometry can be obtained by explicitly embedding the $(d+1)$-dimensional Schr\"{o}dinger algebra into the $so(d+1,2)$ algebra, which is the conformal algebra for a relativistic system in $(d+2)$-dimensions. The Killing vectors that generate the $sl(2, {\mathbb R})$ sub-algebra of the full Schr\"{o}dinger algebra can be obtained as\cite{Blau_2009}:
\begin{eqnarray}
    {\cal K}_D = - i \left(2 t \partial_t + \vec{x}\partial_{\vec{x}} + r \partial_r \right) \ , \quad {\cal K}_H = - i \partial_t \ , \quad {\cal K}_C = - i \left(t^2 \partial_t + t \vec{x} \partial_{\vec{x}} + t r \partial_r - \frac{r^2 + \vec{x}^2}{2} \partial_\xi \right)  \ ,
\end{eqnarray}
which satisfy the algebra:
\begin{eqnarray}
    \left[{\cal K}_D, {\cal K}_H \right] = 2 i {\cal K}_H \ , \quad  \left[{\cal K}_D, {\cal K}_C \right] = -2 i {\cal K}_C  \ , \quad \left[{\cal K}_H, {\cal K}_C \right] = - i {\cal K}_D \ ,
\end{eqnarray}
which is the $sl(2,{\mathbb R})$ algebra. Given this, following the treatment in \cite{hd2}, let us calculate the norm of the general Killing vector: ${\cal K}_X = \alpha_+ {\cal K}_H + \alpha_- {\cal K}_C + \gamma {\cal K}_D$. This yields:
\begin{eqnarray}
||\mathcal{K}_X\|^2
=
\frac{L^2}{r^4}
\left[
-\beta^2
\left(\alpha_{+}+2\gamma t+\alpha_{-}t^2\right)^2
+
\left(\gamma^2-\alpha_{+}\alpha_{-}\right)
r^2\left(x^2+r^2\right)
\right] \ .
\end{eqnarray}
The first term above is always negative, and therefore, to have a real root of $||\mathcal{K}_X\|^2=0$, we must require: $\alpha_+ \alpha_- - \gamma^2 < 0$. It is actually easy to show that this is both a necessary and a sufficient condition for the existence of a root. The algebraic equation essentially becomes:
\begin{eqnarray}
   \left(\alpha_{+}+2\gamma t+\alpha_{-}t^2\right)^2 = (\gamma^2 - \alpha_+ \alpha_-)r^2 (x^2 + r^2) \ . \label{solkill}
\end{eqnarray}
Since both LHS and RHS are positive definite, and this is a quadratic equation for $r^2$, it is easy to see that this equation always has two real roots for r: one positive and one negative. The positive root is the physical one.

Note that a solution to the equation (\ref{solkill}) yields a hypersurface on which the Killing vector ${\cal K}_X$ becomes null. Therefore, it divides the geometry into two regions: one in which ${\cal K}_X$ generates time-like flows and the other in which space-like flows are generated by it. To further check whether the Killing norm vanishing hypersurface is null, we need to evaluate the norm of the unit-normal to this hypersurface. Towards that, let us write down the equation of the hypersurface as:
\begin{eqnarray}
    F(t, r, \vec{x}) = 0 \quad \implies \quad dF = n_\mu dx^\mu \ ,
\end{eqnarray}
where $n_\mu$ defines the components of the normal to this hypersurface. An explicit calculation yields:
\begin{eqnarray}
&&    n_t = - 2 f(t) f'(t) \ , \quad f(t) = \alpha_+ + 2 \gamma t + \alpha_- t^2 \ , \\
&& n_{\vec{x}} = 2 \left(\gamma^2 - \alpha_+ \alpha_- \right) r^2 \vec{x} \ , \quad n_r = 2 r\left(\gamma^2 - \alpha_+ \alpha_- \right) \left( \vec{x}^2 + 2 r^2 \right) \ . 
\end{eqnarray}
Now, using the metric in (\ref{schrmet}), the norm of this vector is obtained to be:
\begin{eqnarray}
    n^2 = g^{\mu\nu} n_\mu n_\nu = 4 \left(\gamma^2 - \alpha_+ \alpha_- \right)^2 r^4 \left[r^2 \vec{x}^2 + \left( \vec{x}^2 + 2 r^2 \right)^2 \right] \ . 
\end{eqnarray}
Therefore, in the hyperbolic class, when $\left(\gamma^2 - \alpha_+ \alpha_- \right)>0$, we obtain $n^2>0$. Hence, the locus of the vanishing norm in (\ref{solkill}) is a time-like hypersurface. Since ${\cal K}_X$ is a natural time-like Killing vector from the perspective of the CFT-evolution, the locus of the vanishing norm, in the bulk, defines an ergosurface. It is also clear from this expression that at $\gamma^2 = \alpha_+ \alpha_-$, {\it i.e.}~for the parabolic class, the norm vanishing hypersurface, provided it exists, indeed becomes null. For the elliptic case, on the other hand, the Killing norm never vanishes.

At the critical line, $\alpha_+\alpha_- = \gamma^2$, the norm is given by
\begin{eqnarray}
   || \mathcal{K}_X\|^2 = - \frac{L^2}{r^4}(\beta^2\alpha_-^2)\left( t- t_0\right)^4 \ , \quad t_0 = - \frac{\gamma}{\alpha_-} \ . 
\end{eqnarray}
Therefore at $t=t_0$, the Killing norm vanishes and yields a null surface and hence a Killing horizon. To further see whether this horizon is extremal, we can calculate the surface gravity on the horizon. Imposing $\gamma^2 = \alpha_+ \alpha_-$, the Killing vector ${\cal K}_X$ takes a simple form:
\begin{eqnarray}
   \left. {\cal K}_X \right|_{t=t_0} = - \frac{\alpha_-}{2}\left(r^2 + \vec{x}^2 \right) \partial_\xi \equiv Q(x, \xi) \partial_\xi \ .\label{kxhor}
\end{eqnarray}
Now, surface gravity is obtained by computing how a Killing vector fails to be affinely parametrized at the horizon. Physically, this corresponds to the local acceleration experienced at the horizon. Denoting the surface gravity by $\kappa$, this is obtained from:
\begin{eqnarray}
    K^\nu \nabla_\nu K^\mu = \kappa K^\mu \ . 
\end{eqnarray}
Using the form of ${\cal K}_X$ at the horizon in (\ref{kxhor}), it is easy to see that:
\begin{eqnarray}
    {\cal K}_X^\nu \nabla_\nu {\cal K}_X = Q \left( \partial_\xi Q\right) \partial_\xi + Q^2 \nabla_{\partial_\xi} \partial_\xi \ .
\end{eqnarray}
Now, since $\partial_\xi Q = 0$ and $\nabla_{\partial_\xi}\partial_\xi = \Gamma_{\xi\xi}^\mu \partial_\mu$, where, the geometry in (\ref{schrmet}) yields $\Gamma_{\xi\xi}^\mu=0$, we immediately obtain that $\kappa=0$ on the $t=t_0$ null surface. Therefore, the parabolic case indeed corresponds to an extremal Killing horizon.

\bibliography{nrfl}